\documentclass[final,3p,times]{elsarticle}
\usepackage{url}
\usepackage[utf8]{inputenc}

\usepackage{amssymb}
\usepackage{amsmath}
\usepackage{hyperref} 
\usepackage[most]{tcolorbox}
\usepackage{booktabs}
\usepackage{array}
\usepackage{wasysym}
\newcommand{\cvF}{\CIRCLE}
\newcommand{\cvH}{\LEFTcircle}
\newcommand{\cvN}{\Circle}
\usepackage[figuresright]{rotating}
\usepackage{pdflscape}

\hypersetup{
    colorlinks=true,        
    linkcolor=blue,         
    citecolor=blue,          
    urlcolor=purple         
}

\usepackage{booktabs}
\usepackage{pifont}   
\usepackage{array}

\newtcolorbox{rqbox}{colback=gray!10, colframe=gray!10, boxrule=0pt, arc=3mm, left=4mm, right=4mm, top=3mm, bottom=3mm}

\newcolumntype{C}[1]{>{\centering\arraybackslash}p{#1}}

\journal{arXiv}

\begin{document}

\begin{frontmatter}



\title{Integrating Machine Learning and Uncertainty Quantification for Real-Time Decision-Making in Digital Twins: A System-Centric Review and Perspective} 


\author[NU]{Yi-Ping Chen}
\author[NU,TXST]{Ying-Kuan Tsai}
\author[NU]{Seul Lee}
\author[NU]{Vispi Karkaria}
\author[NU]{Dex Doksoo Lee}
\author[NU]{Wei Chen\corref{cor1}}

\affiliation[NU]{organization={Department of Mechanical Engineering, Northwestern University },
            addressline={2145 Sheridan Rd.}, 
            city={Evanston},
            postcode={60208}, 
            state={Illinois},
            country={USA}}

\affiliation[TXST]{organization={Ingram School of Engineering, Texas State University },
            addressline={601 University Dr.}, 
            city={San Marcos},
            postcode={78666}, 
            state={Texas},
            country={USA}}

\cortext[cor1]{Corresponding Author. email: weichen@northwestern.edu}
\begin{abstract}

Digital Twins (DT) are evolving from passive virtual representations into closed-loop systems that continuously synchronize with physical assets and drive real-time decisions. Yet predictive modeling, machine learning (ML), uncertainty quantification (UQ), and decision-making have largely been studied as separate capabilities, leaving limited understanding of how they should be integrated and maintained during online operation. This narrative review, scoped by the functional requirements of real-time decision-making rather than by Digital Twin terminology alone, examines how such decisions are generated and sustained in engineering Digital Twins. We first examine foundational decision-making architectures and representative engineering applications, followed by methods that address their limitations across the decision loop. We identify optimization-based, policy-based, and knowledge-based approaches as the major pathways from virtual representation to action. Across them, ML plays three roles in decision-making: enabling efficient surrogate modeling, transforming heterogeneous physical feedback into decision-relevant information, and accelerating or directly generating decisions. UQ supports uncertainty-aware decisions and model adaptation, while continual validation, drift detection, and adaptation help maintain reliable operation as the physical system evolves. Together, these findings underscore the need for a shift from a model-centric to a system-centric perspective on Digital Twins, where its performance depends on the interactions with sensing, state estimation, computation, decision-making, and adaptation rather than on its standalone performance. Therefore, future research should shift from isolated algorithm development toward system-level co-design, scalable decision-making, and approaches to design, monitor, and data-efficient continual adaptation and validation for AI-enabled autonomous systems. (See the \href{https://yiping514.github.io/ML_UQ_DT/#structure}{knowledge graph} to this manuscript.)

\end{abstract}




\end{frontmatter}

\tableofcontents

\section{Introduction}

\subsection{Digital Twin: Evolving Toward Autonomous Decision-Making}

The Digital Twin (DT) paradigm has emerged as a transformative framework for digital and autonomous engineering. The concept was introduced by Grieves in the early 2000s in the context of product lifecycle management~\cite{grieves2016digital}. One influential early definition was later articulated by NASA as \emph{an integrated multi-physics, multiscale, probabilistic simulation of an as-built vehicle or system that uses the best available physical models, sensor updates, fleet history, and so on, to mirror the life of its corresponding flying twin} \cite{glaessgen2012digital}. Since then, the concept has broadened considerably and has been applied across a wide range of engineering activities, including simulation-based design, data-driven design, system monitoring, prediction, and control. While this expansion reflects the growing scope of Digital Twin technologies, it has also blurred the boundary of what constitutes a Digital Twin. In particular, it remains unclear what fundamentally distinguishes a Digital Twin from simulation, digital models, monitoring systems, and other cyber-physical frameworks, and what unique role Digital Twins can play in future engineering systems.

\begin{figure}[h!]
    \centering
    \includegraphics[width=0.8\linewidth]{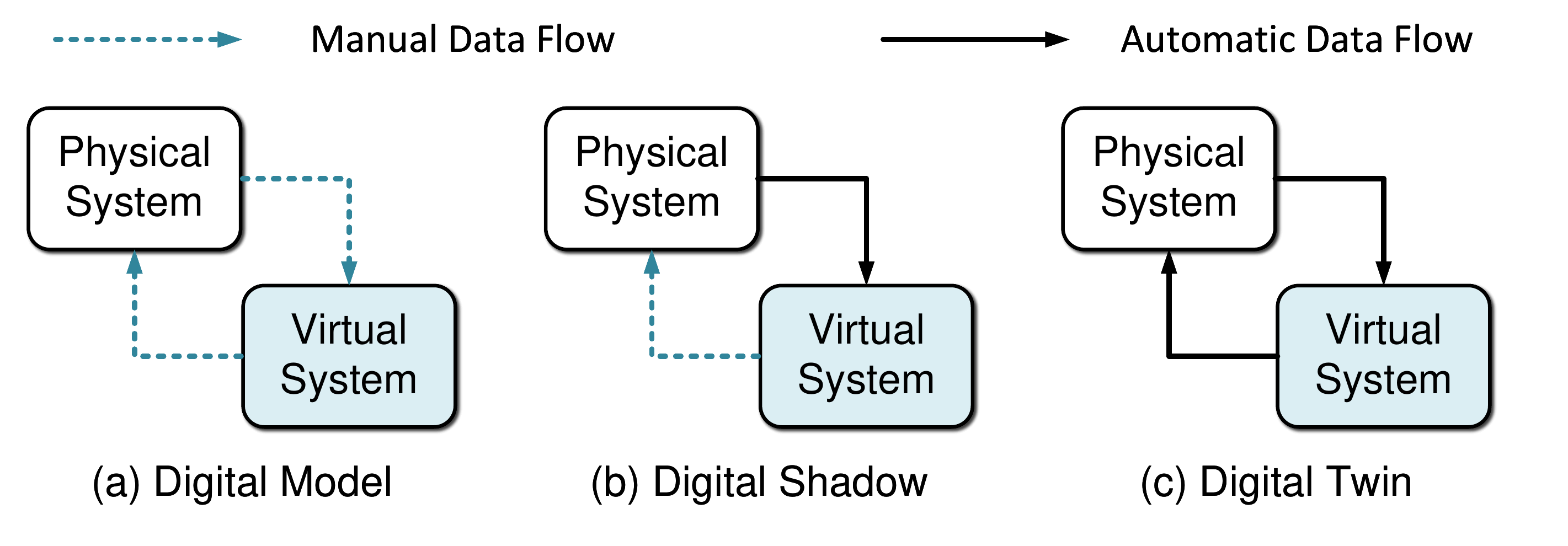}
    \caption{Types of digital model integrations, following~\cite{kritzinger2018digital}. (a) Data in a Digital Model is exchanged manually because changes in the physical system's state are not directly mirrored in the virtual model, and vice versa. (b) Data flows automatically from the physical object to the digital, but manual updates are needed in the opposite direction, so changes in the physical object appear in the digital copy, but not the reverse. (c) There is an automatic bidirectional data flow between the physical and digital objects, so changes in either one directly trigger updates in the other.}
    \label{fig:dt_def}
\end{figure}

To establish a consistent scope for this review, we adopt the definition proposed by the National Academies of Sciences, Engineering, and Medicine, which defines a Digital Twin as \emph{"a set of virtual information constructs that mimics a natural, engineered, or social system. It is dynamically updated by its physical counterpart, possesses predictive power, and drives decisions to create real-world value through two-way interactions."}~\cite{national2023foundational} This definition distinguishes a Digital Twin from a static digital model by emphasizing a closed information and decision loop between the physical and digital systems: observations from the physical system continuously update the Digital Twin, its computational models predict system behavior, and these predictions inform decisions that can affect subsequent physical-system operation~\cite{national2023foundational, galvez2025decision}. As illustrated in Figure~\ref{fig:dt_def}, this bidirectional loop is what separates a Digital Twin from a digital model, where data are exchanged manually, and from a digital shadow, where data flow automatically only from the physical to the virtual system~\cite{kritzinger2018digital}. In essence, a Digital Twin integrates physical-system data with computational models, including physics-based simulations, data-driven surrogates, and hybrid representations, not only to forecast system evolution and explore what-if scenarios, but ultimately to support decisions that create real-world value~\cite{thelen2022comprehensive}.

Viewed from this perspective, the distinguishing potential of Digital Twins lies in transforming computational models from passive analytical tools into operational instruments that remain synchronized with evolving physical systems and continuously support decision-making, with machine learning (ML) and artificial intelligence (AI) as the major enablers that make real-time applications possible. With applications ranging from manufacturing~\cite{karkaria2025optimization, Lee_2026} and healthcare~\cite{vallee2023digital} to urban and transportation systems~\cite{jafari2023review}, Digital Twins have enabled monitoring, prediction, optimization, and control across different engineering scales, with potential benefits in performance, reliability, resilience, and resource efficiency~\cite{es2024methods}.

Realizing this potential, however, requires more than constructing an accurate virtual representation. A useful distinction is between offline and online functions of the Digital Twin~\cite{karkaria2025optimization, grieves2016digital}. \emph{Offline} functions are performed outside system operation, such as design optimization, model training, and calibration, whereas \emph{online} functions use current system information to update the Digital Twin and support or generate operational decisions without interrupting system operation. Real-time operation therefore requires these online functions to be completed within the actionable time scale of the physical system. Following~\cite{thelen2023comprehensive}, the term \emph{real-time} is defined as \emph{the minimum computational speed required to achieve seamless and uninterrupted optimization, prediction, and control of the system of interest}. From this perspective, a key role that distinguishes the Digital Twin from previous paradigms is to continuously synchronize with the physical system, rapidly evaluate possible outcomes, and translate this information into timely decisions throughout the system life cycle. The required computational speed is ultimately determined by the time scale of the system being controlled.

Accordingly, constructing an online Digital Twin for real-time decision-making introduces requirements beyond those of offline applications~\cite{thelen2022comprehensive, zhang2025probabilistic}. The challenge is not only to ensure that the virtual model can be evaluated fast enough within the decision-making algorithm, but also to process system feedback, estimate decision-relevant states, account for uncertainty, and return actions to the physical system within computational constraints. In addition, the Digital Twin must remain synchronized with the evolving physical system throughout deployment, requiring continual uncertainty assessment, validation, and model adaptation. These requirements make online Digital Twin development a system design problem in which sensing, state estimation, predictive modeling, uncertainty quantification (UQ), decision synthesis, and adaptation must operate coherently to support reliable decisions over the system life cycle. We refer to this perspective as \emph{system-centric}: each method is assessed not in isolation, but by the requirements it imposes on, and the value it contributes to, and the closed-loop system as a whole.

\subsection{Existing Work}

Advances in ML and AI have accelerated the growth of engineering applications under the umbrella of Digital Twins, and the field has accordingly produced numerous review papers, each emphasizing different aspects. Rather than attempting an exhaustive survey of all Digital Twin reviews, Table~\ref{tab:review_comparison} summarizes representative reviews based on their relevance to the methodological scope of this review, including Digital Twin modeling, ML integration, decision-making, and continual adaptation. 

The review papers can be roughly divided into three categories. First, general surveys provide broad overviews of Digital Twin trends, enabling technologies, and industrial applications~\cite{tao2018digital, kritzinger2018digital, fuller2020digital, mihai2022digital, ai2025advances}, and complementary community roadmaps have articulated future directions for Digital Twin research and development spanning verification and validation, AI, and cyber-physical system design~\cite{vanbossuyt2025future}. A second body of work narrows onto categorizing methods and explaining the taxonomies of particular functions or entities of Digital Twins. Some reviews focus on surrogate and reduced-order modeling for real-time analysis~\cite{thelen2022comprehensive, es2024methods, li2024digital}, optimization~\cite{karkaria2025optimization}, UQ~\cite{thelen2023comprehensive}, and simulation-based decision support through discrete-event and agent-based modeling~\cite{dos2022decision, galvez2025decision}. Other works review Digital Twin applications in specific domains, such as building operation~\cite{cespedes2024review}, transportation~\cite{jafari2023review}, and healthcare~\cite{sun2023digital, vallee2023digital}, or Digital Twins for predictive maintenance~\cite{chen2023advance}. More recently, a third category has examined the integration of ML and AI into Digital Twins, ranging from broad surveys of the role of AI, ML, and big data in digital twinning~\cite{rathore2021role, kreuzer2024artificial, huang2021survey} to reviews centered on reinforcement learning~\cite{mohammed2026digital}, large language models (LLMs), and agentic frameworks~\cite{sajadieh2025simulation, liu2025survey}.

\begin{sidewaystable}[p]
\centering
\caption{Comparison of related Digital Twin reviews with this work.}
\label{tab:review_comparison}

\footnotesize
\renewcommand{\arraystretch}{1}
\setlength{\tabcolsep}{3pt}

\begin{tabular}{
>{\raggedright\arraybackslash}p{0.1\linewidth}
*{6}{>{\centering\arraybackslash}p{0.07\linewidth}}
>{\raggedright\arraybackslash}p{0.4\linewidth}
}
\toprule

\textbf{Reference}
& \textbf{Surrogate Modeling}
& \textbf{Perception \& Reasoning}
& \textbf{Decision Making}
& \textbf{Uncertainty Quantification}
& \textbf{Continual Validation \& Adaptation}
& \textbf{Real-Time Decision-Making}
& \textbf{Primary Focus}
\\

\midrule

\multicolumn{8}{l}{\textit{\textbf{General and conceptual surveys}}} \\
\addlinespace[2pt]

Kritzinger et al.~\cite{kritzinger2018digital} (2018)
& \cvN & \cvN & \cvN & \cvN & \cvN & \cvH
& Distinguishes digital models, digital shadows, and Digital Twins according to the level and direction of physical--digital data integration.
\\

Tao et al.~\cite{tao2018digital} (2019)
& \cvH & \cvH & \cvH & \cvN & \cvN & \cvH
& Reviews the state of Digital Twins in industry, including enabling technologies, implementation, and applications.
\\

Fuller et al.~\cite{fuller2020digital} (2020)
& \cvH & \cvH & \cvH & \cvH & \cvN & \cvH
& Surveys enabling technologies, applications, challenges, and open research directions for Digital Twins across domains.
\\

Hu et al.~\cite{hu2021digital} (2021)
& \cvH & \cvH & \cvH & \cvN & \cvN & \cvH
& Reviews Digital Twin definitions, enabling technologies, lifecycle applications, and interactions between engineered systems and their environments.
\\

Mihai et al.~\cite{mihai2022digital} (2022)
& \cvH & \cvH & \cvH & \cvN & \cvN & \cvH
& Provides a broad survey of Digital Twin enabling technologies, design objectives, applications, challenges, and future trends.
\\

\addlinespace[4pt]

\multicolumn{8}{l}{\textit{\textbf{Computational and modeling method reviews}}} \\
\addlinespace[2pt]

Rasheed et al.~\cite{rasheed2020digital} (2020)
& \cvF & \cvH & \cvH & \cvH & \cvH & \cvH
& Examines how Digital Twins should be constructed from a modeling perspective, including reduced-order modeling and data assimilation.
\\

Thelen et al.\ Part~1~\cite{thelen2022comprehensive} (2022)
& \cvF & \cvF & \cvH & \cvH & \cvH & \cvH
& Reviews Digital Twin modeling and twinning technologies across physical-to-virtual and virtual-to-physical processes.
\\

Thelen et al.\ Part~2~\cite{thelen2023comprehensive} (2023)
& \cvH & \cvH & \cvF & \cvF & \cvH & \cvH
& Examines how UQ and optimization support modeling, data acquisition, and predictive decision-making in Digital Twins.
\\

Es-haghi et al.~\cite{es2024methods} (2024)
& \cvF & \cvN & \cvH & \cvN & \cvN & \cvF
& Examines how reduced-order and machine-learning models can reduce computational cost for real-time Digital Twin analysis.
\\

\addlinespace[4pt]

\multicolumn{8}{l}{\textit{\textbf{AI/ML integration reviews}}} \\
\addlinespace[2pt]

Rathore et al.~\cite{rathore2021role} (2021)
& \cvH & \cvH & \cvH & \cvN & \cvN & \cvH
& Reviews the roles of AI, ML, and big data as enabling technologies for digital twinning.
\\

Huang et al.~\cite{huang2021survey} (2021)
& \cvH & \cvH & \cvH & \cvN & \cvH & \cvH
& Surveys AI-enabled Digital Twins for smart manufacturing and robotics, including learning, optimization, control, and reinforcement learning.
\\

Liu et al.~\cite{liu2025survey} (2025)
& \cvN & \cvF & \cvF & \cvH & \cvH & \cvH
& Reviews Cognitive Digital Twins and the role of LLMs in perception, knowledge integration, reasoning, and autonomous decision-making.
\\

\addlinespace[4pt]

\multicolumn{8}{l}{\textit{\textbf{Decision- and application-oriented reviews}}} \\
\addlinespace[2pt]

Galvez et al.~\cite{galvez2025decision} (2025)
& \cvH & \cvN & \cvF & \cvN & \cvH & \cvH
& Examines decision support within Digital Twins for manufacturing systems, including feedback, autonomy, and real-time operation.
\\

Chen et al.~\cite{chen2023advance} (2023)
& \cvH & \cvF & \cvH & \cvH & \cvH & \cvH
& Reviews how ML and Digital Twins support condition monitoring, diagnosis, prognosis, and predictive maintenance.
\\

Vall\'ee~\cite{vallee2023digital} (2023)
& \cvH & \cvH & \cvH & \cvH & \cvN & \cvH
& Reviews healthcare Digital Twins for monitoring, diagnosis, prediction, treatment, and personalized decision support.
\\

Karkaria et al.~\cite{karkaria2025optimization} (2025)
& \cvF & \cvH & \cvF & \cvH & \cvH & \cvF
& Examines how AI and optimization can enhance modeling, monitoring, optimization, and control in manufacturing Digital Twins.
\\

\addlinespace[4pt]

\multicolumn{8}{l}{\textit{\textbf{System-oriented reviews}}} \\
\addlinespace[2pt]

VanDerHorn \& Mahadevan~\cite{vanderhorn2021digital} (2021)
& \cvH & \cvH & \cvH & \cvH & \cvH & \cvH
& Develops a general characterization and implementation framework for Digital Twins, including physical-to-virtual and virtual-to-physical interactions.
\\

Kreuzer et al.~\cite{kreuzer2024artificial} (2024)
& \cvH & \cvH & \cvH & \cvN & \cvH & \cvF
& Systematically examines AI-integrated Digital Twin implementations and whether they achieve physical-to-virtual and virtual-to-physical connectivity.
\\

Sajadieh \& Noh~\cite{sajadieh2025simulation} (2025)
& \cvH & \cvH & \cvF & \cvH & \cvH & \cvF
& Reviews how AI can advance Digital Twins from simulation and monitoring toward decision-making, continuous learning, and autonomy.
\\

\midrule

\textbf{This work} (2026)
& \cvF
& \cvF
& \cvF
& \cvF
& \cvF
& \cvF
& \textbf{Examines how real-time decisions are generated and sustained throughout the Digital Twin decision loop by integrating perception, predictive modeling, decision synthesis, uncertainty management, continual validation, and adaptation.}
\\

\bottomrule
\end{tabular}

\vspace{5pt}

\begin{minipage}{0.96\linewidth}
\scriptsize
\cvF~Dedicated or systematic treatment;
\quad
\cvH~substantive but secondary treatment;
\quad
\cvN~not substantively addressed.

Review categories indicate the primary perspective of each review; some reviews span more than one category.
\end{minipage}

\end{sidewaystable}

Several recent reviews overlap with the scope outlined in Section~1.1, but none centers on how ML-enabled components work together to support online, decision-making in Digital Twins. Thelen et al. provide the most comprehensive method-level review of Digital Twin enabling technologies, covering modeling and physical-to-virtual and virtual-to-physical twinning in Part~I~\cite{thelen2022comprehensive} and UQ and optimization in Part~II~\cite{thelen2023comprehensive}. Although they also frame the Digital Twin as a closed loop, the review is organized around enabling technologies and modeling methods, with UQ focused primarily on model uncertainty and calibration and ML treated mainly as a modeling approach rather than from the perspective of system integration. Sajadieh and Noh examine the progression of Digital Twins toward autonomy by organizing AI contributions across life-cycle phases, but their application-level classification does not examine method selection or the technical realization within decision-making~\cite{sajadieh2025simulation}. Kreuzer et al. support the idea of the needing more emphasis on real-time decision-making from a review perspective, showing that fewer than 10\% of the existing studies on Digital Twins demonstrate a virtual-to-physical feedback loop, but their review primarily records whether such integration exists rather than technical components~\cite{kreuzer2024artificial}.

Apart from Thelen et al.~\cite{thelen2022comprehensive, thelen2023comprehensive}, the existing review closest to our motivation is that of G\'alvez del Postigo Gallego et al.~\cite{galvez2025decision}, who regard decision support as integral to a Digital Twin and as what closes its loop. Their review approaches decision-making from an integration and implementation standpoint. It examines whether feedback reaches the physical system to realize the bidirectional interaction, through which industrial interfaces it is delivered, and whether a human or the system acts on it, and it classifies manufacturing applications by the type of decision supported. While it identifies interoperability, latency, and the computational cost of intensive simulations as key barriers to real-time feedback, it leaves room for further examination of the methods by which decision support systems can be constructed under these constraints. Likewise, ML appears mainly through applications such as scheduling and recommendation via reinforcement learning, without examining what capability it provides within the decision loop. Finally, decision support is defined as aiding human decision-makers, with automated feedback treated as a largely unrealized goal, and trust is framed as operators' willingness to act on recommendations and the interpretability of simulation outputs. The technical reliability of decisions in online operation as the physical system changes remains outside its scope.

Across these reviews, three gaps remain:

\begin{enumerate}
    \item \textbf{Real-time decision-making has not been examined from a method-level and system-level perspective.} A broader collection of decision-making approaches is needed to clarify how the virtual-to-physical connection can be established under different system requirements and operating conditions. Moreover, although the components of a Digital Twin share a common computational budget and errors can propagate across them once integrated, these components are typically developed and evaluated in isolation. For example, surrogate models are commonly benchmarked on offline predictive accuracy rather than on their contribution to downstream decisions.

    \item \textbf{The role of ML across the Digital Twin decision loop remains narrowly examined.} Existing reviews have largely focused on surrogate modeling and on introducing individual methods~\cite{rasheed2020digital, thelen2022comprehensive, es2024methods}, with less attention to other functions ML can serve in the decision loop, such as state estimation and decision synthesis. The practical considerations involved in selecting, constructing, validating, and maintaining ML models that continuously interact with physical systems have also received limited attention. Meanwhile, recent advances from predictive models toward generative, knowledge-enabled, and agentic AI create new opportunities for expanding the capabilities of Digital Twins, including their role in decision-making.

    \item \textbf{The operational role of UQ in online Digital Twins applications remains insufficiently examined.} Although UQ methods have been discussed in existing reviews, greater attention is needed to how UQ supports uncertainty-aware decision-making, continual validation, and model adaptation, as well as how ML models can be efficiently updated during operation when they become the building blocks for Digital Twins.
\end{enumerate}

These gaps motivate a distinct review centered on online Digital Twins, with real-time decision-making as its main focus and ML as an enabling technology for autonomous and reliable closed-loop operation.

\subsection{Contributions}

In this work, we examine practical considerations and potential solutions for integrating surrogate modeling, ML, and UQ into real-time decision-making for Digital Twins. The review is decision-centric in its goal and system-centric in its analysis: rather than treating Digital Twin components individually, it examines how they interact as an integrated decision-making system throughout online operation.

This review builds on several complementary foundations in the existing literature rather than providing another comprehensive survey of Digital Twins. As a continuation of our previous work~\cite{karkaria2025optimization}, which highlighted the need to integrate decision-making, UQ, and adaptation in AI-enabled Digital Twins, in this work, we develop this perspective into a broader method-level synthesis. Building on the comprehensive review by Thelen et al. \cite{thelen2022comprehensive, thelen2023comprehensive}, where individual methods are introduced in detail, we extend the scope to examine practical considerations for integrating these components from a system perspective. Finally, this review complements the perspective of Galvez del Postigo Gallego et al.~\cite{galvez2025decision} by providing a more detailed method-level and system-centric review of how decision support can be realized in Digital Twins.

Accordingly, the main contributions of this review are threefold:
\begin{enumerate}
    \item \textbf{A method-level review of real-time decision-making for Digital Twins from a system perspective.} We categorize the major decision-making approaches used to close the physical-to-virtual-to-physical loop and identify the requirements they impose on sensing, state estimation, predictive modeling, computation, and feedback. This provides a system perspective of how these components must work together to support real-time decision-making.

    \item \textbf{An expanded view of ML across the Digital Twin decision loop.} Beyond its established role in surrogate modeling, we examine how ML supports perception, reasoning, and decision synthesis, including emerging generative models and agentic AI, along with the practical considerations for deploying these models in continuous interaction with physical systems. This perspective positions ML as an enabling technology across multiple stages of the decision loop, assessed by its effect on downstream decisions rather than on isolated predictive accuracy.

    \item \textbf{A lifecycle perspective on Digital Twins in online operation.} We examine how UQ, continual validation, drift detection, and model adaptation work together to assess whether a Digital Twin remains reliable for decision-making and to determine when and how it should be updated as the physical system evolves, treating these mechanisms as interdependent components of the decision loop rather than as standalone analyses.
\end{enumerate}

\subsection{Review Scope and Organization}

This work adopts a narrative review rather than a formal systematic review or bibliometric analysis. The selection criterion on literature is defined by the functional requirements of real-time Digital Twin decision-making rather than by the use of the term \emph{Digital Twin} alone. Many studies labeled as Digital Twins focus on offline analysis, prediction, or domain-specific implementations without a virtual-to-physical decision loop~\cite{kreuzer2024artificial}, while relevant methods often originate from control, optimization, system identification, ML, and UQ without being explicitly framed as Digital Twin research. Restricting the literature to Digital Twin terminology would therefore miss the methodological foundations required by this review.

The literature is selected in two stages. We first consider studies that establish the major decision-making architectures in which virtual models support decision-making algorithms. Foundational studies establish the formulations, assumptions, and capabilities of each method family, and representative engineering applications illustrate their use in Digital Twins. We then examine methods that address their principal limitations in model representation, processing of physical-system feedback, online decision synthesis, uncertainty management, and continual adaptation.  Methods from outside the Digital Twin literature are included when they directly support these decision-making functions. The resulting coverage is intended to be representative rather than exhaustive, with the goal of identifying method capabilities, assumptions, tradeoffs, and system-level dependencies rather than quantifying publication trends.

\begin{figure}[h!]
    \centering
    \includegraphics[width=1\linewidth]{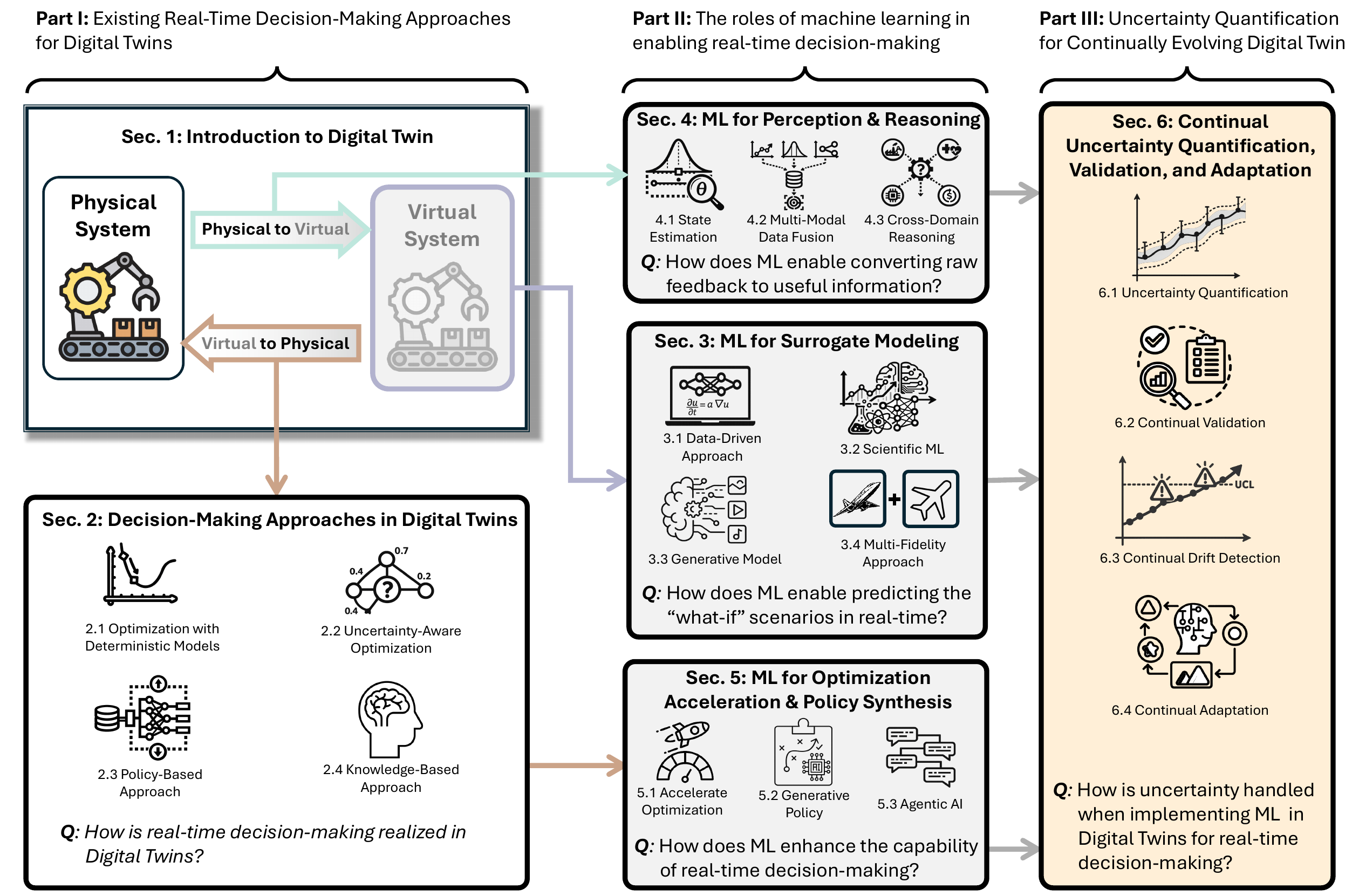}
    \caption{The structure of this paper.}
    \label{fig:structure}
\end{figure}

The remainder of this paper is organized into three parts, as illustrated in Figure~\ref{fig:structure}, each corresponding to one of the contributions in Section~1.3. \textbf{Part~I} establishes the decision-centric foundation of the review. Section~\ref{sec:decision_making} categorizes the major approaches for real-time Digital Twin decision-making, including optimization-based, policy-based, and knowledge-based methods, and identifies their key limitations. \textbf{Part~II} examines how ML can address these limitations across the decision loop. Section~\ref{sec:surrogate} focuses on predicting the time evolution of system responses through surrogate modeling, Section~\ref{sec:perception&reasoning} examines how physical feedback is transformed into decision-relevant information through perception and reasoning, and Section~\ref{sec:ml_decision} considers how ML can accelerate or directly support decision synthesis. \textbf{Part~III} focuses on sustaining the reliability and trustworthiness of these capabilities during operation. Section~\ref{sec:uncertainty} examines UQ, continual validation, drift detection, and adaptation as interconnected mechanisms for responding to uncertainty and lifelong system evolution. Building on these three parts, Section~\ref{sec:future} discusses future research opportunities toward more integrated, adaptive, and trustworthy Digital Twins, and Section~\ref{sec:closure} concludes the paper.

\section{Real-Time Decision-Making Approaches in Digital Twins}
\label{sec:decision_making}

As most of the existing review papers focus on the predictive capability of the Digital Twins rather than the decision-making aspect, we are interested in answering the question, \emph{"How is real-time decision-making realized in Digital Twins?"} In this section, we review and categorize existing methods used for the real-time decision-making of Digital Twins and how the virtual representations of the physical systems are leveraged through a decision-making process to support instant, "fast-enough" decisions. Such methods are categorized into optimization with deterministic models (Section~\ref{sec:deterministic_optimization}) and uncertainty-aware optimization (Section~\ref{sec:opt_with_prob}) where an optimization problem is constantly solved during operation; policy-based approaches (Section~\ref{sec:policy_based}) that map the current state directly into decisions, and knowledge-based approaches (Section~\ref{sec:knowledge-based}) that embed rules or domain knowledge into decision-making processes. In each category, we first introduce the formulation of the decision-making methods, and then provide engineering applications as demonstrations of how these methods are implemented in Digital Twins. Stemming from the existing work, we discuss and conclude how existing decision-making approaches could be limited for online purposes, and thereby discuss the associated implementation challenges and the opportunities for addressing them with ML in Section~\ref{sec:decision_limitations}. These limitations also motivate the problem statement from Section~\ref{sec:surrogate} to \ref{sec:ml_decision} where solutions through ML approaches are reviewed. 

\subsection{Optimization with Deterministic Models}
\label{sec:deterministic_optimization}
Optimization with deterministic models addresses decision-making problems in which a Digital Twin uses a predictive model to evaluate the consequences of candidate actions and select the action that optimizes a prescribed objective under known constraints. Unlike uncertainty-aware optimization, which explicitly represents uncertainty through probabilistic distributions, stochastic transitions, or chance constraints, deterministic optimization relies on a nominal prediction of system behavior and selects decisions based on their expected performance under that forecast. This section first introduces the general formulation of deterministic model-based online optimization in Digital Twins, then introduces model predictive control as a special way of realizing this approach.

\subsubsection{Online Parametric Optimization}

Within deterministic model-based decision-making, online parametric optimization repeatedly solves an optimization problem, whose objective function can be evaluated using the virtual twin that simulates the "what-if" scenarios conditioned on the design and the current state. The relationships between the design variables and the system performance criteria are usually specified explicitly, which may represent process settings, controller parameters, maintenance actions, treatment variables, or resource allocations \cite{ward2021machining, he2019data, chaudhuri2023predictive, vallee2023digital, villalonga2021decision}. As the virtual twin is synchronized with the physical system, iteratively solving the optimization problem provides the best action to reflect the current operating condition. This enables candidate decisions to be evaluated virtually before physical execution and provides a proactive mechanism for selecting actions subject to operational constraints.

A general formulation is
\begin{subequations}
\begin{align}
\min_{\mathbf{u}} & \quad J({\mathbf{x}},\mathbf{u}) \label{eq:det_opt_obj} \\
\mathrm{s.t.} & \quad {\mathbf{x}} = \hat{f}(\mathbf{x}_0,\mathbf{u}) \label{eq:det_opt_model} \\
& \quad \mathbf{g}({\mathbf{x}},\mathbf{u}) \le \mathbf{0} \label{eq:det_opt_ineq} \\
& \quad \mathbf{h}({\mathbf{x}},\mathbf{u}) = \mathbf{0} , \label{eq:det_opt_eq}
\end{align}
\label{eq:det_optimization}
\end{subequations}
where $\mathbf{u}$ denotes the decision variables, $J$ is the objective, $\mathbf{g}$ and $\mathbf{h}$ are inequality and equality constraints, $\mathbf{x}_0$ is the current state, ${\mathbf{x}}$ is the predicted response, and $\hat{f}$ is the model representing the virtual twin. For online purpose where $\hat{f}$ describes the system evolution, commonly as
\begin{equation}
\mathbf{x}_{k+1} = \hat{f}(\mathbf{x}_k,\mathbf{u}_k),
\end{equation}
where the state may include current measurements, input-output histories, or latent variables \cite{thelen2022comprehensive}. The model may come from physics-based simulation, reduced-order modeling, system-identified models, or data-driven surrogates. The Digital Twin receives the current state, evaluates candidate decisions, applies the optimized decision, and repeats the loop as new feedback arrives.

Real-time deterministic optimization has been applied across a broad range of Digital Twin applications. In manufacturing, the Digital Twin can predict process responses under candidate settings and optimize variables such as feed rates, machine parameters, or toolpaths during operation \cite{ward2021machining,niaki2016wavelet}. In feedback systems, the optimized variables may instead be controller gains or other policy parameters, allowing an existing control structure to be adjusted online as operating conditions change \cite{he2019data,lewis2012optimal,viola2021parallel}. Similar formulations have been applied to structural health management, healthcare, and resource allocation, where candidate operational or treatment decisions are evaluated using the current virtual model \cite{he2025digital,chaudhuri2023predictive,vallee2023digital,villalonga2021decision}. Across these applications, the common principle is that the Digital Twin predicts the consequences of candidate actions, while the optimizer selects a feasible decision for execution by the physical system or a human operator.

 \subsubsection{Model Predictive Control}

Among online optimization approaches, model predictive control (MPC) is a special realization of model-based optimization for sequential control decisions. It narrows the generic parametric optimization formulation mentioned in the previous section to a time-ordered, discrete sequence of future control inputs and explicitly optimizes how the system should evolve over a prediction horizon. This horizon-based structure is what distinguishes MPC from general online parameter optimization. Rather than selecting one set of parameters based on a predicted response, MPC repeatedly solves a finite-horizon control problem, applies only the first input, receives new feedback, and re-solves the problem at the next time step. It therefore combines open-loop look-ahead planning with closed-loop feedback correction. When Digital Twins are equipped with the capability of multi-step-ahead predictions from the virtual twin, they can evaluate future consequences before physical execution, proactively avoid overshooting and constraint violations, while real-time sensing keeps the prediction horizon synchronized with the physical system \cite{rawlings2020model, schwenzer2021review, thelen2022comprehensive}.  Note that in this section we refer MPC as nominal MPC where only the nominal prediction is used for optimization; variants of MPC, such as robust and stochastic MPC, will be introduced in Section~\ref{sec:rmpc}.

Using a reference tracking problem as an example, MPC can be written as the following finite-horizon optimization problem,
\begin{subequations}
\begin{align}
    \min_{\mathbf{u}} ~& J(\mathbf{u},\mathbf{x},\mathbf{r}) = \sum_{i=0}^{N-1} \left[ \|\mathbf{x}_{k+i+1}-\mathbf{r}_{k+i+1}\|_{\mathbf{Q}}^2 + \|\mathbf{u}_{k+i}\|_{\mathbf{R}}^2 \right], \label{eq:MPC_objective} \\
    \mathrm{s.t.} \quad 
    & \mathbf{x}_{k+i+1} = \hat{f}(\mathbf{x}_{k+i},\mathbf{u}_{k+i}), \quad \forall i\in\mathbb{N}_{[0,N-1]},\label{eq:MPC_constraints_b}\\
    & \mathbf{x}_{k+i} \in \mathbb{X}, \quad \forall i\in\mathbb{N}_{[1,N]},\label{eq:MPC_constraints_c}\\
    & \mathbf{u}_{k+i} \in \mathbb{U}, \quad \forall i\in\mathbb{N}_{[0,N-1]},\label{eq:MPC_constraints_d} \\
    & \mathbf{g}(\mathbf{x}_{k+i+1},\mathbf{u}_{k+i}) \le \mathbf{0}, \quad \forall i\in\mathbb{N}_{[0,N-1]}, \label{eq:MPC_constraint_e} \\
    & \mathbf{h}(\mathbf{x}_{k+i+1},\mathbf{u}_{k+i}) = \mathbf{0}, \quad \forall i\in\mathbb{N}_{[0,N-1]}. \label{eq:MPC_constraint_f}
    \end{align}
    \label{eq:MPC_constraints}
\end{subequations}
Here, $\mathbf{u}$ is the control-input sequence, $\mathbf{r}$ is the reference trajectory, $\hat{f}$ is the predictive model, $\mathbb{X}$ and $\mathbb{U}$ are feasible state and input sets, and $\mathbf{g}$ and $\mathbf{h}$ denote additional inequality and equality constraints. The subscript denotes the time step under the discrete system representation, and the notation $\mathbf{x}_{k+i+1}$ denotes open-loop predictions initialized at step $k$. The first term in the objective penalizes tracking error, while the second penalizes control effort.

\begin{figure}[h!]
\centering
\includegraphics[width=0.9\linewidth]{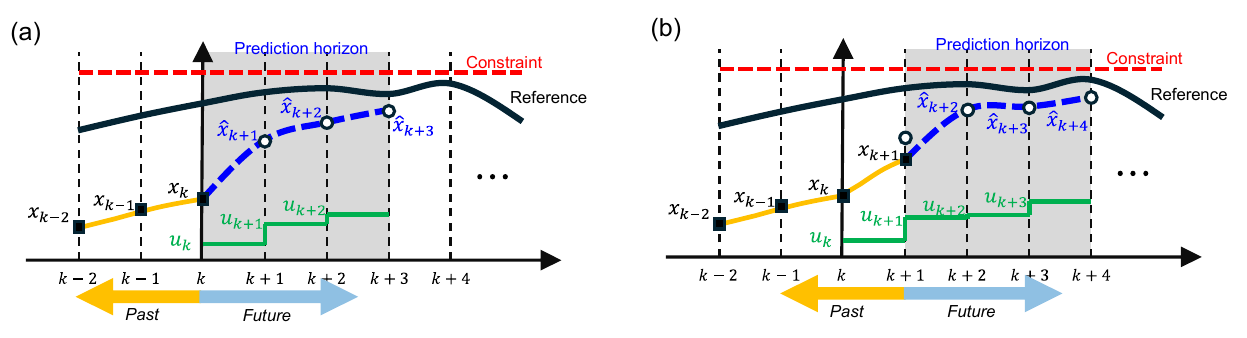}
\caption{Schematic illustration of the receding-horizon operation of model predictive control (MPC). (a) At the current time step, future system behavior is predicted over a finite horizon and the control sequence is optimized to track the reference while satisfying operational constraints. (b) After the first control action is implemented and new measurements become available, the prediction horizon shifts forward for one step, and the optimization is repeated based on the updated system state.}
\label{fig:MPC}
\end{figure}

As shown in Figure~\ref{fig:MPC}, MPC optimizes a sequence $\mathbf{u}_k,\ldots,\mathbf{u}_{k+N-1}$ over a finite horizon, while only $\mathbf{u}_k$ is applied to the physical system at the current time step. At the next time step, the Digital Twin receives updated state feedback, shifts the horizon forward, and solves the same optimization problem again. This repeated prediction, optimization, execution, and feedback update allows MPC to combine model-based planning with closed-loop correction.

While MPC was first proposed as a low-level controller by the control community \cite{muske1993model}, it has become one of the most popular decision-making approaches in Digital Twin applications. It is not only suited for a variety of models with predictive capabilities, including physics-based models and data-driven models, \cite{thelen2022comprehensive, chen2025real, mcclellan2022physics}, but can also be seamlessly integrated into Digital Twins due to its flexibility in the formulation of objectives and constraints. This capability enhances the robustness and adaptability of Digital Twins, enabling them to support a wide range of operational tasks such as manufacturing process control \cite{chen2025adaptivedigitaltwinsheet}, energy management \cite{adeola2025ai}, and navigation \cite{mcclellan2022physics}. Ultimately, the incorporation of MPC allows Digital Twins to go beyond static optimization or reactive control, empowering them with the foresight and agility needed to operate reliably in real-world environments against uncertainties \cite{gunasegaram2024machine, thelen2022comprehensive}.

\subsection{Uncertainty-Aware Optimization}
\label{sec:opt_with_prob}

Uncertainty-aware optimization addresses problems in which a Digital Twin must solve for actions under uncertain future system behavior. Its formulation is strongly conditioned on how the virtual twin represents uncertainty in system evolution. Whereas a deterministic surrogate provides a nominal prediction of the future state, an uncertainty-aware model may instead represent system evolution through probabilistic state transitions, bounded uncertainty sets, or predictive distributions. For example, the stochastic state evolution described by $p(\mathbf{x}_{k+1}\mid\mathbf{x}_k,\mathbf{u}_k)$ allows an optimizer to evaluate the expected consequences of different possible next states rather than relying on a single nominal prediction. Such formulations are useful for engineering problems involving degradation, disturbances, or variable operating conditions, where nominal predictions alone may not adequately characterize the consequences or risk of a candidate action \cite{agrell2023optimal}. Uncertainty-aware optimization therefore incorporates these representations into the solution process to balance uncertainty representation and probabilistic optimization. This section reviews two representative approaches: dynamic programming and optimal control (Section~\ref{sec:dp}), which use uncertain state transitions to evaluate long-term expected consequences, and robust and stochastic MPC (Section~\ref{sec:rmpc}), which incorporate bounded or probabilistic uncertainty into receding-horizon optimization.
 
\subsubsection{Dynamic Programming and Optimal Control}
\label{sec:dp}

Dynamic programming (DP) is naturally connected to Digital Twins because it uses the virtual twin to evaluate how current decisions influence long-term system evolution and provides a systematic framework for synthesizing sequential decisions \cite{shapiro1993mathematical,busoniu2017reinforcement}. While MPC repeatedly solves a finite-horizon optimization problem online and applies only the first action, DP uses Bellman's principle of optimality to recursively evaluate the long-term value of states and actions, allowing classical methods such as value iteration and policy iteration to leverage the Digital Twin's virtual model and transition dynamics to synthesize sequential policies without perturbing the physical system \cite{zhao2022variants,gejo2022development,frantzen2022digital,zheng2023policy,karkaria2025optimization}. To connect Digital Twins with DP for decision-making, it must provide a state-transition model that describes how the system evolves under candidate decisions. Depending on the application, this representation may take the form of a state-space model or a probabilistic model that provides transition probabilities among the states. 

Decision-making through DP brings three advantages to Digital Twins. First, DP supports discrete, hybrid, and combinatorial decision spaces, such as maintenance scheduling, replacement planning, and route selection, where gradient-based optimization may be inefficient. Second, DP evaluates cumulative future consequences through a value function, rather than only optimizing a receding finite-horizon control sequence. Third, DP can be formulated over stochastic state transitions, where the next state and reward follow a probability distribution rather than a single deterministic prediction. DP is therefore probabilistic when applied to Markov Decision Processes (MDPs). In this setting, the Bellman recursion computes the expected future cost or reward over possible next states, allowing uncertainty to enter the decision rule. Deterministic DP is recovered as a special case when the transition distribution collapses to a single next state.

Formally, the decision problem can be posed on a discrete-time controlled Markov process. Let $\mathbf{x}_k \in \mathbb{X}$ denote the system state at stage $k$, let $\mathbf{u}_k \in \mathbb{U}$ denote an admissible decision, and let the state evolve according to
\begin{equation}
\mathbf{x}_{k+1}
\sim
p(\cdot\mid\mathbf{x}_k,\mathbf{u}_k),
\end{equation}
where $p(\cdot\mid\mathbf{x}_k,\mathbf{u}_k)$ is the transition model provided by the virtual representation of the physical system. Each transition incurs a stage cost $l(\mathbf{x}_k,\mathbf{u}_k,\mathbf{x}_{k+1})$, and the objective is to minimize the expected cumulative cost over the decision horizon. Bellman's principle of optimality states that any tail of an optimal policy is itself optimal for the subproblem initiated at the corresponding intermediate state, which converts the $N$-stage optimization into the backward recursion
\begin{equation}
J_N(\mathbf{x}) = l_N(\mathbf{x}), \qquad
J_k(\mathbf{x}) = \min_{\mathbf{u} \in \mathbb{U}}
\mathbb{E}_{\mathbf{x}' \sim p(\cdot \mid \mathbf{x},\mathbf{u})}
\!\left[\, l(\mathbf{x},\mathbf{u},\mathbf{x}') + J_{k+1}(\mathbf{x}') \,\right],
\quad k = N-1,\dots,0 ,
\label{eq:dp_recursion}
\end{equation}
where $\mathbf{x}'$ is used to represent the notation of next state for simplification, $J_k(\mathbf{x})$ is the optimal cost-to-go, i.e., the minimum expected cost
accumulated from state $\mathbf{x}$ at stage $k$ to the end of the horizon and $l_N$ is
the terminal cost. For infinite-horizon problems with discount factor
$\gamma \in (0,1)$, the recursion in Equation~\eqref{eq:dp_recursion} becomes the
stationary fixed-point condition known as the Bellman optimality equation\cite{bertsekas2012dynamic},
\begin{equation}
J^{\star}(\mathbf{x}) \;=\;
\min_{\mathbf{u} \in \mathbb{U}}
\mathbb{E}_{\mathbf{x}' \sim p(\cdot \mid \mathbf{x},\mathbf{u})}
\!\left[\, l(\mathbf{x},\mathbf{u},\mathbf{x}') + \gamma\, J^{\star}(\mathbf{x}') \,\right],
\label{eq:bellman_optimality}
\end{equation}
from which the optimal decision rule is recovered greedily as
\begin{equation}
\pi^{\star}(\mathbf{x}) \in
\arg\min_{\mathbf{u} \in \mathbb{U}}
\mathbb{E}_{\mathbf{x}' \sim p(\cdot \mid \mathbf{x},\mathbf{u})}
\!\left[\, l(\mathbf{x},\mathbf{u},\mathbf{x}') + \gamma\, J^{\star}(\mathbf{x}') \,\right].
\label{eq:greedy_policy}
\end{equation}
Here, the expectation operator is the mechanism by which uncertainty in the state transition enters the decision rule. Deterministic dynamic programming is recovered when $p(\cdot\mid\mathbf{x},\mathbf{u})$ collapses to a Dirac measure concentrated at a deterministic transition $\mathbf{x}'=f(\mathbf{x},\mathbf{u})$. The minimization is taken over an arbitrary admissible set $\mathbb{U}$, which may be discrete, hybrid, or combinatorial.

\begin{table}[h!]
\centering
\caption{Instantiation of the dynamic programming ingredients in
Eqs.~\eqref{eq:dp_recursion}--\eqref{eq:greedy_policy} across representative
Digital Twin deployments.}
\label{tab:dp_engineering_mapping}
\small
\begin{tabular}{@{}p{2.4cm} p{3.0cm} p{2.8cm} p{3.2cm} p{3.4cm}@{}}
\toprule
\textbf{Deployment} & \textbf{State} & \textbf{Decision} &
\textbf{Stage cost $l$} & \textbf{Role of the Digital Twin} \\
\midrule
Wastewater aeration \cite{heo2025real} &
Dissolved oxygen, nitrite-to-ammonium ratio, influent load &
Aeration setpoint (near-continuous) &
Blower energy plus penalty on effluent limit violation &
Calibrated biokinetic process model predicting the effect of aeration on
nitrogen species \\
\addlinespace
Grinding feed-rate cycle \cite{pereverzev2019designing} &
Stock removal, wheel wear, thermal and force state &
Feed rate per removal increment (discretized) &
Cycle time, with defect-inducing conditions entering as constraints &
Virtual sensing of burn, chatter, and dimensional error along candidate cycles \\
\addlinespace
Tire retreading and rotation \cite{karkaria2025digital} &
Per-tire tread depth, casing condition, position in fleet &
Retread, rotate, or retain (combinatorial) &
Negative retread capability subject to safety and operational constraints &
Learned degradation model driven by live pressure, load, and temperature data \\
\addlinespace
UAV health-aware maneuvering \cite{kapteyn2021probabilistic} &
Structural health state and predicted wing-strain quantities of interest &
2g or 3g maneuver &
Structural-health reward combined with a reward favoring faster maneuvers &
Probabilistic graphical Digital Twin that assimilates strain measurements and models the evolution of structural health under candidate maneuvers \\
\bottomrule
\end{tabular}
\end{table}

Engineering deployments illustrate this synergy across domains and decision timescales, as summarized in Table~\ref{tab:dp_engineering_mapping}. In wastewater treatment, policy-iteration DP uses a calibrated biokinetic model to evaluate how candidate aeration setpoints affect the process state and iteratively improves the aeration policy by balancing blower energy consumption against effluent-limit violations \cite{heo2025real}. In manufacturing, DP has been used to design time-optimal feed-rate cycles for grinding processes. Candidate machining cycles are simulated within the Digital Twin to detect defect-inducing conditions, which are then incorporated as constraints and used to re-optimize the cycle until a defect-free, time-optimal profile is obtained \cite{pereverzev2019designing}. Kapteyn et al.~\cite{kapteyn2021probabilistic} use a probabilistic graph for a structural Digital Twin of an unmanned aerial vehicle. The probabilistic graphical model represents uncertain structural health and its evolution under candidate maneuvers, and the induced planning problem is approximated as an MDP and solved offline using value iteration to obtain a health-dependent policy that selects between more aggressive and conservative maneuvers as the estimated structural condition evolves. Across these cases, DP is useful when the Digital Twin can represent system evolution through a state-transition model and must convert evolving asset states into sequential policies over discrete or mixed decisions rather than only optimize continuous inputs over a short receding horizon.
 
\subsubsection{Robust and Stochastic Model Predictive Control}
\label{sec:rmpc}

Robust MPC (RMPC) and stochastic MPC (SMPC) are introduced after DP because they address a different form of uncertainty-aware optimization. While DP handles uncertainty by evaluating expected long-term consequences over probabilistic state transitions, RMPC and SMPC retain the receding-horizon structure of MPC and incorporate uncertainty directly into the online optimization problem. From the modeling perspective, uncertainty in future state trajectories may be represented through bounded disturbance sets, error tubes, or probability distributions over disturbances, states, or model errors \cite{chen2025uncertainty,tsai2023control}. From the optimization perspective, RMPC seeks decisions that remain feasible for all disturbances within a prescribed uncertainty set, whereas SMPC controls the probability of constraint violation under a probabilistic uncertainty model \cite{mayne2005robust,mesbah2016stochastic}. Under classical formulations based on linear state-space models, known uncertainty forms, and convex constraints, these uncertainty representations can often be reformulated into computationally tractable optimization problems that support online receding-horizon control. 

\begin{figure}[h!]
\centering
\includegraphics[width=0.4\linewidth]{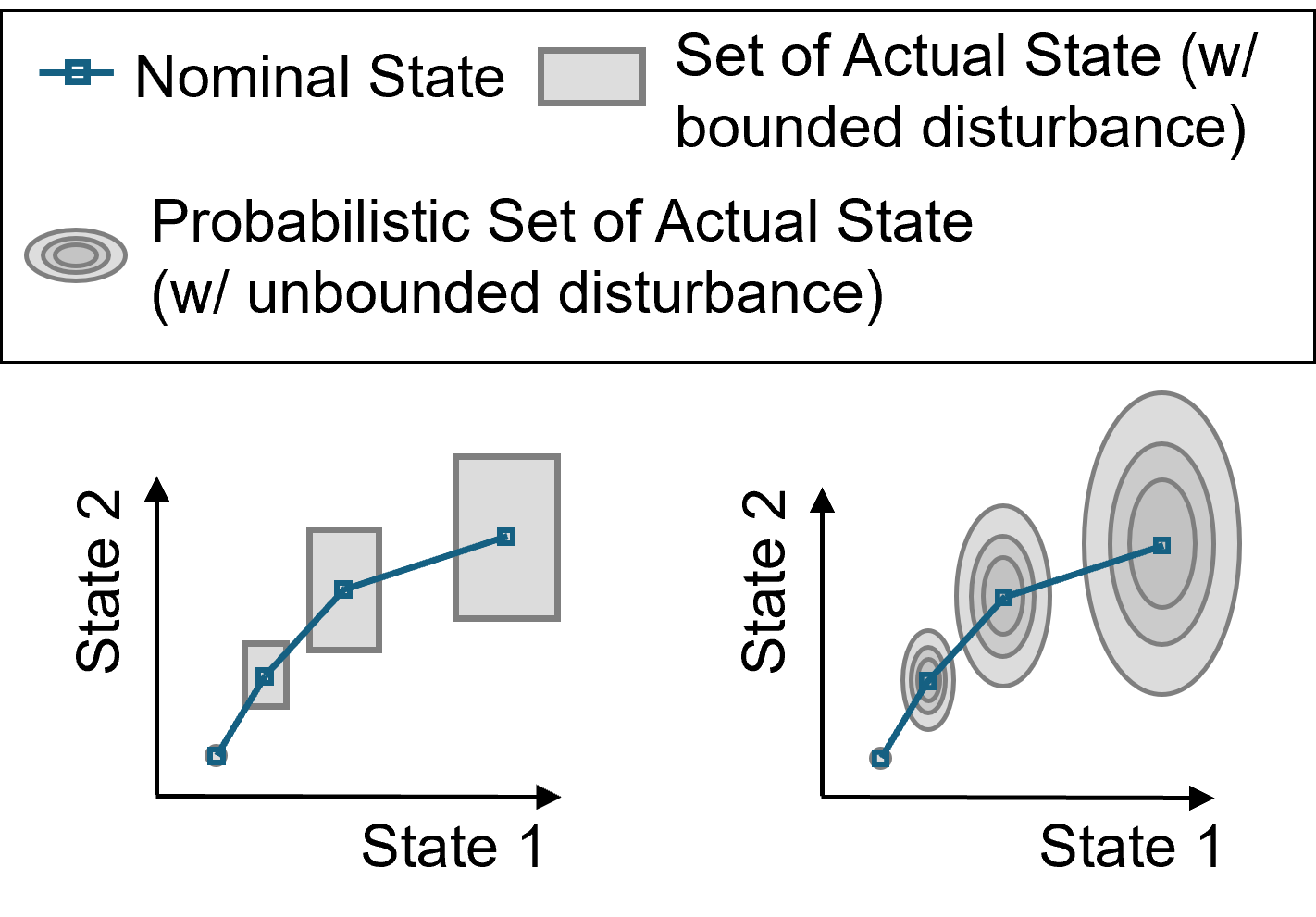}
\caption{Conceptual comparison of nominal states and actual states under bounded (left) and unbounded (right) disturbances.}
\label{fig:States_bounded_unbounded}
\end{figure}

Figure~\ref{fig:States_bounded_unbounded} illustrates the evolution of actual states around a nominal trajectory predicted by a model. Under bounded disturbances, the possible actual states remain within finite reachable sets around the nominal prediction. Under disturbances with unbounded support, the actual state cannot be guaranteed to remain inside any finite set and is instead characterized by a probability distribution or confidence region. This distinction motivates two uncertainty-aware extensions of MPC.

A general discrete-time dynamical system under disturbance can be written as
\begin{equation}
\mathbf{x}_{k+1}
=
f(\mathbf{x}_k,\mathbf{u}_k,\mathbf{w}_k),
\end{equation}
where $\mathbf{x}_k$ is the state, $\mathbf{u}_k$ is the control input, and $\mathbf{w}_k$ represents an uncertain disturbance. RMPC and SMPC differ primarily in how $\mathbf{w}_k$ is represented and incorporated into the optimization problem. In RMPC, the disturbance is assumed to belong to a bounded uncertainty set $\mathbb{W}$, and the controller seeks decisions that remain feasible for all admissible disturbance realizations. Min--max RMPC explicitly optimizes against the worst disturbance sequence, while tube-based RMPC constructs a bounded region around a nominal trajectory and uses feedback to keep the actual state within this region \cite{lofberg2003minimax,langson2004robust,gao2014tube,tsai2023robust}.

In SMPC, the disturbance is instead modeled probabilistically, $\mathbf{w}_k\sim p(\mathbf{w})$, allowing uncertainty in future trajectories to be incorporated through expected objectives and chance constraints. A general formulation is
\begin{subequations}
\begin{align}
\min_{\mathbf{u}} \quad &
\mathbb{E}\left[
\sum_{i=0}^{N-1}
l(\mathbf{x}_{k+i},\mathbf{u}_{k+i})
+
l_N(\mathbf{x}_{k+N})
\right] \\
\mathrm{s.t.} \quad
&
\mathbf{x}_{k+i+1}
=
\hat{f}(\mathbf{x}_{k+i},
\mathbf{u}_{k+i},
\mathbf{w}_{k+i}),
\qquad
\mathbf{w}_{k+i}\sim p(\mathbf{w}),
\quad
\forall i\in\mathbb{N}_{[0,N-1]}, \\
&
\mathbb{P}\left(
g_j(\mathbf{x}_{k+i},\mathbf{u}_{k+i})
\leq 0
\right)
\geq 1-\epsilon_j,
\quad
\forall i\in\mathbb{N}_{[0,N-1]},
\quad
\forall j, \\
&
\mathbb{P}\left(
\mathbf{x}_{k+i}\in\mathbb{X}
\right)
\geq 1-\epsilon_x,
\quad
\forall i\in\mathbb{N}_{[1,N]}, \\
&
\mathbf{u}_{k+i}\in\mathbb{U},
\quad
\forall i\in\mathbb{N}_{[0,N-1]} .
\end{align}
\label{eq:smpc_general}
\end{subequations}
Here, $\epsilon_j$ and $\epsilon_x$ denote the allowable maximum probabilities of violating the $j$-th stage constraint and the state admissibility set constraint, respectively. The expectation accounts for the distribution of possible future trajectories, while the chance constraints translate predictive uncertainty into an explicit risk requirement. In contrast, RMPC replaces the probabilistic description with a bounded uncertainty set and requires the relevant constraints to hold for all admissible realizations.

Several strategies are available for solving or approximating the resulting stochastic optimization problem online. Sampling-based methods \cite{kim2022smooth,tao2022path} approximate probabilistic constraints through scenario generation. Constraint-tightening and stochastic tube methods \cite{lorenzen2016constraint,tsai2025control} convert uncertainty into tightened deterministic constraints, while stochastic programming formulations \cite{de2005stochastic,chamanbaz2025stochastic} incorporate probabilistic objectives and constraints directly into the optimization. Compared with worst-case RMPC, SMPC can reduce conservatism by permitting a small and explicitly quantified probability of constraint violation when a probabilistic uncertainty model is available.

RMPC and SMPC have been demonstrated across diverse engineering systems for uncertainty-aware decision-making. In automotive control, SMPC can learn probabilistic driver-behavior models online to maintain safety distance from the leading vehicle and optimize power requested by the motor, improving adaptive cruise control and hybrid-vehicle energy management~\cite{bichi2010stochastic}. In building and smart-energy systems, probabilistic forecasts of weather, renewable generation, and demand are incorporated through chance constraints to balance energy efficiency, occupant comfort, and operational risk~\cite{sokoler2012stochastic,heirung2018stochastic}. RMPC has also been extended to uncertain nonlinear systems, including quadrotor dynamics, by constructing trajectory-dependent tubes that preserve constraint satisfaction without requiring prohibitively expensive worst-case optimization~\cite{kohler2020computationally}. In complex processes such as metal additive manufacturing, accurate physics-based models are exceedingly difficult to obtain because of strong nonlinearity, multiscale phenomena, and process variability. Data-driven MPC responds by learning the dynamics directly from operational data, aligning naturally with the emphasis of Digital Twin on real-time prediction and solvability \cite{hewing2019cautious,drgona2022differentiable,jung2023model,wu2019real,huang2022lstm}. Recent work extends this alignment to UQ, using deep learning to infer disturbance tubes from data \cite{fan2020deep,chen2025uncertainty} so that the controller's robustness margins reflect the learned, potentially state-dependent uncertainty rather than a fixed design-time assumption. In this configuration, the Digital Twin supplies not only the prediction inside the MPC loop but also the uncertainty set that the robust formulation defends against. Overall, the effectiveness of both RMPC and SMPC depends on the fidelity of the system and uncertainty models, for which the Digital Twins can play a role in providing probabilistic state prediction, which will be addressed more in Section~\ref{sec:aleatoric_uncertainty}.

\subsection{Policy-Based Approach}
\label{sec:policy_based}

Policy-based approaches address repeated sequential decision-making problems in which control actions must be generated rapidly from the current system state, often under computational constraints that make online optimization impractical~\cite{gonzalez2024neural}. Rather than using the Digital Twin model to solve a new optimization problem at every decision instant, these methods construct a policy that maps observed or estimated states directly to actions. This shifts much of the computational burden from online operation to offline analysis, simulation, or training, while still allowing the resulting decision rule to be evaluated within the closed-loop Digital Twin. Policy-based approaches therefore differ from the optimization-based methods discussed in Sections~\ref{sec:deterministic_optimization} and \ref{sec:opt_with_prob} in where and how the decision problem is solved. Optimization-based methods compute each action through an explicit online search over predicted system behavior, whereas policy-based methods encode the result of prior optimization or learning into a precomputed state-to-action mapping. This section reviews two main classes of policy-based decision-making. The first treats the policy as an approximate optimizer, including explicit model predictive control and neural policies trained to reproduce or directly minimize an optimization objective. The second considers reinforcement learning, in which policies are improved through interaction with a physical or virtual environment using observed rewards, with or without an explicit model of the system dynamics.

\subsubsection{Policy as Approximate Optimizer}
\label{sec:approximate_optimizer}

\begin{figure}[h!]
    \centering
    \includegraphics[width=0.5\linewidth]{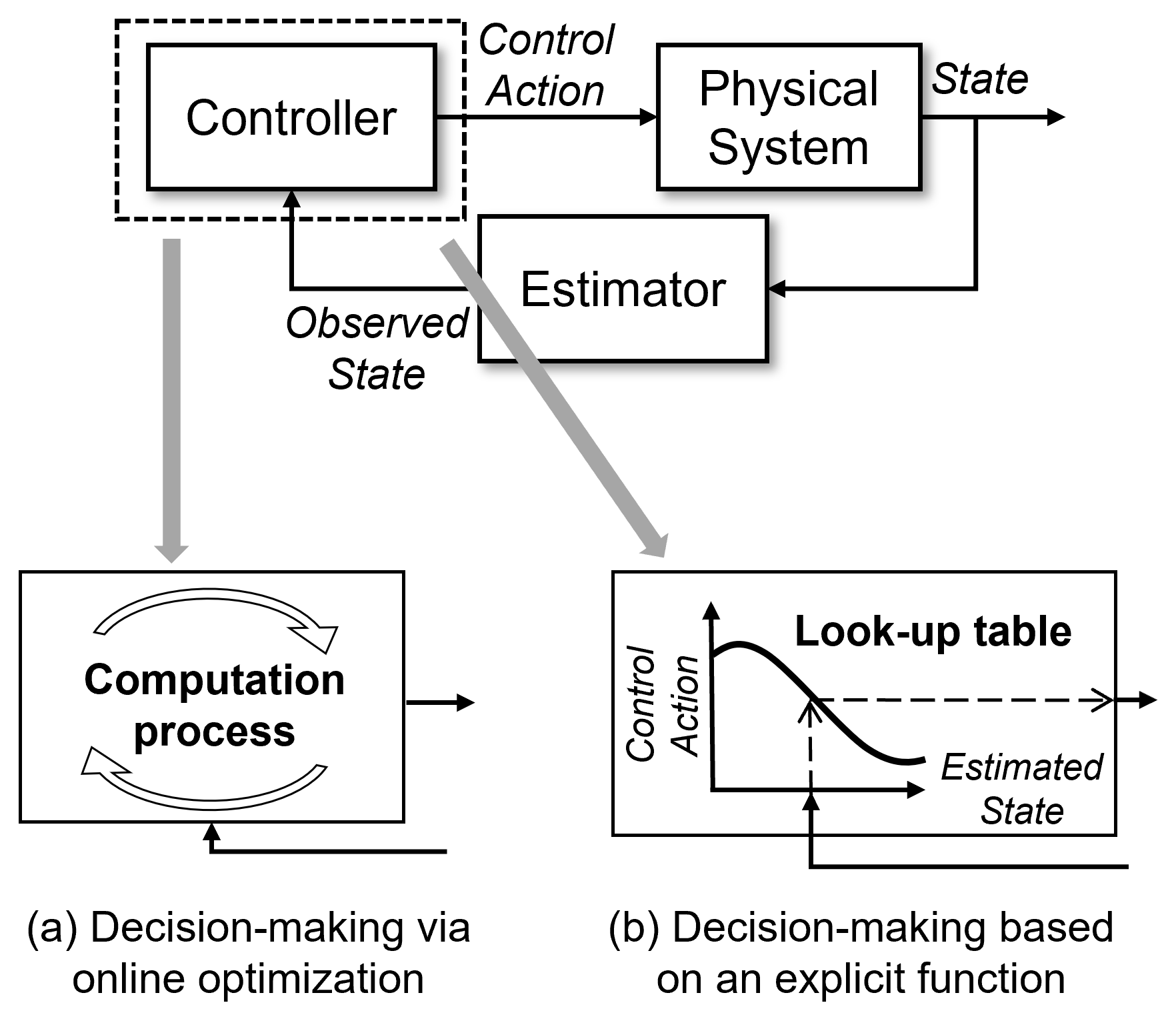}
    \caption{Comparison between (a) online optimization-based decision-making and (b) policy-based decision-making using a precomputed function.}
    \label{fig:online_policy}
\end{figure}

Approximate-optimizer policies preserve the objective structure of an established optimization-based controller while replacing its online numerical solution with a learned or precomputed control law, as illustrated in Figure~\ref{fig:online_policy}. They are therefore enabling solutions for applications when a reliable optimizer can be formulated and evaluated offline, but cannot satisfy the latency, memory, or computing constraints imposed during operation. This situation arises in embedded controllers, distributed infrastructure, and systems that require decisions at rates faster than a full model predictive control problem can be solved~\cite{chan2021deep,karg2018deep}. The Digital Twin supports this approach by supplying the prediction model, simulated trajectories, operating conditions, and constraint evaluations required to construct and validate the policy. The resulting mapping can also be inspected to study how control actions vary across the state space and where the controller approaches active constraints~\cite{tsai2022design}.

Approximate optimizers can be divided into analytical explicit policies, supervised controller imitation, and direct policy optimization. Explicit MPC (EMPC) derives an analytical solution to a parametric optimization problem. For linear dynamics, quadratic objectives, and polyhedral constraints, the feasible state space is partitioned into critical regions, within each of which the optimal control action is affine:
\begin{equation}
\pi_{\mathrm{EMPC}}(\mathbf{x}) = F_j \mathbf{x} + G_j,
\qquad
\mathbf{x}\in\mathcal{R}_j,
\end{equation}
where $\mathcal{R}_j$ is the $j$th critical region, and $F_j$ and $G_j$ are the precomputed gain matrix and offset vector associated with the active region $j$~\cite{bemporad2002explicit}. Online evaluation then consists of identifying the region containing the current state and applying its associated control law.

Supervised controller imitation replaces the piecewise analytical law with a parameterized function $\pi_\theta$. A teacher optimizer is executed offline to evaluate the optimal control policy $\mathbf{u}^*$ at sampled states $\mathbf{x}$, and to construct the tuples of state-to-action pairs as a training set:
\begin{equation}
\mathcal{D} = \left\{ (\textbf{x}_i, \textbf{u}_i^*) \right\}_{i=1}^{N_s},
\qquad
\textbf{u}_i^*=\pi^*(\textbf{x}_i),
\end{equation}
and the policy model $\pi_\theta$ is trained through supervised learning, such as mean squared error (MSE):
\begin{equation}
\theta^* = \arg\min_\theta
\frac{1}{N_s} \sum_{i=1}^{N_s}
\left| \pi_\theta(\textbf{x}_i)-\textbf{u}_i^* \right|_2^2.
\end{equation}
Since exhaustively solving the optimizer over the operation domain is prohibitive, samples should cover regions where they are most informative and frequently visited. These can be represented as the states visited in closed-loop operation. For example, \cite{chan2021deep} samples along feasible closed-loop trajectories, while \cite{hertneck2018learning} injects disturbances $d_k$ to also cover states reachable under perturbations.

Different from supervised policy approximation that requires a pre-existing control law to run the system for collecting closed-loop data for training the policy model, direct policy optimization does not require optimizer-generated labels, but directly approximates the optimal policy from the system model. Instead, the policy is trained through simulated closed-loop trajectories generated by a prediction model $\hat{f}_\phi$
\begin{equation}
\mathbf{x}_{k+1} = \hat{f}_\phi \left( \mathbf{x}_k, \pi_\theta(\mathbf{x}_k) \right),
\end{equation}
with the policy model $\pi_\theta$ obtained from an MPC-inspired loss
\begin{equation}
\theta^* = \arg\min_\theta \left[ \sum_{k=0}^{N-1} l \left( \mathbf{x}_k, \pi_\theta(\mathbf{x}_k) \right) + l_N(\mathbf{x}_N) + \rho \sum_{k=0}^{N-1} \mathcal{P} \left( \mathbf{x}_k, \pi_\theta(\mathbf{x}_k) \right) \right],
\end{equation}
where the first two terms of the loss function are exactly the stage cost and the terminal cost introduced in the previous section, and $\mathcal{P}$ is a penalty term that converts hard constraints into soft constraints and penalizes violations of state and input constraints~\cite{ahmed1998neural,akesson2006neural}. A representative application is the Differentiable Predictive Control (DPC), which extends this formulation by representing both the policy and the prediction model as differentiable components, which permits end-to-end gradient propagation through the closed-loop trajectory~\cite{drgona2022differentiable}.

Across engineering applications, approximated policies address the computational and latency limitations of repeatedly solving complex optimization problems during online operation, especially for large-scale, complex systems. In power systems, they encode coupled decisions over generation, storage, and network states into a continuous policy, avoiding a new constrained solve whenever operating conditions change~\cite{king2022koopman}. In robotics, they compile trajectory optimization or predictive control into a fixed-cost policy that meets control-rate requirements despite nonlinear dynamics and variable solver time~\cite{viljoen2024differentiable}. In large-scale infrastructure, local or hierarchical policies retain system-level coordination while reducing the computation and communication required by repeated centralized optimization~\cite{tumu2024differentiable}. In each case, the Digital Twin is used offline to generate and evaluate closed-loop trajectories under representative operating conditions and disturbances. Once deployed, the policy computes control actions directly from the current state, thereby reducing online computational cost and decision latency.

\subsubsection{Reinforcement Learning}
\label{sec:RL}

The approaches discussed in Section \ref{sec:approximate_optimizer} reduce real-time computational burden by learning a direct mapping from system conditions to the solution of a predefined optimization problem. However, this formulation still assumes that the underlying decision problem (including its objective, constraints, and optimization structure) can be specified in advance. Many engineering decision problems involve long-term and coupled consequences that are difficult to represent through a fixed optimization objective or a predetermined decision procedure. As system scale and complexity increase, solving the associated optimization problem either online or offline may also become computationally intractable. Reinforcement learning (RL) addresses this challenge by making sequential decisions through interaction with an environment so as to maximize expected cumulative reward, thereby accounting for how current actions influence both immediate performance and future operating conditions~\cite{sutton1998reinforcement}. RL methods can be characterized along two complementary dimensions: value-based, policy-based, and actor-critic methods differ in how the decision rule is represented and learned, whereas model-free and model-based methods differ in whether an explicit model of the environment dynamics is used for learning or planning. This formulation is suited to sequential decision-making under uncertain or evolving conditions, where desirable behavior can be evaluated from system responses but cannot be fully prescribed in advance.

\begin{figure}[h!]
    \centering
    \includegraphics[width=0.75\linewidth]{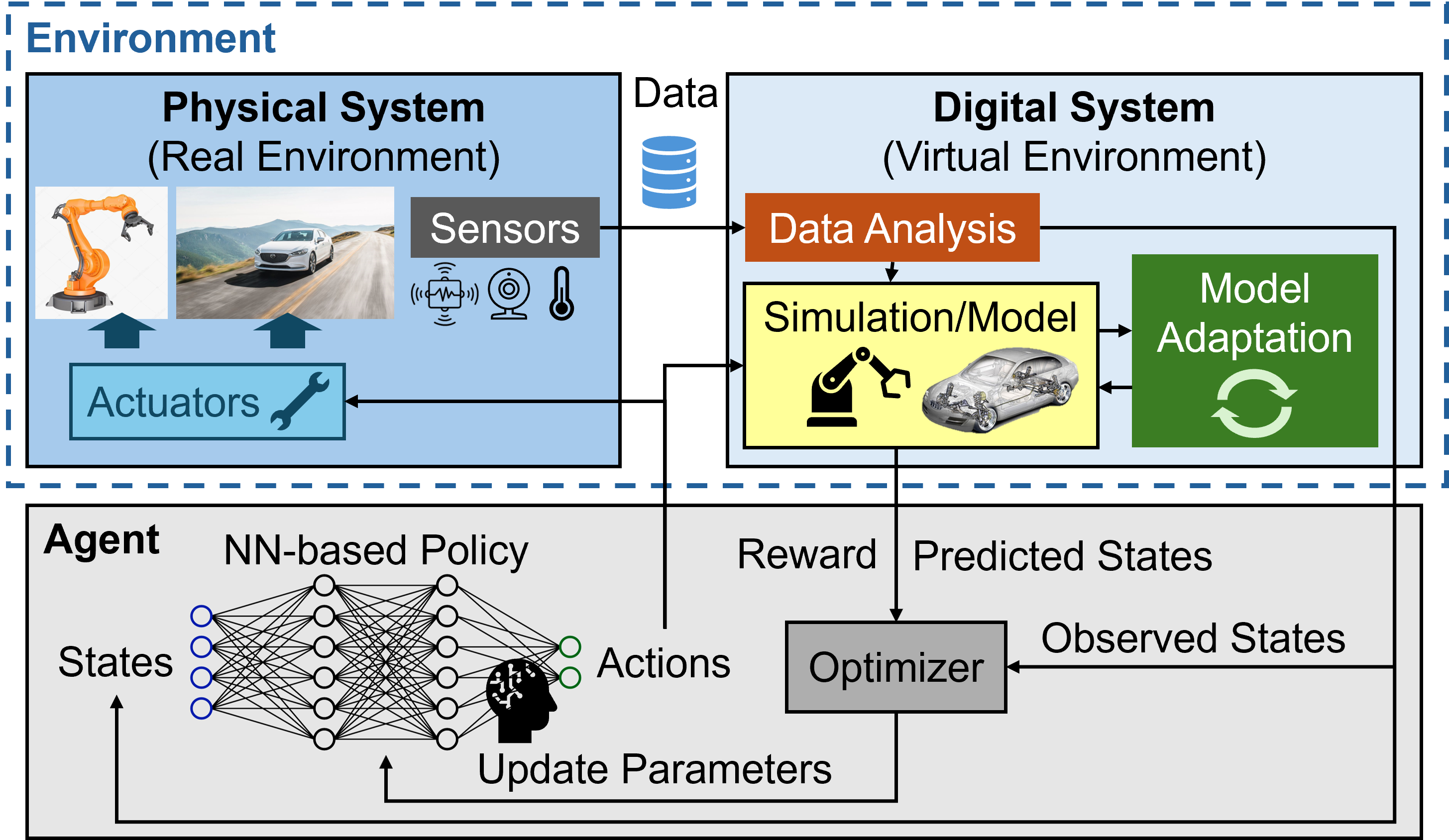}
    \caption{Conceptual framework of Deep Reinforcement Learning (DRL) integrated with a Digital Twin system. It highlights the interaction between the agent and the real and virtual environments, allowing the agent to learn optimal actions using feedback from both predicted and observed outcomes for intelligent decision-making and continuous adaptation.}
    \label{fig:RL_DT_framework}
\end{figure}

From a Digital Twin perspective, the distinction between model-free and model-based RL is particularly important. Model-free RL does not require an explicit dynamics model for policy or value-function learning, although a Digital Twin may still provide a simulated environment in which interactions can be generated more safely and efficiently. The integration becomes more intrinsic in model-based RL, where a known or learned model of the system is explicitly used to predict future states, evaluate candidate actions, generate simulated experience, or support policy improvement~\cite{lee2022digital,kuo2025deep}. As shown in Figure~\ref{fig:RL_DT_framework}, operational data from the physical system update the virtual environment, while the RL agent interacts with both predicted and observed system responses to improve its policy. The updated policy is then deployed to the physical system, closing the loop among sensing, modeling, learning, and control. This structure is valuable when exploration on the physical system is costly, slow, or unsafe. In such settings, Digital Twins allow candidate actions, disturbances, and failure conditions to be evaluated using the digital model before the policy is deployed to the physical system. Feedback from the physical system can also be used to update the digital model and reduce the mismatch between the digital representation and its physical counterpart~\cite{xia2021digital,alam2021physics,cheng2024toward,tsai2025multi}. By integrating the physical system, the Digital Twin, and the RL agent into a continuously evolving learning loop, model-based RL extends the role of the Digital Twin from a predictive representation to an adaptive environment for policy learning under uncertain and changing dynamics~\cite{liu2022digital}.

To maximize the expected cumulative reward over time, the framework of finite MDPs is introduced, comprising a finite set of states \( \mathcal{X} \), a finite set of actions \( \mathcal{U} \), a transition probability function \( p(\mathbf{x}', r \mid \mathbf{x}, \mathbf{u}) \), and a discount factor \( \gamma \in (0, 1) \) that weights future rewards. Within this setting, the behavior of an agent is governed by a policy \( \pi(\mathbf{u} \mid \mathbf{x}) \), which specifies the probability of selecting action \( \mathbf{u} \) in state \( \mathbf{x} \). The long-term desirability of states and actions is captured by the value function \( v_\pi(\mathbf{x}) \) and the action-value function \( q_\pi(\mathbf{x}, \mathbf{u}) \), defined as the expected return under policy \( \pi \), and these satisfy the following recursive identity known as the Bellman equation:
\begin{equation}
v_\pi(\mathbf{x}) = \sum_{\mathbf{u}} \pi(\mathbf{u} \mid \mathbf{x}) \sum_{\mathbf{x}', r} p(\mathbf{x}', r \mid \mathbf{x}, \mathbf{u}) \left[ r + \gamma v_\pi(\mathbf{x}') \right],
\end{equation}
where $(\cdot)'$ is used to represent the next state for simplification. To compute the highest expected return achievable from each state, the Bellman optimality equation is introduced as a refinement of the Bellman equation. It characterizes the optimal value function \( v_*(\mathbf{x}) = \max_\pi v_\pi(\mathbf{x}) \) via the following recursive form:
\begin{equation}
v_*(\mathbf{x}) = \max_{\mathbf{u}} \sum_{\mathbf{x}', r} p(\mathbf{x}', r \mid \mathbf{x}, \mathbf{u}) \left[ r + \gamma v_*(\mathbf{x}') \right].
\end{equation}
In practice, RL often operates with the Q-function, which explicitly fixes the initial action and evaluates future return assuming a policy \( \pi \). The optimal Q-function satisfies the following Bellman optimality equation:
\begin{equation}
q_*(\mathbf{x}, \mathbf{u}) = \sum_{\mathbf{x}', r} p(\mathbf{x}', r \mid \mathbf{x}, \mathbf{u}) \left[ r + \gamma \max_{\mathbf{u}'} q_*(\mathbf{x}', \mathbf{u}') \right].
\end{equation}

The MDP is a standard mathematical foundation for RL \cite{sutton1998reinforcement}, but it is not a strict requirement for all applications. When the available observations do not fully characterize the physical state, the problem can instead be formulated as a partially observable MDP (POMDP) \cite{kaelbling1998planning}, or the state representation can be augmented using historical observations or recurrent representations \cite{hausknecht2015deep}, learned latent states \cite{hafner2019learning}, or state estimates maintained and updated by the Digital Twin \cite{kapteyn2021probabilistic}.
 
A key distinction from the policy-as-approximate-optimizer approaches in Section~\ref{sec:approximate_optimizer} is that the computational challenge in RL arises not only from the size of an individual optimization problem, but also from the scale of sequential coupling across states, actions, and time. A current decision changes the future state of the system and therefore affects the decisions and rewards that can be obtained later. Dynamic programming represents this dependence explicitly through Bellman recursion, but requires a transition model and becomes intractable as the state--action space grows. Model-free RL estimates these updates from sampled interactions. Q-learning and temporal-difference learning therefore replace exact expectation and state-transition enumeration with incremental updates from observed states, actions, and rewards~\cite{watkins1992qlearning,sutton1988learning}. This allows the policy to improve from operational data even when the transition model is unknown or imperfect. For continuous and high-dimensional engineering problems, deep reinforcement learning replaces tabular value functions and policies with neural function approximators. Deep Q-networks approximate the action-value function for discrete actions~\cite{mnih2015human}, while actor-critic methods learn a parameterized policy guided by a learned value function~\cite{konda1999actor}. These methods extend RL from finite state-action spaces to the continuous domains common in Digital Twin applications.

RL-enabled Digital Twins have been applied to engineering systems that require adaptive decisions under changing and uncertain operating conditions. For example, in a manufacturing Digital Twin for process control, the state $\mathbf{x}_k$ may consist of measured or estimated temperatures, machine conditions, and process variables; the action $\mathbf{u}_k$ may adjust operating parameters such as power, speed, or feed rate; and the reward \(r_k\) may balance product quality, productivity, and operating cost. The Digital Twin provides the environment for predicting how these actions affect future process states, while the policy determines the control action to apply from the current state. The Digital Twin serves as the environment in which state transitions and consequences of candidate actions can be simulated, while the learned policy maps the current estimated state to an appropriate control action. Beyond process control, RL-enabled Digital Twins have also been used for predictive maintenance, anomaly response, and robotic manipulation~\cite{fuller2020digital,gautam2025iiot,alam2021physics,liu2022digital}. In logistics and supply chains, it enables routing, inventory, and allocation policies to respond to disruptions and demand variation over long planning horizons~\cite{martin2022reinforcement,abideen2021digital,hubbs2020or}. In smart cities and communication networks, RL coordinates traffic, energy, bandwidth, and distributed resources using an updated representation of the system state~\cite{salunke2023reinforcement,cakir2024ai,dai2020deep,zhang2024digital,liu2024digital}. In autonomous vehicles, synchronized Digital Twins provide virtual environments for learning and evaluating maneuvering policies under varying road, traffic, and environmental conditions~\cite{tsai2025multi,tsai2026digital,tsai2026reinforcement}. Across these applications, the Digital Twin supplies predicted and observed system responses, while RL uses this feedback to refine policies for adaptive sequential decision-making.

\subsection{Knowledge-Based Approach}
\label{sec:knowledge-based}
An artisan does not determine each action by solving an explicit optimization problem or evaluating a learned policy. Instead, decisions are formed by interpreting visual, tactile, and contextual cues through accumulated experience and domain knowledge. When an outcome deviates from expectations, the artisan revises the underlying judgment and adapts future actions. Knowledge-based approaches seek to reproduce this form of decision-making in Digital Twins for problems in which decisions are a result of reasoning and deduction of information. They integrate heterogeneous observations with expert rules, semantic relationships, and prior experience to produce decisions that are context-aware, interpretable, and responsive to unfamiliar conditions.

This paradigm differs from synthesizing actions from predictive models or learned state-to-action mappings for actions through optimization-based and policy-based approaches discussed earlier. Rather, knowledge-based methods reason over explicit representations of engineering knowledge and qualitative information \cite{d2022cognitive}. Their central goal is to convert human expertise and cross-domain information into machine-readable structures that support repeatable inference and decision-making in cases where the decisions themselves cannot be directly solved or inferred via qualitative descriptions. This section first reviews fuzzy logic inference, which translates linguistic rules and imprecise observations into quantitative actions. It then examines Cognitive Digital Twins, where ontologies, knowledge graphs, semantic reasoning, and learning mechanisms are integrated to support higher-level perception, reasoning, adaptation, and autonomous decision-making.

\subsubsection{Fuzzy Logic Inference}

Digital Twin decision-making often requires the use of expert knowledge, heuristics, and qualitative judgments that cannot be expressed fully through numerical models or optimization objectives. Fuzzy logic provides a structured way to encode this knowledge through linguistic variables, membership functions, and rule-based inference~\cite{mendel1995fuzzy,mohammadi2021survey}. Unlike Boolean logic, which assigns each condition to a discrete state, fuzzy logic represents partial membership over the interval $[0,1]$. This allows imprecise or subjective descriptions of quantities, such as \emph{low} temperature, \emph{moderate} wear, or \emph{high} risk, to be translated into quantitative estimates or control actions. In Digital Twins, fuzzy logic can therefore complement data-driven and physics-based models when information is uncertain, incomplete, or dependent on human experience.

\begin{figure}[h!]
\centering
\includegraphics[width=1\linewidth]{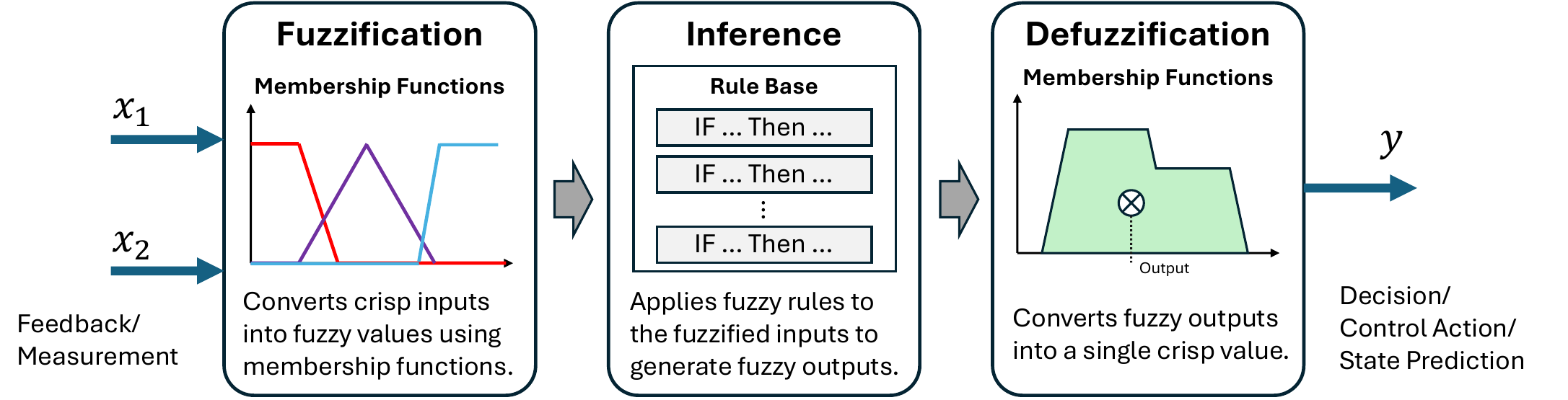}
\caption{Illustration of a fuzzy inference system.}
\label{fig:fuzzy}
\end{figure}

A fuzzy inference system maps crisp inputs to crisp outputs through fuzzification, inference, and defuzzification, as shown in Figure~\ref{fig:fuzzy}. A crisp input is a precise numerical measurement or estimate, such as a temperature of $80^\circ$C or a tool-wear value of $0.2$ mm. During fuzzification, each crisp input is converted into a fuzzy representation by assigning degrees of membership to linguistic categories such as low, medium, and high. These membership values lie between 0 and 1 and allow a single input to belong partially to several categories. The inference stage then evaluates a rule base composed of statements such as ``IF temperature is high THEN fan speed is fast.’’ These rules may be obtained from domain experts, empirical observations, or data-driven learning. Defuzzification converts the aggregated fuzzy conclusion into a numerical output using methods such as the centroid or weighted average.

Mamdani and Sugeno systems are two common fuzzy inference systems~\cite{dhimish2018comparing}. Mamdani systems represent the output of each fuzzy rule using fuzzy sets and obtain a crisp output through defuzzification, commonly using the centroid method, whereas Sugeno systems represent each rule output using a constant or a function of the inputs and combine the rule outputs through a weighted average. When embedded in a feedback loop, the fuzzy inference system becomes a fuzzy logic controller. Its overlapping membership functions support smooth transitions across operating regions, while the rule base can encode nonlinear behavior, disturbance responses, and operating constraints without requiring an exact analytical model. In a Digital Twin, the crisp inputs may come from physical sensors, virtual sensors, inferred states, or model predictions, allowing fuzzy inference to translate heterogeneous system information into interpretable estimates or control actions.

In Digital Twins, fuzzy logic supports both virtual sensing and real-time decision-making. For flood assessment, fuzzy rules combine hydrological and meteorological variables to estimate a flood vulnerability index when the relationships among contributing factors are uncertain~\cite{manocha2023digital}. For electric vehicle motor health monitoring and machine tool prognosis, fuzzy inference maps operating conditions, vibration, temperature, and runtime data to health indicators and remaining-life estimates~\cite{venkatesan2019health,luo2019digital}. These applications use expert knowledge to infer quantities that are difficult to measure directly.

When implemented in Digital Twins for decision-making, fuzzy logic controllers can use the outputs from the virtual twin as decision inputs. In CNC machining, predicted surface roughness from the virtual model is evaluated by a fuzzy controller to adjust process parameters toward the target condition~\cite{seid2024adaptive}. In hydropower control, a fuzzy correction layer supplements an optimization-based controller when surrogate predictions are affected by nonlinearities or environmental uncertainty~\cite{zeng2025integrating}. The Expert Twin framework follows a related approach by combining user knowledge and fuzzy rules to support interpretable manufacturing decisions~\cite{monek2025expert,jungmann2024towards}. Compared with standalone fuzzy controllers, Digital Twin integration provides access to virtual sensors, inferred states, historical records, and predicted responses, allowing the rule base to act on richer system information.

\subsubsection{Decision-Making via Cognitive Digital Twin}
\label{sec:CDT}

\begin{figure}[h!]
\centering
\includegraphics[width=0.9\linewidth]{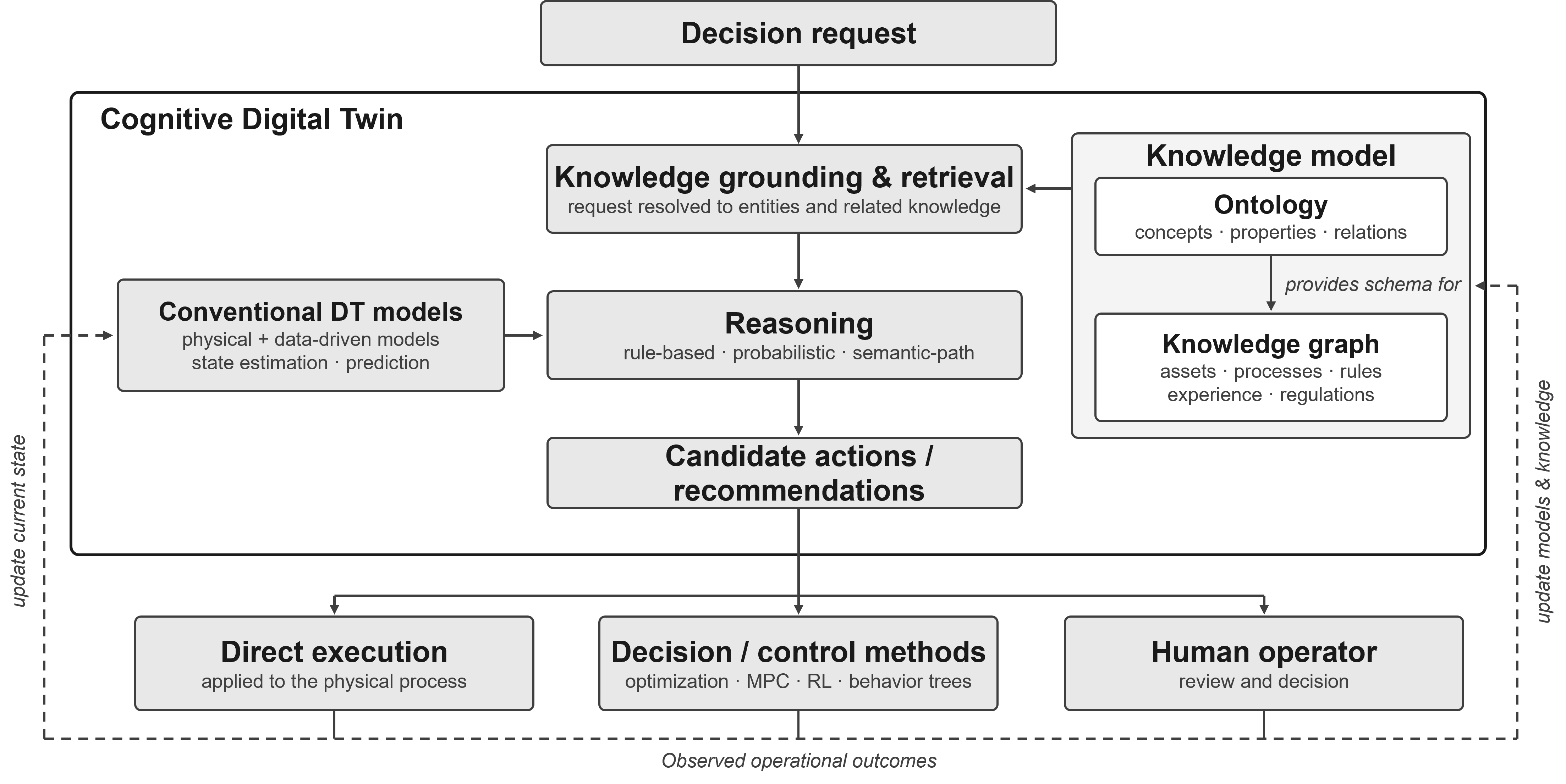}
\caption{Overview of knowledge-based decision-making in a Cognitive Digital Twin. The Cognitive Digital Twin adds an explicit knowledge model and a reasoning mechanism to the predictive models of a conventional Digital Twin.}
\label{fig:cdt}
\end{figure}

Conventional Digital Twins support monitoring, prediction, and optimization, but they remain limited when decisions require semantic understanding, cross-domain knowledge, or adaptation to situations not represented in historical data. Cognitive Digital Twins extend this paradigm by integrating explicit knowledge, contextual reasoning, and learning mechanisms into the digital representation of a physical system~\cite{d2022cognitive,zheng2022emergence,el2022development}. Their objective is not only to estimate system states or optimize predefined objectives, but also to interpret the current situation, relate observations to prior experience and engineering rules, anticipate consequences, and recommend actions under uncertain or unfamiliar conditions~\cite{xu2024cognitive,al2021cognitive,ji2023cognitive,ji2024developing}. This capability is relevant when decisions depend on regulations, design intent, multi-modal sensing, operational history, and knowledge transferred across related tasks or systems~\cite{abburu2020cognitwin,su2024knowledge,yitmen2021adapted}. An illustration of the Cognitive Digital Twin is shown in Figure~\ref{fig:cdt}.

A Cognitive Digital Twin combines physical and data-driven models with a knowledge model and a reasoning mechanism. The knowledge model is commonly organized through an ontology, which defines the concepts, properties, and relationships within and across engineering domains~\cite{steinmetz2018internet,barth2020systematization}. The ontology acts as a shared semantic vocabulary that links sensors, predictive models, physical assets, processes, regulations, and prior experience, thereby supporting interoperability and context awareness~\cite{savic2023towards,su2024knowledge}.

Knowledge graphs provide a computational representation of this ontology through entities and relations~\cite{akroyd2021universal,ramonell2023knowledge}. Decision-making begins by mapping a decision request or observed condition to relevant entities, retrieving connected knowledge, and applying rule-based, probabilistic, or semantic path reasoning to infer candidate actions~\cite{su2024knowledge}. Knowledge graph embeddings extend this process by mapping entities and relations into continuous vector spaces for link prediction and knowledge completion~\cite{ji2021survey,rossi2021knowledge}. Temporal knowledge graphs attach time information to facts and relations so that the Cognitive Digital Twin can reason about system evolution and event dependencies~\cite{wang2023survey_tkg}. Large language models can further support natural-language queries and reasoning, while the knowledge graph provides factual grounding and explicit constraints~\cite{pan2024unifying}. The resulting recommendations may be executed directly or supplied to optimization, model predictive control, reinforcement learning, behavior trees, or human operators~\cite{yitmen2021adapted,sivaneasan2024cognitive,xu2024cognitive}. Operational feedback then updates the physical and knowledge models, allowing the Cognitive Digital Twin to revise relationships, rules, and decision logic as the system evolves~\cite{zhou2020knowledge}.

In practical applications, Cognitive Digital Twins are most relevant when decisions require the integration of heterogeneous engineering knowledge rather than the evaluation of a single predictive model. In tolerance design, knowledge graphs connect CAD models, failure analyses, process diagrams, decision tables, and regulatory information. Graph-ranking methods can then retrieve relevant prior designs and support iterative evaluation of manufacturability, compliance, and cost~\cite{zhang2017graph}. In aero-engine blade manufacturing, a Cognitive Digital Twin integrates geometric visualization, process knowledge, regulations, and decision elements within a knowledge graph. Knowledge matching and semantic path reasoning are used to interpret real-time tool and part conditions and recommend process parameters consistent with manufacturing requirements~\cite{su2024knowledge}. This differs from conventional process optimization because decisions are derived from relationships among manufacturing objects, rules, and operational context rather than from a single numerical objective. As shown in an example in multi-robot manufacturing, Cognitive Digital Twins combine behavior trees, multi-modal perception, reinforcement learning, and transfer learning to decompose tasks, classify process states, plan motions, and transfer knowledge between related assembly operations~\cite{xu2024cognitive}. A related robotic production-line framework integrates geometric, physical, behavioral, and rule models with an ontology expressed in the Ontology Web Language. The system extracts relevant rules from assigned tasks and updates control responses as production conditions change~\cite{ding2025cognitive}. These cases illustrate how Cognitive Digital Twins coordinate several models and knowledge sources to support decisions that require task reasoning, semantic interpretation, and cross-domain integration.

Beyond knowledge-based reasoning, Digital Twins with reasoning capability also offer opportunities to integrate a broader umbrella of decision-making methods, such as preference modeling and utility functions, that depend on subjective judgments. Existing studies have begun incorporating predefined utility functions or human preference feedback into Digital Twin-based decision support \cite{lee2021digital,lehmann2026semantic}. However, these approaches remain largely focused on predefined preferences and offline evaluation rather than continuous real-time decision-making. Nevertheless, future developments in semantic reasoning via LLMs could enable Digital Twins to interpret qualitative requirements and human preferences while leveraging their predictive models to simulate and evaluate the consequences of alternative decisions. This integration could extend Digital Twin decision-making beyond predefined numerical objectives toward more context-aware, preference-driven, and adaptive decision-making.

\subsection{Roles of ML in Enabling Real-Time Decision-Making for Digital Twins}
\label{sec:decision_limitations}

Overall, real-time decision-making in Digital Twins can be realized through optimization-based, policy-based, and knowledge-based approaches, which differ in how the virtual representation is translated into actions and where the decision computation is performed. These approaches also impose different requirements on predictive models, uncertainty representation, computational cost, interpretability, and online adaptability, as summarized in Table~\ref{tab:decision_making_comparison}. 

\begin{table}[h!]
\centering
\caption{Comparison of representative real-time decision-making approaches for Digital Twins.}
\label{tab:decision_making_comparison}
\scriptsize
\begin{tabular}{@{}
>{\raggedright\arraybackslash}p{1.3cm}
>{\raggedright\arraybackslash}p{2.3cm}
>{\raggedright\arraybackslash}p{2.7cm}
>{\raggedright\arraybackslash}p{2.1cm}
>{\raggedright\arraybackslash}p{2.5cm}
>{\raggedright\arraybackslash}p{3.0cm}
@{}}
\toprule
\textbf{Approach} &
\textbf{Decision generation} &
\textbf{Required Digital Twin representation} &
\textbf{Decision horizon} &
\textbf{Treatment of uncertainty} &
\textbf{Online operation} \\
\midrule

Online parametric optimization &
Numerically optimizes decision variables for the current operating condition &
Predictive model relating candidate decisions to system responses and objectives &
Single decision or application-dependent horizon &
Typically uses nominal predictions without explicit uncertainty representation &
Repeated numerical optimization as new system states become available \\

\addlinespace

Nominal MPC &
Optimizes a sequence of future control actions and applies the first action &
Multi-step predictive dynamical model with state feedback &
Finite receding horizon &
Uses nominal future trajectories &
Prediction and constrained optimization are repeated at each control step \\

\addlinespace

Dynamic programming &
Uses Bellman recursion to construct value functions and sequential decision policies &
State-transition model, which may be deterministic or probabilistic &
Finite or infinite horizon &
Can explicitly account for probabilistic state transitions through expected future cost or reward &
Policy synthesis can be computationally intensive, while execution can use a precomputed state-to-action policy \\

\addlinespace

Robust MPC &
Optimizes a receding-horizon control sequence that remains feasible under prescribed disturbances &
Predictive dynamical model and bounded uncertainty set &
Finite receding horizon &
Represents uncertainty using bounded sets, tubes, or worst-case disturbances &
Repeated robust optimization or constraint tightening is performed online \\

\addlinespace

Stochastic MPC &
Optimizes expected performance subject to probabilistic constraints &
Predictive dynamical model and probability distributions of disturbances or future states &
Finite receding horizon &
Represents uncertainty probabilistically through expected objectives and chance constraints &
Repeated stochastic or deterministically reformulated optimization is performed online \\

\addlinespace

Approximate optimizer &
Evaluates a precomputed or learned mapping from the current state to an action &
Offline optimizer, predictive model, or simulated trajectories used to construct the policy &
Typically reproduces a finite-horizon or static optimization policy &
Usually inherits uncertainty treatment from the optimization problem or training data rather than quantifying it explicitly &
Direct evaluation of the policy avoids repeated numerical optimization \\

\addlinespace

Reinforcement learning &
Selects actions using a policy learned from interaction and cumulative reward &
Physical or virtual environment; an explicit predictive model is optional &
Long-term sequential &
Accounts for stochastic transitions through interaction or an MDP formulation, although uncertainty need not be explicitly quantified &
Direct policy evaluation after training enables rapid online decisions; learning may also continue from operational feedback \\

\addlinespace

Fuzzy inference &
Evaluates linguistic rules and membership functions to infer numerical decisions &
Rule base, membership functions, and crisp measurements, estimates, or predictions &
Primarily current-state decision &
Represents imprecise or qualitative information through partial membership rather than probabilistic uncertainty &
Rule evaluation and fuzzy inference are performed directly from current inputs \\

\addlinespace

Cognitive Digital Twin &
Reasons over explicit knowledge, semantic relationships, rules, and contextual information to infer or recommend actions &
Ontology, knowledge graph, reasoning mechanism, and potentially predictive or data-driven models &
Application-dependent &
Can combine qualitative, rule-based, semantic, or probabilistic representations of uncertain information &
Retrieves relevant knowledge and performs reasoning as new observations and decision requests arise \\

\bottomrule
\end{tabular}
\end{table}

Across the literature, successful demonstrations of Digital Twins that support effective real-time decision-making show that ML is an enabler for resolving computational bottlenecks to make online execution possible. Specifically, ML plays three major roles:

\begin{itemize}
    \item \textbf{Constructing predictive surrogate models that balance fidelity with computational efficiency}: Decision methods that repeatedly evaluate system dynamics, such as MPC, require models that balance predictive fidelity with computational efficiency. High-fidelity simulations may be too costly for online use, while simplified models can omit nonlinearities and coupled phenomena relevant to action selection~\cite{thelen2022comprehensive,benner2015survey,he2024toward}. ML can help manage this trade-off through computationally efficient surrogate models that approximate decision-relevant dynamics with sufficient predictive fidelity, and further support downstream decision-making ~\cite{es2024methods,zhao2023surrogate}. Section~\ref{sec:surrogate} reviews these approaches and their suitability for downstream decision tasks.

    \item \textbf{Transforming unstructured, noisy, and multi-modal observations into decision-relevant information}: Decision-making methods generally require the current system state, operating context, and other relevant quantities in a structured form, whereas physical systems are often observed through noisy, incomplete, delayed, and heterogeneous data streams. ML can bridge this gap by extracting and integrating decision-relevant information from structured/unstructured and multi-modal data through learning-based state estimation, virtual sensing, representation learning, and multi-modal fusion. Moreover, it provides reasoning capability on historical context and can convert knowledge into machine-executable knowledge. Section~\ref{sec:perception&reasoning} reviews these approaches as the informational interface that transforms physical observations into the states and conditioning information required for subsequent prediction and decision-making.

    \item \textbf{Accelerating decision synthesis through solver assistance or learned policies}: Even when sufficiently fast predictive models are available, methods such as MPC, DP, and probabilistic decision frameworks may remain computationally prohibitive due to the repeated optimization, inference, or policy evaluation required to determine actions online. ML can reduce this computational burden by assisting conventional solvers through trajectory prediction, warm-starting solvers, active-set prediction, value-function approximation, and learned search strategies, or by learning policies that directly generate candidate actions or trajectories. Section~\ref{sec:ml_decision} reviews these approaches and examines how ML can accelerate decision synthesis while retaining the objectives, constraints, and decision structure of the underlying formulation.
\end{itemize}
    
This perspective also implies that ML is not a necessary component of a Digital Twin when the required prediction, information processing, and decision synthesis can already be performed within the available operational time. When ML is introduced to overcome these limitations, however, its predictions and decisions must be continuously assessed for reliability under changing operating conditions. Section~\ref{sec:uncertainty} therefore examines verification, validation, and UQ, together with model adaptation, as complementary capabilities for maintaining trustworthy real-time decision-making for ML-enabled Digital Twins.

\section{Machine Learning in Surrogate Modeling for Real-Time Decision-Making}
\label{sec:surrogate}

Modeling is central to Digital Twins because it provides a computational representation of the physical system for predicting its behavior and evaluating candidate decisions. For complex engineering systems, however, these capabilities face a trade-off between model fidelity and computational efficiency. The governing physics may be partially known or difficult to represent accurately, while high-fidelity simulations may be too computationally expensive to support online decisions \cite{thelen2022comprehensive}. This challenge becomes particularly acute when decision-making requires repeated model evaluations within the optimization, rollout, and inference methods discussed in Section~\ref{sec:decision_making}. ML-based surrogate models offer a way to mitigate this trade-off by learning system responses from physics-based simulations, operational data, or both, and approximating these responses at substantially lower computational cost. Their value for real-time decision-making, however, depends on how well they satisfy the requirements of the decision-making methods in which they are embedded.

Existing reviews provide extensive coverage of surrogate-modeling methods and their applications. Thelen et al.~\cite{thelen2022comprehensive} survey surrogate-modeling methods and frameworks, while Es-haghi et al.~\cite{es2024methods} emphasize methods that support real-time applications. Yu et al.~\cite{yu2024learning} review data-driven approaches for modeling dynamical systems, Kumar et al.~\cite{kumar2023state} categorize reduced-order modeling methods, and Barkanyi et al.~\cite{barkanyi2021modelling} examine surrogate-model construction and validation for Digital Twins. Complementing these surveys, the National Academies report on foundational research gaps for Digital Twins~\cite{national2023foundational} identifies predictive modeling at actionable speed and scale as a core research need. These reviews establish a foundation for developing accurate and computationally efficient surrogates, but the requirements imposed by their use in closed-loop decision-making warrant further examination.

Motivated by this need, the review in this section aims to answer the question: "\emph{How can ML-based surrogate models mitigate the trade-off between fidelity and computational efficiency to enable real-time decision-making?}" We take the decision-making methods in Section~\ref{sec:decision_making} as the starting point and identify the surrogate-modeling requirements they impose, the resulting design considerations, and the ML approaches that can address them. Note that the focus in this section is on forward prediction of temporal "what-if" scenarios, i.e., predicting how system dynamics evolve under different candidate inputs or actions. Surrogate models for state estimation and inverse problems are discussed separately in Section~\ref{sec:perception&reasoning}.

When a Digital Twin is viewed as a closed-loop system operating in real time for decision support, the requirements of its surrogate model differ from those of surrogate models developed for offline analysis, which are mainly restricted by computational speed, sensor availability, and connection with downstream decision-making. Based collectively on the literature, four requirements can be identified for surrogate models intended to support real-time Digital Twins applications \cite{thelen2022comprehensive,es2024methods,galvez2025decision, national2023foundational}:

\begin{enumerate}
    \item The surrogate should provide a dynamic state representation that can be updated from system feedback and interfaced with the control and optimization methods discussed in Section~2.

    \item Its internal representation should retain system behavior that is relevant to the available observations and control actions, thereby connecting model states to quantities measured by sensors and influenced by actuators.

    \item It should represent the system dynamics with sufficient fidelity over the operating conditions relevant to downstream decision-making, including nonlinear behavior when it affects the resulting decisions.

    \item It should remain computationally efficient enough to support the repeated simulations and model evaluations required for real-time decision-making.
\end{enumerate}

Whether these requirements can be satisfied in practice depends not only on the architecture of the surrogate model, but also on the information available during development and operation and on how the surrogate is ultimately used by the decision layer. Several practical considerations therefore influence surrogate selection and construction.

\begin{itemize}

    \item \textbf{Available Data Regime and Distribution.}
    The source, quality, and coverage of the available data determine what behavior a surrogate can reliably learn. Simulation data can often provide synchronized and fully observed states, whereas experimental data may be noisy, incomplete, asynchronously sampled, and limited to indirect measurements \cite{kirkpatrick2017overcoming,wu2019local,kapteyn2022design}. Moreover, operational data collected under feedback may cover only a narrow portion of the feasible domain, leaving abnormal, transient, or near-constraint conditions poorly represented \cite{celestini2024transformer,feng2023model,desai2023enhanced}. The available data should therefore be assessed according to their informativeness for the intended operating region, rather than their volume alone.

    \item \textbf{State Observability and Representation.}
    The state representation used by the surrogate should retain the information required for prediction and decision-making while remaining observable from available measurements and computationally tractable. Physical systems are often only partially observable because of limitations in sensor placement, so decision-relevant states may need to be estimated or inferred from indirect measurements and their history \cite{feng2023model,kapteyn2022design}. This challenge becomes more pronounced for nonlinear and spatially dependent systems, where directly representing high-dimensional fields or multi-modal measurements may also be prohibitively expensive for real-time operation \cite{zhang2025potential,chen2025continuous,chakrabarty2023moving,san2023decentralized,zhou2024multimodal,kaigom2025deep}. Conversely, excessive reduction can remove information needed to characterize the system response or support downstream control. The appropriate representation should therefore balance observability, physical detail, dimensionality, and the information required by the intended decision task.

    \item \textbf{Temporal Representation and Rollout Stability.}
    The performance of the surrogates should be evaluated over the prediction horizon required by the decision method rather than solely through one-step accuracy. Small prediction errors can accumulate during rollout, while differences between training and deployment sampling resolutions can further degrade trajectory predictions \cite{de2025challenges,thelen2023comprehensive,schar2025surrogate}. The training objective should therefore be capable of mitigating error accumulation, and the temporal resolution should be synchronized during deployment.

    \item \textbf{Compatibility with Online Solving.}
    Surrogate modeling and decision synthesis should be considered jointly because model structure directly affects the complexity of downstream numerical solution. A highly accurate surrogate may still be unsuitable if its dimensionality, nonlinearity, or differentiation requirements prevent decisions from being synthesized at an actionable frequency. Model expressiveness must therefore be balanced against computational tractability and, where relevant, stability and feasibility requirements \cite{yu2024learning}.

\end{itemize}

Overall, conventional data-driven surrogates (Section \ref{sec:surrgate_exp_sim}) provide the computational efficiency and flexibility required for real-time Digital Twin decision-making, but their performance remains fundamentally dependent on the information contained in the training data, leaving challenges in data scarcity, partial observability, extrapolation, and the representation of uncertain or non-unique system responses. The following sections progressively extend this data-driven foundation by introducing additional forms of information and predictive capability. Scientific machine learning (Section~\ref{sec:surrogate:SciML}) incorporates mechanistic knowledge into surrogate construction, providing physical inductive biases that can improve data efficiency, physical consistency, and generalization. Generative models (Section~\ref{sec:surrogate_gen}) then extend surrogate modeling beyond deterministic input--output mappings by learning conditional response distributions, enabling the representation of stochastic, multi-modal, or non-unique system behavior. Finally, multi-fidelity modeling (Section~\ref{sec:surrogate_MF}) addresses the broader problem of information integration by combining data from sources with different levels of accuracy, resolution, and cost, including simulations, experiments, and operational measurements. Their implications and compatibility with real-time Digital Twin decision-making are further discussed in Section~\ref{sec:remaining_gap_surrogate}. Together, these approaches represent a progression toward surrogate models that can exploit broader sources of knowledge and provide richer predictive capabilities beyond what can be learned from a single data source alone.

\subsection{Conventional Data-Driven Methods for Surrogate Modeling}
\label{sec:surrgate_exp_sim}

\begin{figure}[h!]
\centering
\includegraphics[width=0.7\linewidth]{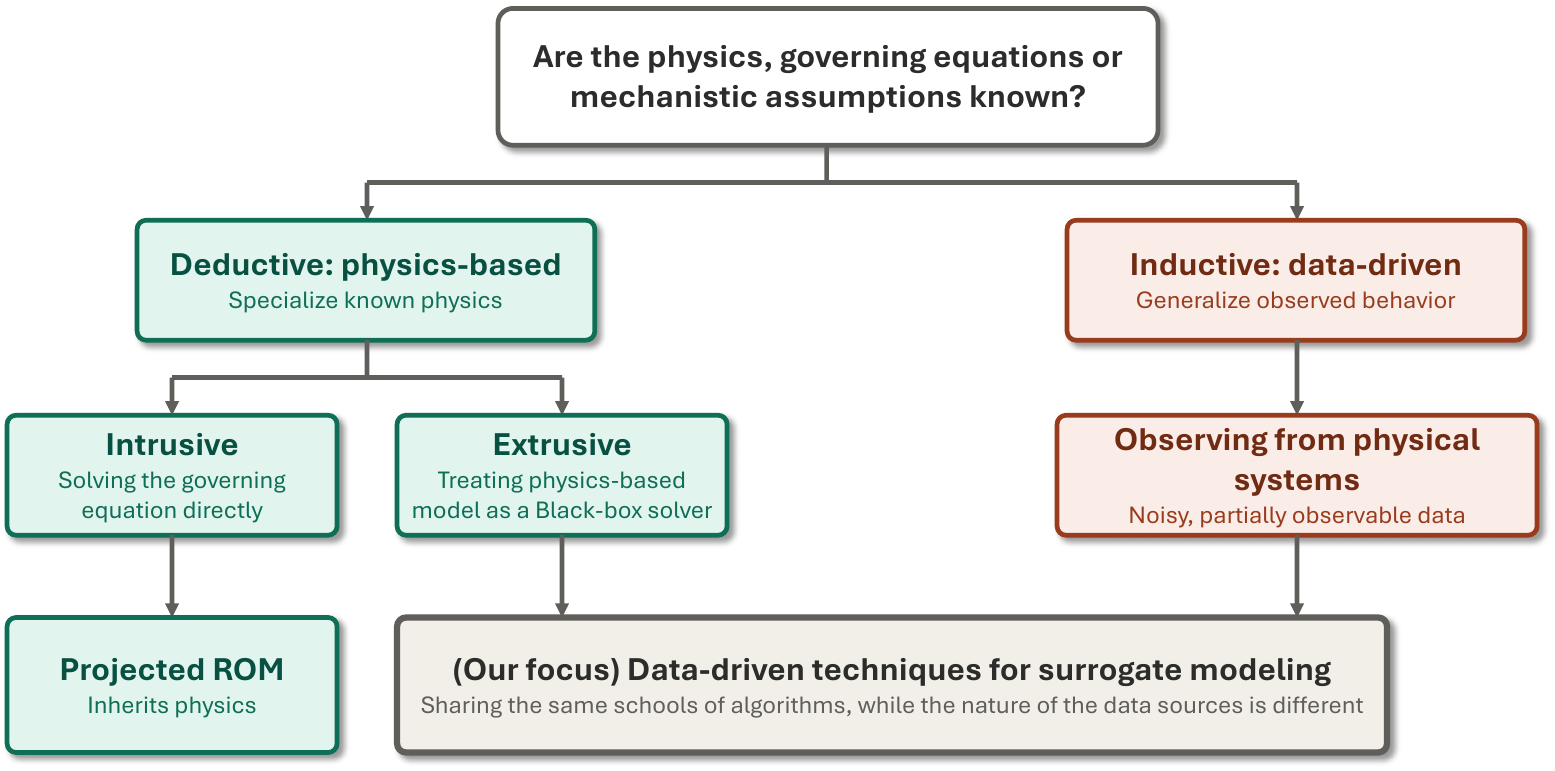}
\caption{Position of data-driven surrogate modeling within the broader modeling landscape. Deductive approaches begin with governing equations, whereas inductive approaches learn from observations. Extrusive physics-based surrogates and purely data-driven surrogates may use the same learning algorithms but differ in whether their training data originate from simulation or from the physical system. This review focuses on the selection and use of these shared data-driven algorithms for real-time decision-making.}
\label{fig:data_driven_surrogate}
\end{figure}

For selecting methods for surrogate modeling, a fundamental distinction concerns the source of knowledge used to construct the surrogate, i.e., from governing equations or learned from data directly. As illustrated in Figure~\ref{fig:data_driven_surrogate}, deductive approaches begin with established governing equations, whereas inductive approaches learn system behavior from observations \cite{kutz2025accelerating}. One data-driven setting arises when a reliable high-fidelity simulator is available but is too computationally expensive for repeated use in optimization, control, inference, or policy evaluation. In this case, the simulator is treated as a black box to generate training data for an extrusive, or non-intrusive, surrogate through a “digital-to-digital” learning process\cite{chen2025targeted,ganguli2020digital,chakraborty2021role}. When the governing equations are unavailable, incomplete, distributed across interacting components, or difficult to formulate, the same learning algorithms can instead be trained directly from sensor or experimental data, as commonly encountered in supply chains, manufacturing systems, and multi-component engineering systems \cite{abideen2021digital,cimino2019review,tao2018digital,abadi2025leveraging,jafari2023review}. Although these two settings differ substantially in the quality, observability, and coverage of their training data, they share many of the same learning algorithms. This subsection therefore focuses on these conventional data-driven methods rather than projection-based ROM methods, which have been reviewed comprehensively in \cite{kumar2023state}.

Across these two settings, representative model families can be roughly divided into statistical system-identification methods \cite{keesman2011system, choi2012arma}, Gaussian processes (GP) \cite{su2017gaussian, bhattacharyya2022uncertainty}, sparse system identification \cite{kaiser2018sparse}, time series models, latent dynamics models \cite{masti2021learning}, and graph-based surrogate models \cite{jiang2025g}. Although the source of training data determines important issues such as noise, observability, and operating-region coverage, model selection also depends on the structure of the states and their compatibility with downstream decision-making. State-space and autoregressive models are natural choices when recursive state updates and temporal prediction are required; probabilistic models are useful when data are limited, or uncertainty must be quantified, and more expressive NN-based models are better suited to strongly nonlinear or high-dimensional systems. Table~\ref{tab:surrogate_families} summarizes the principal model families and their properties for the Digital Twin decision loop.

\begin{table}[h!]
\centering
\caption{Representative data-driven surrogate modeling families and their properties for real-time decision-making in Digital Twins}
\label{tab:surrogate_families}
\renewcommand{\arraystretch}{1.3}
\setlength{\tabcolsep}{3pt}
\footnotesize
\begin{tabular}{
>{\raggedright\arraybackslash}p{0.17\textwidth}
>{\raggedright\arraybackslash}p{0.16\textwidth}
>{\raggedright\arraybackslash}p{0.18\textwidth}
>{\raggedright\arraybackslash}p{0.23\textwidth}
>{\raggedright\arraybackslash}p{0.20\textwidth}}
\toprule
\textbf{Family} &
\textbf{Learning basis/data requirement} &
\textbf{Representation} &
\textbf{Strengths for the decision-making} &
\textbf{Key limitations} \\
\midrule

Parametric system identification, ARX/NARX
\cite{keesman2011system,thelen2022comprehensive}
&
Time-series simulation or sensor data
&
Explicit states or autoregressive histories
&
Interpretable dynamics, natural feedback updates, and direct integration with MPC and filtering
&
Limited flexibility for highly nonlinear or high-dimensional dynamics
\\

Gaussian processes
\cite{su2017gaussian,bhattacharyya2022uncertainty}
&
Typically modest paired datasets
&
Nonparametric input--output or state-transition mapping
&
Data-efficient learning with uncertainty estimates for risk-aware decisions
&
Computational scaling with dataset size and input dimension
\\

Sparse identification, SINDy
\cite{kaheman2020sindy}
&
Measurements and a candidate function library
&
Sparse explicit dynamical equations
&
Interpretable dynamics with compact models for prediction and diagnosis
&
Sensitive to noise, state observability, and library specification
\\

Latent dynamics models, autoencoders
\cite{wu2023digital,zhu2023reduced,hassanpour2020integrating,masti2021learning}
&
High-dimensional simulation or sensor data
&
Reduced coordinates or learned latent states
&
Reduces prediction and optimization dimension while retaining dominant system information
&
Reduced states may omit decision-relevant information or exhibit unstable latent dynamics
\\

Neural network (NN) dynamical models, NARX encoders, recurrent neural network and time-series models
\cite{zhao2023surrogate,drgona2022differentiable,chen2025uncertainty,masti2021learning}
&
Moderate to large time-series datasets
&
Explicit, history-dependent, or latent states
&
Expressive nonlinear dynamics and flexible multi-step prediction
&
Data-intensive, limited extrapolation, and weak rollout or stability guarantees
\\

DMD and Koopman operator models
\cite{koo2023digital,marti2025optimization,chen2025adaptivedigitaltwinsheet}
&
Simulation or sensor trajectories
&
Linear dynamics in reduced or lifted coordinates
&
Can formulate downstream optimization as convex optimization and enable efficient MPC
&
Performance depends strongly on the selected observables or embedding
\\

Graph-based surrogate models
\cite{jiang2025g,lin2025g,ciklamini2025enhancing,zacarias2025effectiveness}
&
Relational, spatial, or component-level data
&
Graph-structured states and interactions
&
Preserves spatial and topological dependencies and supports scalable structured prediction
&
Requires appropriate graph construction and may increase training complexity
\\

\bottomrule
\end{tabular}
\end{table}

The model families in Table~\ref{tab:surrogate_families} provide different predictive structures, but selecting a family alone does not resolve the requirements and practical considerations mentioned in the beginning of this section. Several design strategies can be applied across these model classes to improve how the surrogate uses available information, represents the system state and dynamics, and interfaces with the decision layer.

\begin{enumerate}

    \item \textbf{Data Collection in Temporally and Spatially Correlated Spaces.}
    A key challenge in constructing surrogate models for dynamical Digital Twins is designing experiments over the correlated state--input space $(\mathbf{x}_k,\mathbf{u}_k)$. Unlike computer experiments, where $\mathbf{y}_k=f(\mathbf{x}_k)$ can often be evaluated directly at a selected query point $\mathbf{x}_k$, samples $(\mathbf{y}_k,\mathbf{x}_k,\mathbf{u}_k)$ from a dynamical system are constrained by system evolution and reachability from the initial condition. Consequently, arbitrary regions of the state--input space may be difficult or costly to access, leading to imbalanced datasets and limited coverage of the online operating regime. Several approaches can improve data collection. Fisher information can quantify the informativeness of candidate trajectories and support optimal experiment design. Information gain can also be incorporated into the MPC objective so that the controller actively drives the system toward informative regions while satisfying operational objectives \cite{abraham2019active}. Physics-based knowledge can further reduce dependence on dense data coverage, as discussed in Section~\ref{sec:surrogate:SciML}. Finally, because offline open-loop and online closed-loop trajectories can exhibit different data distributions \cite{celestini2024transformer}, online surrogate adaptation may be needed to maintain predictive fidelity, as discussed in Section~\ref{sec:uncertainty}.

    \item \textbf{Handling Partial Observability and Limited Information.}
    When the available measurements do not fully characterize the system state, a surrogate can compensate for missing information either through explicit state estimation, discussed in Section~\ref{sec:perception_state}, or by incorporating measurement history into the predictive model \cite{fu2020learning}. Finite memory models therefore use past observations and inputs to recover temporally dependent information that is unavailable from instantaneous measurements alone \cite{drgona2022differentiable,karkaria2025asno,chen2025real}. Alternatively, optimal experimental design and sensor placement can improve the informativeness of the collected data by increasing observability and information gain \cite{karnik2024constrained,kapteyn2022design,tian2024novel,jacoby2020digital}.

    \item \textbf{Learning Decision-Relevant State Representations.}
    Once sufficient information is available, the next question is how it should be represented within the surrogate. For high-dimensional observations, principal component analysis (PCA), dynamic mode decomposition (DMD), and encoder-based latent dynamical models can extract reduced coordinates that retain dominant system behavior while lowering the dimension of the prediction problem \cite{wu2023digital,zhu2023reduced,hassanpour2020integrating,koo2023digital,marti2025optimization,suk2016state,paniagua2024nonlinear}. For example, a sheet metal forming Digital Twin uses PCA to represent the deformation field through a small number of modal coefficients, whose evolution is modeled with a Koopman operator for online MPC \cite{chen2025adaptivedigitaltwinsheet}. When reconstruction of the broader state is unnecessary, targeted modeling can instead focus on quantities directly relevant to sensing, constraints, and decision-making \cite{chen2025targeted}. In directed energy deposition, melt pool temperature and depth serve as such decision-relevant states and are predicted using a TiDE-based NARX model for laser power control \cite{chen2025real,das2023long}. For relational systems, graph-based representations provide another alternative by preserving spatial and topological dependencies \cite{jiang2025g,lin2025g,ciklamini2025enhancing,zacarias2025effectiveness}.

    \item \textbf{Learning Stable Rollout Dynamics.}
    An appropriate state representation must also remain predictive over the multi-step-ahead rollout required by the decision layer. Dynamical surrogates can therefore be trained using multi-step objectives that penalize errors over recursive rollout rather than optimizing only one-step prediction accuracy \cite{masti2021learning,boulle2023mathematical,chen2025adaptivedigitaltwinsheet}. The temporal parameterization of the surrogate also matters. Discrete-time models such as NARX and recurrent networks integrate naturally with many decision methods, but can accumulate errors and remain tied to the sampling resolution used during training \cite{schar2025surrogate}. Continuous-time or resolution-independent formulations, including functional NARX and neural ODEs, provide alternatives when predictions must remain consistent across different temporal resolutions \cite{schar2025surrogate,zheng2025physics,chen2025continuous,kannapinn2024digital}. Related scientific machine learning formulations are discussed further in Section~\ref{sec:surrogate:SciML}.

    \item \textbf{Selecting Surrogate Structures for Computationally Efficient Decision-Making}
    Beyond predictive performance, the structure of a surrogate can be designed to enhance the efficiency and performance of downstream decisions. This is relevant for MPC and other online optimization methods, where dimensionality, nonlinearity, and gradient computation directly affect solution speed. Koopman-based methods, for example, seek a lifted representation in which nonlinear dynamics evolve approximately linearly, while recent operator-theoretic approaches further seek representations that induce linear or convex optimization structures directly from data \cite{lusch2018deep,mezic2013analysis,houska2025convex}. Such representation introduces a tradeoff between predictive expressiveness and computational tractability, since restrictive structures may reduce fidelity while highly expressive nonlinear models can increase temporal optimization cost \cite{yu2024learning}. When repeated numerical optimization remains prohibitive, learned policies or ML-enhanced optimization can instead be used to accelerate decision synthesis, as discussed in Section~\ref{sec:accelerate_solving}.
\end{enumerate}

Overall, conventional data-driven surrogates offer the computational efficiency and flexibility required for real-time Digital Twin decision-making, but their reliability remains strongly tied to the quality, quantity, the coverage of the training data, as well as the data source itself, including the fidelity or the complexity of the simulation/experiment. When mechanistic knowledge of the physical system is available, it provides an additional source of information that can reduce this dependence on data alone. This motivates scientific machine learning approaches, introduced in the next section, that incorporate physical knowledge and structure into surrogate modeling to improve properties such as data efficiency, physical consistency, and generalization.

\subsection{Scientific Machine Learning}
\label{sec:surrogate:SciML}

Scientific machine learning (SciML) extends conventional data-driven surrogate modeling by incorporating scientific structure into the learning problem, either through mechanistic knowledge or through representations tailored to the mathematical structure of physical systems. In contrast to the conventional methods discussed in Section~\ref{sec:surrgate_exp_sim}, SciML can exploit governing equations, conservation laws, boundary conditions, symmetries, constitutive relations, or the functional structure of solution operators~\cite{karniadakis2021physics,cuomo2022scientific}. These structures are particularly useful when observations are sparse or when the quantities required for decision-making are high-dimensional physical fields that are difficult to represent using conventional finite-dimensional mappings.

\begin{figure}[h!]
    \centering
    \includegraphics[width=0.8\linewidth]{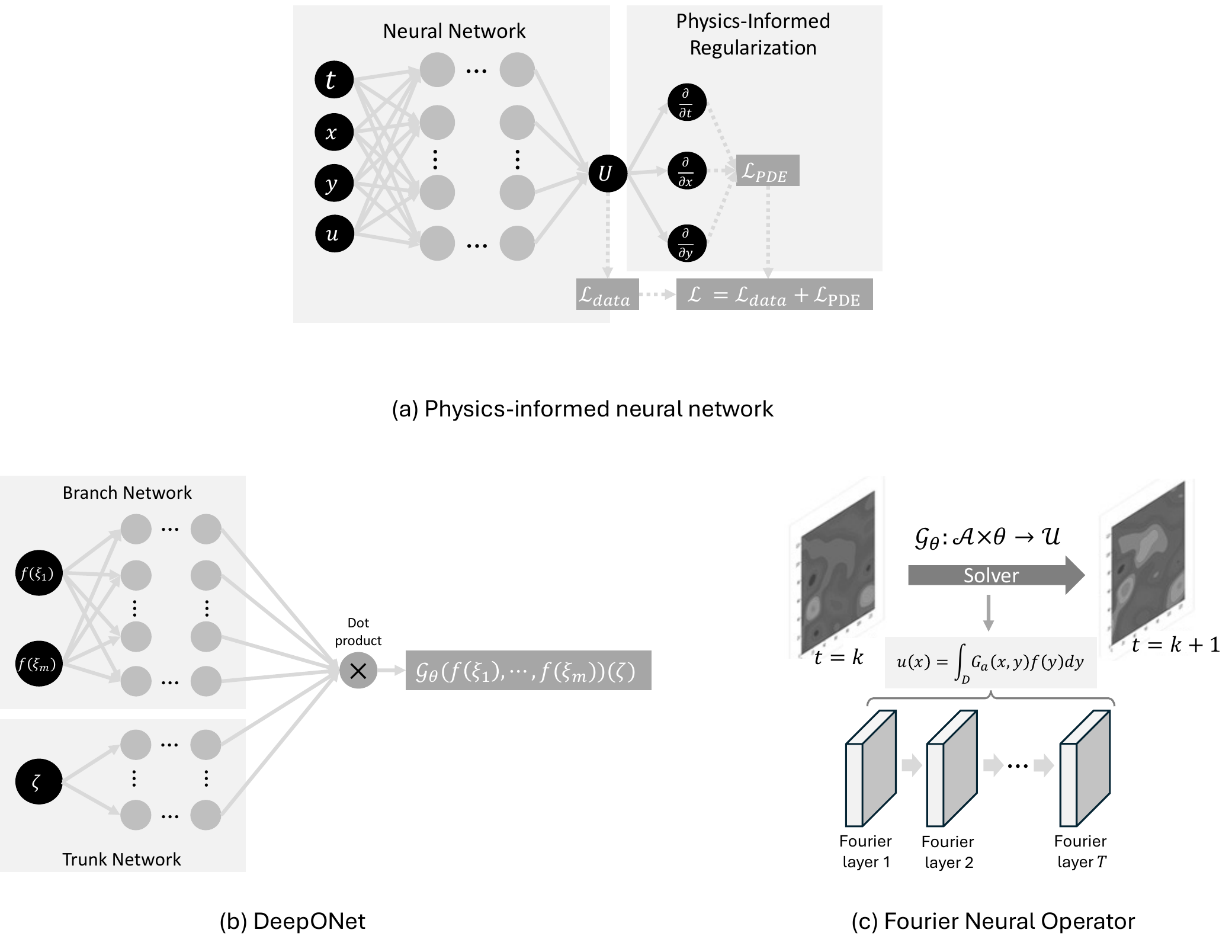}
    \caption{Representative scientific machine learning approaches for surrogate modeling. (a) Physics-informed neural network (PINN), where the neural network maps spatial and temporal coordinates $(t,x,y)$, as well as the control input $u$ to the predicted field $U$. Training minimizes a composite objective, $\mathcal{L}=\lambda_d\mathcal{L}_{\mathrm{data}}+\lambda_f\mathcal{L}_{\mathrm{PDE}}$, combining agreement with observations, governing equations, and/or initial conditions. (b) Deep Operator Network (DeepONet), where a branch network encodes sampled values of the input function, $f(\xi_1),\ldots,f(\xi_m)$, and a trunk network encodes the query coordinate $\zeta$. Their outputs are combined to approximate $\mathcal{G}_{\theta}(f)(\zeta)$. (c) Fourier neural operator (FNO), which learns field-to-field mappings through Fourier layers that parameterize integral operators in the frequency domain, enabling rapid prediction without repeated numerical solution.}
    \label{fig:sciml}
\end{figure}

SciML methods have been reviewed extensively, and readers seeking broader taxonomies and methodological details are referred to~\cite{karniadakis2021physics,cuomo2022scientific,azizzadenesheli2024neural,yu2024learning}. Here, we focus on physics-informed neural networks (PINNs) and neural operators because they represent two complementary pathways particularly relevant to Digital Twin surrogate modeling. PINNs primarily follow a deductive pathway in which known governing equations, boundary conditions, or conservation laws are incorporated into the learning objective and data are used to infer unknown states, parameters, or solutions~\cite{raissi2019physics,lu2021deepxde}. It relaxes the ill-posed statistical learning problem by constraining the admissible hypothesis space toward functions that approximately satisfy the governing physical laws, with equation residuals and boundary or initial condition errors penalized alongside data mismatch.
As illustrated in Figure~\ref{fig:sciml}(a), the relative weights of these terms determine the balance between physical consistency and agreement with observations~\cite{karniadakis2021physics}.

Neural operators follow a different pathway by learning mappings between functions rather than between finite-dimensional input and output vectors. Represented here by DeepONet~\cite{lu2021learning} and the Fourier neural operator (FNO)~\cite{li2020fourier}, they approximate a solution operator $\mathcal{G}$ with a learned operator $\mathcal{G}_{\theta}$ that maps input functions, such as an initial condition, boundary condition, coefficient field, or control input, to a corresponding solution field. In DeepONet, a branch network encodes samples of the input function while a trunk network encodes the query coordinate, and the predicted solution at that location is obtained through the inner product of their output vectors, optionally with an additive bias, as shown in Figure~\ref{fig:sciml}(b). FNO instead applies learned transformations to a truncated set of Fourier modes, allowing spatially distributed mappings to be learned efficiently. Once trained, these models can emulate a family of PDE solutions through forward prediction rather than solving a single problem instance. When the architecture is discretization-independent, the learned operator can also be evaluated at spatial or temporal locations that differ from those used during training. The distinction between these pathways is not absolute, as physics-informed operator learning can augment neural operator training with governing equations and boundary residuals, combining operator learning with secondary loss terms~\cite{wang2021learning}.

\begin{figure}[h!]
    \centering
    \includegraphics[width=0.95\linewidth]{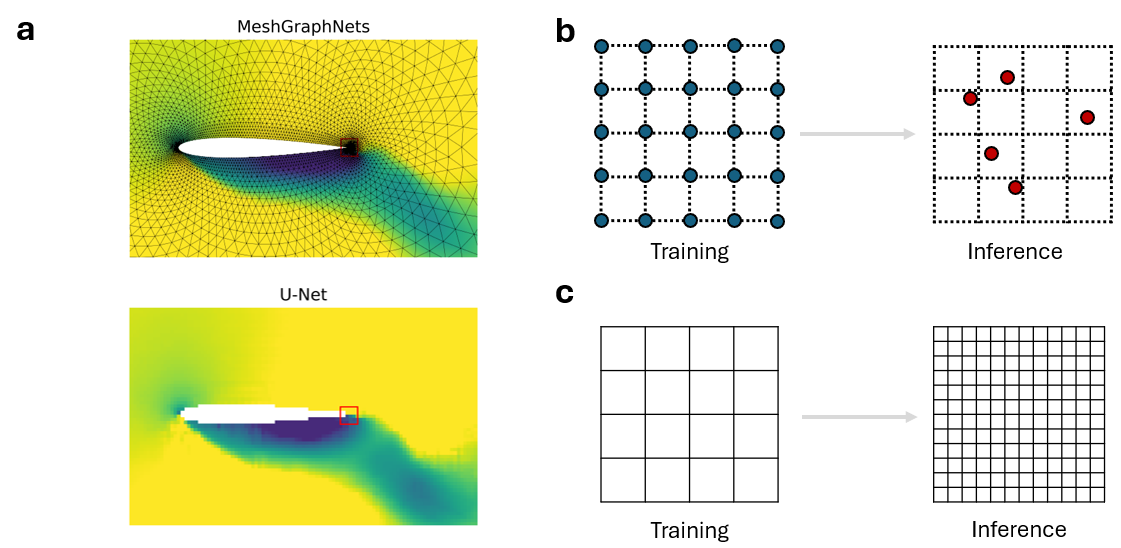}    
    \caption{Merits of SciML models as a full-field surrogate for DT purposes. (a) Learning from non-uniform discretization (meshes) with an example of the airfoil benchmark (reconstructed from ref.~\cite{pfaff2020learning} using the public data and code). Some SciML models can take non-uniform meshes (bottom), while conventional image processing ML models are fed with pixelated images (bottom, 128$\times$128). (b) Queries at any arbitrary points beyond the training meshes. This supports the ``continuous observability''. (c) Low-to-high resolution transfer across uniform grids, a simple instantiation of discretization invariance.}
    \label{fig:sciml_benefit}
\end{figure}

These formulations address several of the practical requirements identified at the beginning of this section, such as the rollout stability and temporal resolution. By leveraging the given/learned functions, the surrogates can yield less error accumulation when rolling out long state predictions and better out-of-distribution (OOD) predictions \cite{karkaria2025asno} because they encode governing relationships that can remain valid across time and changing operating conditions. By combining sparse or noisy measurements with physical constraints, PINNs and neural operators can infer states and spatial fields that are not directly measured, and support prediction of high-dimensional fields while avoiding the repeated solution of the underlying governing equations. Their ability to evaluate solution fields at query locations or resolutions different from the training discretization can further support decision-making when the required spatial or temporal representation mismatches with the available sensing or simulation grid~\cite{lu2021learning,li2020fourier,kovachki2023neural}. With the full-field and discretization-free prediction capability, they can also support virtual sensing and state estimation~\cite{liu2024deep,iranshahi2025digital}, which is a feature that the SciML models heavily rely on in Section~\ref{sec:perception_state}. When mechanistic regularization is incorporated into the learning objective or model structure, it can also improve sample efficiency and discourage physically inconsistent predictions. These features may improve generalization beyond the observed data when the encoded physics remains valid in the new operating regime~\cite{kobayashi2024deep}.  These properties make SciML attractive when a Digital Twin must rapidly predict spatially distributed physical responses under changing operating conditions. Representative benefits are summarized in Figure~\ref{fig:sciml_benefit}.

The features of SciML, including improved OOD generalization, greater data efficiency during training, more effective representation of complex spatial and temporal dynamics, and enhanced physical consistency through the incorporation of governing laws and constraints, can improve decision-making when these models are embedded directly in the decision loop. By enabling more reliable prediction with stable long-horizon rollout, efficient use of data and improved spatial representation, they can not only enhance the performance of MPC but also RL for better generalization. For example, Nicodemus et al.~\cite{nicodemus2022physics} use a PINN to approximate system flow maps for MPC, reducing the amount of trajectory data required for surrogate construction while supporting prediction under sparse observations. De et al.~\cite{de2025deep} develop a multi-step DeepONet that maps the current state and candidate control sequence directly to a future output trajectory in a single evaluation, reducing recursive error accumulation and the computational cost of MPC. Zhao et al.~\cite{zhao2025physics} combine a physics-informed neural operator (PINO) with reinforcement learning for turbulent flow control, enabling model-based policy evaluation across Reynolds numbers and effective drag reduction at an unseen operating condition.

Despite these advantages, SciML introduces its own practical limitations. Physical residuals improve consistency but do not eliminate the sim-to-real gap, particularly when the assumed governing equations omit relevant physics or cease to describe the physical system accurately. Standard PINNs are also commonly trained for a particular problem instance, so changes in boundary conditions, loads, or material parameters may require retraining unless these quantities are explicitly parameterized. Furthermore, training physics-informed models can be challenging because data-fitting, governing-equation, and boundary-condition losses may produce imbalanced gradients, leading to slow or unsuccessful convergence even when the underlying physics is correctly specified. Neural operators are better suited to rapid evaluation across a family of conditions, but their reliability remains dependent on the range of functions and operating regimes represented during training. 

The interaction between physical structure and online adaptation creates an additional challenge for Digital Twin deployment. Under distribution shift, the constitutive priors given by the physics can support generalization when the encoded physics remains valid, but may hinder adaptation when discrepancies arise from parameter drift, changing boundary conditions, or missing physics. This difficulty is amplified when a SciML model trained offline on dense or full field data must be updated using sparse, noisy, and partially observable measurements, which may be insufficient to serve as ground truth and support fine-tuning and adaptation. Full field prediction is therefore both a strength and a challenge: it could provide rich spatio-temporal information that may enhance decision-making, but many inferred quantities cannot be directly validated from limited sensor measurements. Reliable deployment consequently requires observability analysis or experimental design to determine which quantities are identifiable, complemented by residual, conservation, or other consistency checks for unobserved states. To better align model complexity with the information available during operation, SciML models could serve as offline teachers whose full-field predictions are distilled into compact surrogates targeting observable, decision-relevant quantities. This approach could reduce the computational burden of inference and adaptation while enabling validation against available measurements, although errors inherited from the teacher would still require assessment through physical observations and consistency checks.

Overall, SciML models provide a means of exploiting mechanistic knowledge and the functional structure of physical systems to improve data efficiency, physical consistency, and high-dimensional prediction, which benefit real-time decision-making. Their effectiveness, however, depends on the validity of the incorporated physics, the coverage of the training conditions, and the observability of the predicted states during deployment. These considerations should therefore be evaluated alongside the predictive and computational advantages of SciML when the models are integrated into the Digital Twin decision loop.

\subsection{Generative Models for Surrogate Modeling}
\label{sec:surrogate_gen}

Although generative models were not originally developed specifically for surrogate modeling, they have increasingly been adopted as probabilistic surrogate models in Digital Twin applications where representing a distribution of possible system responses is more appropriate than predicting a single deterministic output. Unlike conventional surrogate models that approximate a deterministic input--output mapping, generative surrogates learn the conditional distribution of possible system responses, enabling Digital Twins to represent multiple plausible states or future trajectories under stochasticity, ambiguity, and incomplete information for downstream decision-making. Specifically, the distribution $\hat{\mathbf{y}}\sim p(\mathbf{y}\mid\mathbf{x})$ is conditioned on system information $\mathbf{x}$, such as current states, control actions, parameters, boundary conditions, or sparse observations.
This formulation is useful when system evolution is stochastic, multi-modal, or not uniquely determined by the available information. In such cases, similar observed inputs may correspond to several plausible states or future outcomes, whereas conventional point-prediction models may average across these possibilities. Generative models can instead preserve distinct response modes and produce multiple candidate trajectories, fields, or system states. When conditioned on system information, these models can also support state reconstruction, missing-data completion, and prediction from incomplete or heterogeneous observations, enabling uncertainty-aware decision-making beyond fixed point-to-point prediction \cite{shao2025generative,huang2024digital}.

This subsection does not aim to provide another general review of generative-model architectures \cite{huang2024digital,chiaro2025generative,lee2025generative}. Instead, it examines their use for surrogate modeling in real-time Digital Twin decision-making, with emphasis on the conditioning information they require, the responses they generate, their inference characteristics, and the decision tasks they support. These distinctions are summarized in Table~\ref{tab:generative_surrogates} for five representative model families: GANs, VAEs, autoregressive Transformers, diffusion models, and normalizing flows.

\begin{table}[h!]
\centering
\caption{Generative model families and their suitability as surrogate models for real-time Digital Twin applications.}
\label{tab:generative_surrogates}
\renewcommand{\arraystretch}{1.25}
\setlength{\tabcolsep}{4pt}
\footnotesize
\begin{tabular}{
>{\raggedright\arraybackslash}p{0.1\textwidth}
>{\raggedright\arraybackslash}p{0.17\textwidth}
>{\raggedright\arraybackslash}p{0.17\textwidth}
>{\raggedright\arraybackslash}p{0.25\textwidth}
>{\raggedright\arraybackslash}p{0.25\textwidth}}
\toprule
\textbf{Model family} &
\textbf{Typical inputs} &
\textbf{Generated outputs} &
\textbf{Suitable Digital Twin uses} &
\textbf{Key limitations} \\
\midrule

Generative adversarial networks (GANs)
\cite{chen2025generative,wang2020deep}
&
System states, control actions, parameters, sensor measurements, and a latent sample
&
Images, spatial fields, fault signatures, or short trajectory segments generated in one forward pass
&
Fast scenario generation, virtual sensing, fault-data synthesis, and repeated evaluation within real-time monitoring or decision loops
&
Mode collapse may omit rare but decision-relevant operating regimes
\\

Variational autoencoders (VAEs)
\cite{chiaro2025generative}
&
High-dimensional, noisy, or multi-modal measurements, including sensor streams, images, and spatial fields
&
Reconstructed measurements, completed states, latent-state estimations
&
State estimation, missing-data completion, anomaly detection, and low-dimensional probabilistic state representation
&
Generated responses may be over-smoothed and contain less detail than GAN outputs
\\

Autoregressive Transformers
\cite{shao2025generative,andreoni2024enhancing}
&
Sequences of past states, control actions, events, and multi-modal observations
&
Future states or trajectories generated sequentially over a variable prediction horizon
&
Long-horizon forecasting, traffic and demand prediction, equipment degradation tracking, and multi-modal sequence modeling
&
Inference cost increases with horizon length, and early prediction errors may propagate through the sequence
\\

Diffusion models
\cite{ho2020denoising}
&
System states, controls, parameters, boundary conditions, sparse observations, and an initial noise sample
&
Complete trajectories, deformation or flow fields, reconstructed states, and rare or counterfactual scenarios
&
Uncertainty-aware planning, state reconstruction, safety testing, and generation of globally coherent multiple outcomes
&
Iterative denoising creates high inference cost and may preclude high-frequency control without accelerated sampling
\\

Physics-informed normalizing flows
\cite{guo2022normalizing}
&
Spatial coordinates, latent samples, and sparse field measurements;
governing equations and boundary conditions incorporated during training
&
Samples of stochastic solution, coefficient, or forcing fields,
including non-Gaussian distributions
&
Probabilistic field reconstruction, virtual sensing, inverse
inference, and uncertainty propagation for decision support
&
Invertibility constrains architecture design; reliability depends
on the validity of the encoded physics and sufficiently informative
measurement snapshots
\\

\bottomrule
\end{tabular}
\end{table}

Generative models can serve as surrogate models for real-time decision-making of Digital Twins through two primary predictive roles. First, they can generate or reconstruct system responses for full-field prediction conditioned on specific measurements, inputs, or states, thereby serving as virtual sensors for state estimation, as reviewed in Section~\ref{sec:perception_state}. Second, generative models can serve as forward predictive models that forecast system evolution from histories of states, sensor measurements, images, control actions, and external conditions, while offering capabilities beyond conventional point-prediction models. For example, VAEs can directly process raw image data and use compressed latent representations to capture underlying dynamics and multi-modal future trajectories \cite{ivanovic2020multimodal, kwon2020combining}, while GANs can synthesize high-fidelity future frames to provide look-ahead information for early operational decisions \cite{shuvo2023empowering}. Autoregressive Transformers instead discretize continuous states and actions into unified sequences and use attention mechanisms to capture long-range temporal dependencies, providing an effective alternative to conventional time-series deep neural networks (DNNs) for trajectory prediction \cite{hu2024solving, park2023simultaneous}. Diffusion models further formulate the future horizon as a joint distribution and generate entire trajectories through joint denoising rather than step-by-step causal rollout, which can reduce error accumulation over long prediction horizons \cite{zhou2024diffusion}. More broadly, these models can generate multiple plausible futures, including trajectories \cite{gomez2020real}, degradation paths \cite{mikolajewska2025generative}, or resource demands, providing a distribution of predictive information for uncertainty- or risk-aware predictive maintenance, resource allocation, safety assessment, and closed-loop control \cite{boskabadi2025industrial, zarifis2025diffusion}. However, the distribution of generated outcomes should not be interpreted directly as calibrated aleatoric uncertainty (which will be discussed in Section~\ref{sec:aleatoric_uncertainty}), since the diversity of generated samples from generative models does not necessarily correspond to the true stochastic variability of the physical system.

A practical limitation of generative models as surrogate is their strong dependence on the coverage and balance of the training distribution. Because these models learn a conditional response distribution rather than a single input--output mapping, inadequate sampling can lead to poor representation of low-probability or sparsely observed response modes, particularly in low-data regimes \cite{ajay2022conditional}. This dependence complicates both offline data collection and online model adaptation, where newly collected high-fidelity data are often limited and concentrated around the current operating regime. Fine-tuning on such locally sampled and imbalanced streaming data may distort the learned distribution or degrade previously captured response modes. This challenge is further amplified for computationally intensive generative architectures, such as GANs and diffusion models, which can be difficult to adapt continuously under real-time constraints. An additional limitation is that statistically plausible generated responses may violate governing equations, conservation laws, or boundary conditions, requiring explicit physical constraints or additional consistency checks before their use in decision-making. Consequently, distribution shifts can affect not only the predicted response but also the inferred likelihood of alternative outcomes. Validation must therefore extend beyond pointwise prediction error to assess distributional fidelity, including properties such as mode coverage, sample diversity, and probabilistic calibration where applicable, increasing the importance of drift detection and continual validation during Digital Twin operation \cite{guzy2021evaluating,xiang2023concept}.

\subsection{Multi-Fidelity Modeling Strategies}
\label{sec:surrogate_MF}

The preceding approaches primarily differ in how system behavior is learned or represented, whereas multi-fidelity (MF) modeling addresses how information from multi-fidelity sources can be combined when no single source provides sufficient coverage, accuracy, and affordability for constructing a surrogate. Low-fidelity (LF) information may come from analytical approximations, simplified simulations, reduced-order models, or inexpensive sensing, while high-fidelity (HF) information may come from detailed numerical simulations or physical experiments. MF modeling uses abundant and inexpensive LF data to reduce the amount of costly HF data required to model the target system \cite{peherstorfer2018survey}. This provides a solution to address a common challenge for Digital Twins in which experimental observations are scarce and high-fidelity simulations are too expensive to densely cover the operating domain. MF modeling can therefore be viewed through two related problems: how heterogeneous sources should be fused to estimate the HF response and how information should be acquired under a limited budget.

\begin{figure}[h!]
    \centering
    \includegraphics[width=0.95\linewidth]{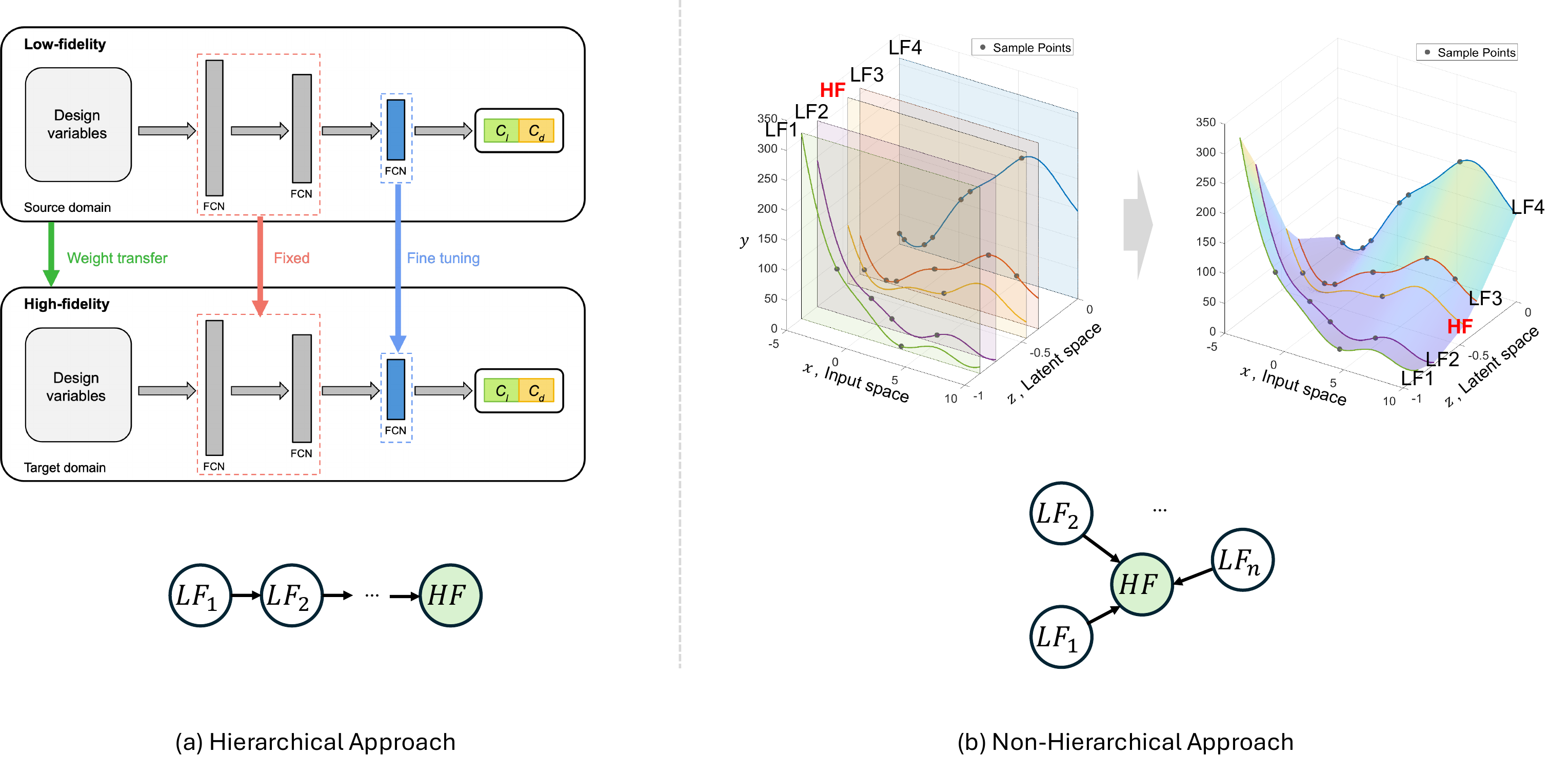}
    \caption{Illustration and examples of hierarchical and non-hierarchical multi-fidelity approaches. (a) An example of a hierarchical approach, where the information from the higher fidelity sources is added to enhance the prediction of the pre-trained MF models using data from lower fidelity sources. The application (modified from \cite{wu2024efficient}) shows that the backbone of the model is trained using LF data, then the last layer is fine-tuned using the HF data. (b) An example of a non-hierarchical approach, where information from all of the LF data sources is simultaneously used to enhance the prediction on the HF source. The application (modified from \cite{chen2024latent}) demonstrates an MF data fusion method to map multi-fidelity data sources on a single surface/manifold via a latent variable Gaussian process through an augmented latent dimension. Then, by conditioning on all the observations from multiple LF data sources, it can improve the prediction accuracy of the HF prediction and reduce the predictive variance.}
    \label{fig:MF}
\end{figure}

The first problem concerns \emph{multi-fidelity data fusion}. The appropriate fusion strategy depends on whether the available information sources possess a consistent fidelity ordering. As illustrated in Figure~\ref{fig:MF}, hierarchical methods assume that sources can be ranked from lower to higher fidelity and transfer or correct information sequentially toward the HF target. A common implementation first trains a model using LF simulation data and subsequently fine-tunes or updates it using a smaller HF dataset. This approach is applicable when LF and HF sources describe the same underlying response and their relationship remains sufficiently consistent across the operating domain~\cite{thelen2022comprehensive,chen2025digital}. Methods for MF modeling under hierarchical settings include transfer learning, bias or residual learning, or the use of generative models~\cite{wu2024online,shi2025diffusion,xu2025self}. A common formulation is to learn a discrepancy function, where the HF response is represented as an LF prediction plus a correction term that captures the systematic difference between the two fidelities~\cite{chakraborty2021role,chinesta2019virtual}. Co-Kriging, residual learning, and bias-correction models follow this principle and have been used to combine simulation predictions with physical measurements for Digital Twin monitoring and prediction~\cite{le2015cokriging,lai2022building,hewing2019cautious,ren2022bias}. Generative models provide another implementation of hierarchical fusion when the correlations between fidelity sources may not be able to be explicitly represented, such as in full-field predictions. For example, a GAN can be first trained using only LF data to learn the underlying features, and then refines the features with scarce HF data through adversarial and supervised learning~\cite{liu2022gan}. Similarly, diffusion-based MF methods first leverage inexpensive simulations as conditioning information during denoising, and use available expensive simulations to guide the reverse diffusion process~\cite{shi2025diffusion}.

Non-hierarchical methods become preferable when more than two information sources have comparable, uncertain, or operating-condition-dependent accuracy, making a fixed fidelity hierarchy difficult to justify. Rather than prescribing a global LF-to-HF ordering, these methods learn relationships among all available sources jointly~\cite{chen2016nonhierarchical}. For example, shared latent representations can map heterogeneous observations into a common feature space from which the target HF response is inferred~\cite{wu2023disentangled,pellegrini2023multi}. Alternatively, augmented latent dimensions can be used to encode correlations among fidelity sources while mapping all the data sources on the same manifold, enabling inference of the HF response conditioned on all available LF samples~\cite{chen2024latent}. Non-hierarchical MF fusion can also be implemented with a GAN-based architecture, in which separate LF models are trained for multiple LF sources, the LF source most relevant to the HF data is selected for transfer learning, and predictions from all LF models are used as auxiliary inputs to the HF generator~\cite{wang2026gtl}. Such formulations are useful when source reliability varies across the operating domain, but they require sufficient data to learn both cross-source relationships and source-specific reliability.

The second problem concerns \emph{multi-fidelity data acquisition}. Once a fusion structure has been specified, sampling decisions determine both where new information is needed and which fidelity should be queried. When models or experiments are available at different costs, uncertainty-aware acquisition criteria can balance the expected information gained from a candidate observation against its cost~\cite{chen2024latent,foumani2023multi}. Inexpensive LF evaluations can therefore be used to explore the operating domain, while HF evaluations are allocated to regions with large uncertainty, poor LF accuracy, or decision-relevant constraints. This allocation is critical in hierarchical settings, where a surrogate is first trained on LF data as a backbone and then refined using targeted HF samples to improve performance in transient, off-nominal, or near-constraint regimes.

MF modeling is therefore complementary to the surrogate modeling approaches discussed above rather than a separate modeling paradigm. Data-driven, SciML, and generative models can all be embedded within an MF framework, while the MF strategy determines how information of different cost, resolution, and fidelity is acquired and combined. For Digital Twins, MF modeling can support three related functions: improving surrogate accuracy when high-fidelity data are limited, allocating data acquisition across fidelity levels, and managing model fidelity at runtime. The latter is important when higher-fidelity models are too expensive to evaluate within the decision deadline, while lower-fidelity models may omit dynamics relevant to the current decision. Runtime fidelity management when Digital Twins are deployed therefore involves selecting or switching models to dynamically satisfy requirements on computational time and prediction accuracy as well as the downstream decision-making.

\subsection{Remaining Gaps in Surrogate Modeling from a System-Centric Perspective}
\label{sec:remaining_gap_surrogate}

Despite substantial progress in surrogate modeling, much of the existing literature remains model-centric, emphasizing predictive accuracy and representation quality while giving less attention to how the surrogate interacts with the broader Digital Twin system. From a system-centric perspective, the suitability of a surrogate also depends on whether its requirements for modeling, validation, and adaptation can be satisfied during operation, how it interacts with other system components, and ultimately how it contributes to overall system performance and capability. From this perspective, we identify three gaps that remain insufficiently addressed in the existing literature:

\begin{itemize}

    \item \textbf{Required model inputs may be unavailable during operation.}
    Many surrogate models rely on inputs that are readily available during offline development, particularly when training data are generated from simulations, but may be estimated, delayed, or unavailable during operation. Examples include boundary conditions, internal states, material properties, and other quantities that cannot be directly measured. The input requirements of a surrogate therefore translate directly into sensing and state-estimation requirements for the Digital Twin. A model that performs accurately offline may be unusable in operation if the required inputs cannot be supplied at the necessary rate and fidelity. Surrogate selection and sensor design should therefore be treated as coupled system-design decisions rather than sequential tasks, motivating the state-estimation and soft-sensing methods discussed in Section~\ref{sec:perception_state}.

    \item \textbf{Predictive accuracy is not always balanced against real-time computational requirements.}
    Surrogate modeling studies commonly benchmark predictive accuracy, while computational cost and its implications for the overall decision loop receive comparatively less attention. In particular, a systematic approach to design the trade-off between model fidelity and decision performance needs to be integrated in model selection criteria. For real-time Digital Twin applications, however, model fidelity should not be maximized independently of computational efficiency. Higher-fidelity models may provide more accurate predictions but remain too expensive for online inference, whereas reduced or hyper-reduced representations can sacrifice some accuracy to achieve substantially faster evaluation~\cite{vlachas2026utility, zhang2025probabilistic}. The appropriate fidelity should therefore be determined at the system level, balancing the accuracy required for the relevant quantities of interest against the computational latency permitted by the downstream decision process. 

    \item \textbf{Validation and adaptation are rarely treated as part of surrogate modeling.}
    In much of the Digital Twin literature, a surrogate model is constructed, validated against a held-out dataset, and then deployed without an explicit mechanism for assessing whether it remains accurate during operation. Synchronization, uncertainty treatment, end-to-end validation, and adaptation are also reported inconsistently across studies~\cite{fridkin2026physics}. Although the physical system continues to evolve after deployment, validation is often treated as a step that concludes model development rather than a capability that the deployed Digital Twin must retain. Adaptation is similarly introduced as a corrective response to model degradation rather than considered jointly with the choice of surrogate representation. This issue becomes particularly important when the surrogate predicts quantities that cannot be directly validated against measurements, including spatial fields at unsensed locations, response distributions when the physical system provides only a single realization, or latent states without direct physical correspondence. Determining when such predictions are no longer reliable, and when the surrogate should be re-examined or updated, is therefore essential for sustained Digital Twin operation.

\end{itemize}

These gaps suggest that surrogate modeling should not be treated as an isolated model development task, but as one component of a broader Digital Twin decision architecture. The limited availability of model inputs motivates the use of ML in perception and reasoning to infer decision-relevant states and information from heterogeneous observations. The tradeoff between model fidelity and computational cost motivates learning-based approaches that accelerate decision synthesis, while the need to maintain model validity as the physical system evolves motivates continual validation and adaptation. The following sections therefore review these complementary capabilities and their roles in enabling sustained Digital Twin operation.

\section{Machine Learning in Perception and Reasoning}
\label{sec:perception&reasoning}

Real-time decision-making in a Digital Twin not only requires predictive models and decision algorithms, but it also requires an internal representation that remains synchronized with the physical system and provides the information needed to interpret its current condition. In this review, \emph{perception} is considered from the perspective of machine perception in AI and intelligent systems, referring to the process of transforming observations of the physical systems into structured, decision-relevant representations, including estimated states and integrated information from heterogeneous sensing modalities. \emph{Reasoning} refers to the subsequent interpretation of these representations together with engineering knowledge and operational context to infer decision-relevant relationships, implications, and constraints. Together, perception and reasoning form the information interface between physical-system feedback and downstream prediction and decision-making. This interface is critical because raw observations are often noisy, incomplete, heterogeneous, or unstructured, and therefore cannot be directly used by surrogate models or decision algorithms. ML provides an important capability for bridging this gap by reconstructing unmeasured states, extracting and fusing information from high-dimensional multi-modal data, and integrating observations with heterogeneous engineering knowledge. This section reviews these roles through three capabilities: state estimation and data assimilation (Section~\ref{sec:perception_state}), multi-modal perception (Section~\ref{sec:perception_multimodal}), and cross-domain knowledge integration and reasoning (Section~\ref{sec:reasoning}).

\subsection{State Estimation through Data Assimilation}
\label{sec:perception_state}

\begin{figure}[h!]
    \centering
    \includegraphics[width=0.6\linewidth]{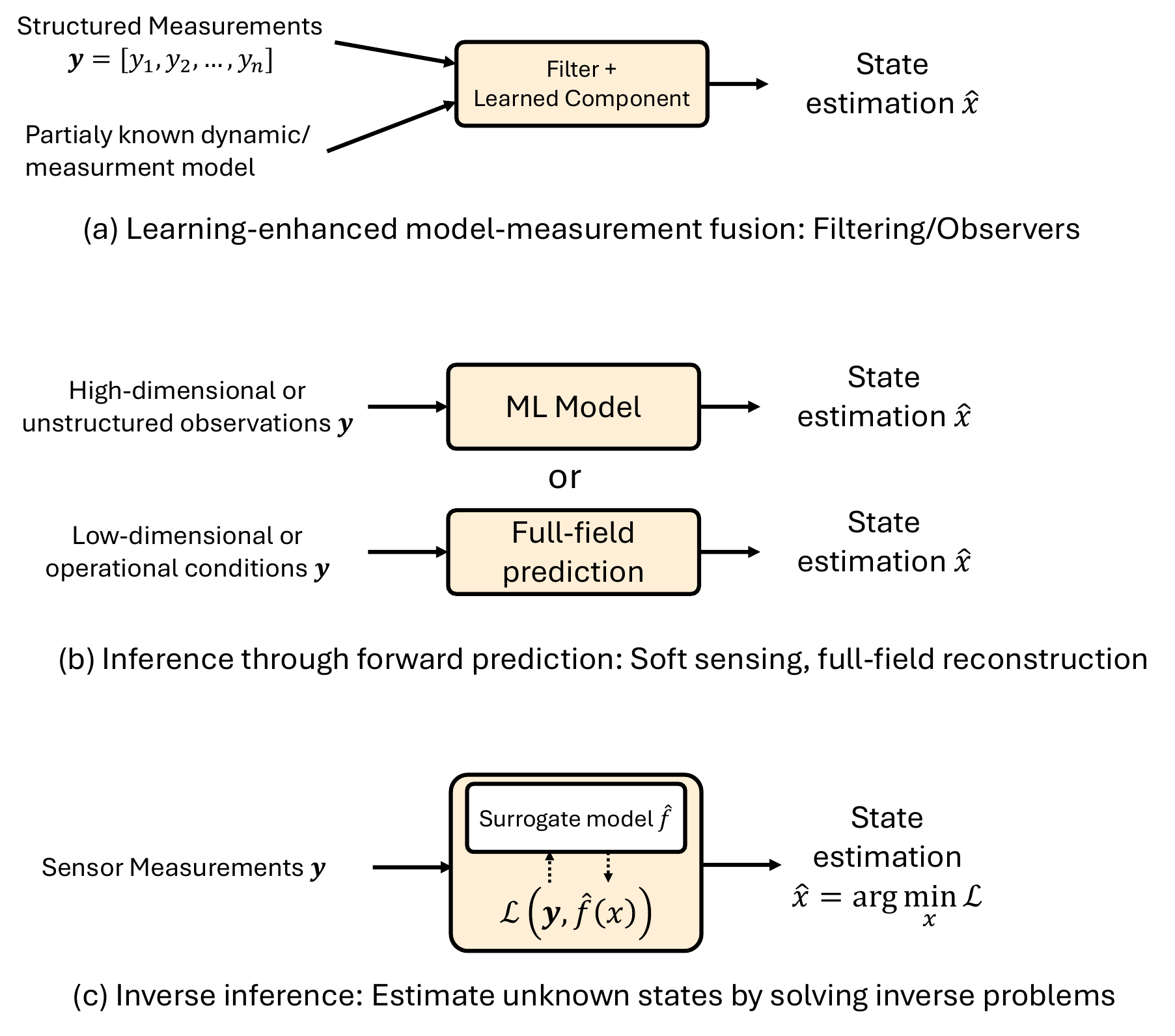}
    \caption{Three ML pathways for data assimilation in Digital Twins: (a) learning-enhanced model-based assimilation, (b) direct soft sensing from observations, and (c) surrogate-enabled inverse estimation through measurement matching.}
    \label{fig:state_estimation}
\end{figure}

State estimation through data assimilation refers to the use of physical observations to infer the internal states and conditioning information required to synchronize a Digital Twin with its physical counterpart. In this review, data assimilation is used in a broad system-level sense to encompass the transformation of available observations into quantities needed for subsequent prediction and decision-making. These quantities may include physical states, latent variables, unknown inputs or disturbances, and slowly varying system parameters that are not directly measured by available sensors. ML can support this process by reconstructing such quantities from incomplete and noisy observations, either by augmenting conventional model-based estimators or by learning the required inference mappings from data. Existing ML-enabled approaches can be broadly viewed through three pathways, as illustrated in Figure~\ref{fig:state_estimation}: learning-enhanced model-based assimilation, direct soft sensing that learns observation-to-state mappings, and inverse estimation that identifies hidden quantities through a predictive model of how they generate the observed response.

The first pathway augments conventional model-based state estimation, especially when the dynamical or measurement models required by the estimator are only partially known. Classical filtering methods provide a natural mechanism for Digital Twin synchronization because they recursively combine model predictions with incoming measurements. Their performance, however, depends on the availability and accuracy of the state-transition and measurement models, as well as the assumed noise statistics. Hybrid Kalman filter and neural network approaches relax these requirements by retaining known model structure while learning uncertain components from data. NNs can, for example, approximate unknown state-space relationships, compensate for unmodeled nonlinearities or systematic estimation errors, or adapt quantities used within the filtering process~\cite{feng2023review}. KalmanNet follows this principle by learning the Kalman gain directly from data while preserving the recursive predict-and-update structure, enabling state estimation when useful dynamical knowledge is available but the underlying model or noise statistics are incomplete or inaccurate~\cite{revach2022kalmannet}. Such hybrid estimators therefore provide a middle ground between fully model-based and fully data-driven assimilation.

\begin{figure}[h!]
    \centering
    \includegraphics[width=0.8\linewidth]{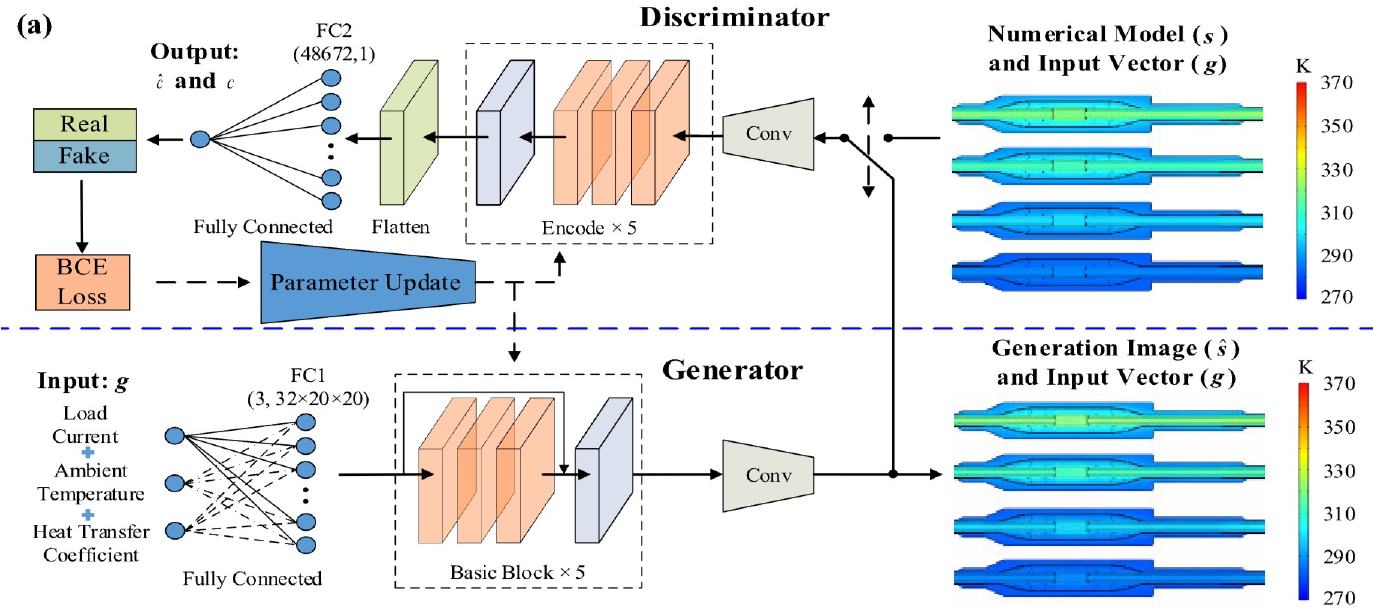}
    \caption{Example of a conditional GAN used to generate full-field temperature distributions for real-time power-system monitoring from load current, ambient temperature, and heat-transfer coefficients as conditioning information (reused from \cite{zhao2023dynamic}).}
    \label{fig:full-field_prediction}
\end{figure}

A second pathway uses ML for direct soft sensing, in which available measurements and operating information are mapped directly to quantities that cannot be measured reliably or economically, thereby providing virtual measurements for Digital Twin synchronization and decision-making. Depending on the quantities to be estimated, soft sensing may reconstruct either compact decision-relevant states or spatially distributed fields, resulting in two different major applications, as illustrated in Figure~\ref{fig:state_estimation}. The first form of soft sensing is to extract or identify specific states of interest (as the dynamical states for predicting the evolution of the systems) from high-dimensional or unstructured data. For structured time-series measurements, regression models, recurrent NNs (RNNs), and long short-term memory (LSTM) can infer variables such as degradation states, operating conditions, material properties, or other decision-relevant quantities from correlated sensor signals~\cite{xing2019dynamic,cao2020scalable,abbas2019performance}. For high-dimensional observations such as images or point clouds, convolutional NNs (CNNs), Transformers, and encoder-based models can extract compact state representations for downstream prediction and control. For example, high-dimensional surface images have been compressed into low-dimensional surface-roughness representations for RL-based control~\cite{knaak2021improving}, while vision-based perception can estimate pose and task states in robotic Digital Twins~\cite{wang2026digital}. When labeled states are scarce, self-supervised, semi-supervised, and unsupervised representation learning can further reduce dependence on exhaustive annotation~\cite{li2021cutpaste,hu2024semi}. A special case is when the surrogate model itself operates in a learned latent space; e.g. through VAE, data assimilation and prediction are closely coupled: an encoder maps physical measurements to the latent state required by the surrogate, while the learned dynamics propagate that state for downstream prediction and decision-making. For example, Baumeister et al.~\cite{baumeister2018deep} use an encoder to infer an unmeasurable latent state from observable laser signals, which is then propagated by a learned dynamical model for MPC. Such architectures illustrate that successful Digital Twin synchronization depends not only on the accuracy of the latent dynamics, but also on whether the required latent state can be reliably reconstructed from available physical measurements. A second form of direct soft sensing reconstructs spatially distributed quantities that cannot be measured comprehensively using physical sensors. ML models, including SciML and generative models, can infer deformation, temperature, stress, or other distributed fields from sparse observations, boundary conditions, and operating variables, thereby providing virtual measurements at inaccessible or prohibitively expensive sensing locations. For example, \cite{mu2024online} combines a vector-quantized variational autoencoder, a generative adversarial network, and an RNN to estimate distortion fields in wire arc additive manufacturing for online planning and process control. Similarly, \cite{zhao2023dynamic} maps load current, ambient temperature, and heat-transfer conditions to cable-joint thermal fields for real-time monitoring, enabling temperatures to be inferred at locations where physical sensors cannot be installed, as illustrated in Figure~\ref{fig:full-field_prediction}. Other examples include conditional GANs for reconstructing communication-network performance fields used in antenna optimization \cite{he2025intelligent} and diffusion models for recovering complete motion or spatial fields from sparse observations \cite{chen2024generative}.

A third pathway formulates data assimilation as an inverse problem, in which hidden quantities such as unknown parameters, disturbances, or spatial fields are inferred by leveraging a surrogate model as the forward model within an inverse inference loop~\cite{wunsch2006discrete}. In contrast to direct soft sensing, which learns an observation-to-state mapping, inverse estimation searches for the hidden quantities whose surrogate-predicted response is most consistent with the available measurements. ML models can support this process by providing computationally efficient forward surrogates for repeated inverse evaluations or by learning probabilistic inverse models that amortize the inference process. For example, Zhang et al.~\cite{zhang2025probabilistic} estimate spatial force fields through a maximum-likelihood inverse problem powered by a real-time reduced-order model. Similarly, by training a lightweight NN as a fast and differentiable surrogate for nonlinear structural behavior, the internal force of a cable-strut structure can be estimated by solving the inverse problem to minimize the discrepancy between the surrogate's prediction and sensor measurements \cite{dong2026real}.  These approaches extend data assimilation beyond conventional state estimation to quantities that cannot be directly sensed, while illustrating how surrogate modeling and ML can make otherwise computationally expensive inverse problems more tractable for Digital Twin operation.

Across these pathways, ML broadens the data assimilation capability of a Digital Twin by relaxing dependence on complete physical models or sensors, reconstructing quantities that cannot be directly measured, and enabling inference from high-dimensional and incomplete observations. The objective is to maintain the internal states and conditioning information required by surrogate models and downstream decision-making methods. In complex systems, however, this information is often distributed across multiple sensing modalities rather than contained in a single data source. The following subsection therefore examines how ML integrates heterogeneous observations to construct a more complete representation of the physical system.

\subsection{Multi-modal Data Fusion}
\label{sec:perception_multimodal}
 
Real-time decision-making in Digital Twins can benefit from multi-modal measurements obtained through feedback from the physical system, which increases information density and enables a more holistic representation for decision-making \cite{zhang2025toward},
as individual sensing modalities often provide only partial or complementary information \cite{yao2023systematic, national2023foundational}. For instance, monitoring a machine tool and adjusting process parameters to extend tool life requires an overarching observation of cutting force, vibration, temperature, and operation history \cite{zhang2025toward}. Before the widespread use of ML, multi-modal perception largely relied on hand-crafted features and predefined statistical relationships among sensing sources, which limited fusion when the data were high-dimensional, nonlinear, or weakly aligned~\cite{zhao2024deep}. ML expands this capability by automatically extracting modality-specific features, learning nonlinear cross-modal relationships, and mapping heterogeneous observations into shared representations that can be used for state estimation and reasoning~\cite{zhao2024deep}. The perception capability of processing multi-modal feedback data therefore extends the Digital Twin from processing individual sensor streams toward constructing integrated awareness for real-time decision-making.

ML realizes this fusion by learning representations and cross-modal relationships directly from heterogeneous data~\cite{zhao2024deep,korthals2019jointly,jiao2024comprehensive,gao2020survey,liu2017heterogeneous}. Conventional fusion methods, such as Kalman filtering~\cite{caron2006gps}, typically require predefined measurement relationships and statistical assumptions, whereas DNN-based multi-modal fusion can be implemented through different architectures and mechanisms. For example, encoder-decoder models map different modalities into shared latent representations, attention mechanisms dynamically emphasize informative modalities or features, generative models can reconstruct missing modalities from available observations, and graph neural networks capture relational dependencies among spatially or functionally connected sensing sources~\cite{zhao2024deep}. These mechanisms allow the fusion process to adapt to heterogeneous and partially observed data.

For Digital Twins, the main value of multi-modal learning is to produce a state representation that is more informative and robust for downstream prediction and decision-making. By mapping/fusing data into a common learned latent representation, it can reduce communication and storage requirements in large distributed systems by transmitting compact features rather than raw sensor streams, reducing the computational overhead for data transmission in real-time applications~\cite{korthals2019jointly}. Fusion can further compensate for noisy or incomplete observations by exploiting complementary information across modalities~\cite{jiao2024comprehensive}. In manufacturing, for example, visual and acoustic measurements can be combined to improve defect detection~\cite{kullu2022deep,xie2025audio}. In the RETROFIT framework~\cite{biehler2024retrofit}, 3D point clouds, in-situ video, and process measurements such as arc voltage, travel speed, and wire feed rate are encoded by modality-specific networks and fused into a unified latent representation that augments Koopman observables for MPC. Related applications combine sensor observations with physics-based predictions for fault diagnosis~\cite{huang2023novel,cai2025study}, or integrate visual and haptic information to estimate task-relevant states for robotic decision-making~\cite{zhou2024multimodal}.

The deployment of multi-modal fusion nevertheless introduces system-level challenges that are particularly relevant to Digital Twins. Multi-modal measurements from different sensors may be spatially and temporally misaligned because they operate at different sampling rates and resolutions~\cite{lahat2015multimodal,jiao2024comprehensive}. Sensor dropout may remove entire modalities during operation, while heterogeneous sensing mechanisms introduce different noise and uncertainty characteristics that can be propagated or amplified through the fusion model~\cite{zhao2022data,jiao2024comprehensive}. Cross-modal relationships may also change as operating conditions evolve, making models trained under a fixed combination of sensors difficult to maintain. Moreover, increasingly complex fusion architectures introduce additional computational cost that can conflict with real-time requirements, and therefore redundancy analysis should be incorporated during the sensor allocation and selection stages~\cite{gao2020survey,xie2025redundancy}. Multi-modal perception in Digital Twins should therefore be evaluated not only by fusion accuracy, but also by its robustness to missing and asynchronous observations, computational efficiency, uncertainty awareness, and ability to remain valid as sensing conditions evolve.

\subsection{Cross-Domain Knowledge Integration and Reasoning}
\label{sec:reasoning}

Sections~\ref{sec:perception_state} and~\ref{sec:perception_multimodal} describe how ML transforms incomplete measurements and multi-modal observations into estimates of the physical system state. However, state estimates and fused features alone may not provide the contextual information required for decision-making. In a machining process, for example, a detected geometric deviation must often be interpreted together with the process plan, the condition of the equipment, and past production records before its implications can be determined. Cross-domain knowledge integration and reasoning provide this additional capability by connecting perceptual outputs with engineering knowledge drawn from different sources and inferring their decision-relevant meaning. Within the Cognitive Digital Twin architecture introduced in Section~\ref{sec:CDT}, this layer therefore bridges the question of \emph{what is happening in the physical system} with \emph{what information implies for subsequent decisions}.

This reasoning capability first requires the Digital Twin to represent knowledge beyond numerical state variables. A common approach is to use ontologies and knowledge graphs to explicitly represent engineering entities, relationships, rules, and constraints, while technical documents, maintenance records, and operational reports contain complementary knowledge in unstructured form. ML expands the use of these sources by learning latent relationships through knowledge graph embeddings and graph neural networks, and by extracting and contextualizing information from unstructured sources through foundation models. These representations provide the knowledge base against which current system observations can be interpreted.

With such a knowledge base in place, the Digital Twin must connect its observations to it, linking estimated states and detected changes to the specific entities and records they concern. Machine-learning methods support this process through semantic labeling, entity and relation extraction, entity linking, and information retrieval. For example, in a Digital Twin for fault diagnosis of a steel-plant bridge crane, segments of vibration, temperature, and current-voltage signals are matched by similarity against sequences aligned with past fault investigation records and alarm logs, linking current observations to descriptions of similar past faults~\cite{SUN202483}. Cross-domain integration extends this process by aligning corresponding concepts across different information sources. An aero-engine blade Digital Twin, for instance, integrates process data, engineering rules, historical cases, sensor observations, and geometric measurements through a manufacturing knowledge graph, allowing observed geometric changes to be connected with machining decisions~\cite{su2024knowledge}. Similar knowledge graphs have linked building information models, process records, and sensing platforms in built-asset Digital Twins~\cite{ramonell2023knowledge}.

Building on these links, ML can support reasoning that extends beyond explicitly recorded information. Knowledge graph embeddings and graph neural networks can infer missing or latent relationships within structured knowledge, while LLMs can retrieve, interpret, and synthesize evidence from unstructured sources. For example, graph embeddings have been used to infer dynamic event links and latent process relationships from manufacturing time-series information~\cite{Zhou18062023}, while retrieval-augmented LLM agents can combine real-time machine observations with operational and maintenance knowledge to examine fault evidence and recommend context-specific responses~\cite{gautam2025iiot}. In this role, reasoning does not replace the decision-making methods reviewed in Section~\ref{sec:decision_making}; rather, it provides the contextualized states, relationships, explanations, and constraints on which those methods can operate.

Because the physical system and its operating context evolve, the knowledge used for reasoning must also remain synchronized with the physical systems. Dynamic information can be updated as new observations arrive while relatively stable engineering knowledge is retained. Computational agents, for example, can continuously incorporate observations into dynamic knowledge graphs~\cite{akroyd2021universal}, while LLM-enabled Digital Twins can retain relatively static knowledge and receive changing system states separately during inference~\cite{Hong2024LLMTwin}. This separation allows the reasoning context to evolve with the physical system without requiring the entire knowledge representation to be repeatedly reconstructed.

The reliability of cross-domain reasoning ultimately depends on the quality and consistency of the information from which it is constructed. Incomplete or conflicting knowledge, incorrect semantic grounding, unsupported inference, and evolving terminology or system relationships can all propagate through the reasoning layer into downstream decisions. Large and continuously updated knowledge structures may also introduce computational costs that conflict with real-time operation. From a system-centric perspective, reasoning should therefore be evaluated not only by its ability to infer useful knowledge, but also by whether its outputs remain traceable, uncertainty-aware, computationally feasible, and consistent with the evolving physical system. 

Taken together, these capabilities form the information interface between the physical system and the decision-making layer of a Digital Twin. ML extends this interface from direct sensing toward reconstruction of unmeasured states, integration of heterogeneous observations, and contextual interpretation of system conditions. Once the required states and knowledge are available, the remaining challenge is to translate them into actions within the computational time scale of operation. This motivates the discussion in the next section.

\section{Machine Learning for Decision Synthesis}
\label{sec:ml_decision}

Building on the decision-making approaches reviewed in Section~\ref{sec:decision_making}, ML plays different roles in progressing decision-making from numerical solving to learned policies, and ultimately to decision synthesis through reasoning and orchestration. Motivated by its role, this section reviews how ML enables this progression in Digital Twin decision-making. In conventional solving-based approaches such as MPC and DP, the virtual twin is repeatedly queried online to evaluate candidate decisions, making the computational cost increasingly demanding as the model fidelity, decision dimension, and problem complexity increase. Aside from using surrogate models as introduced in Section~\ref{sec:surrogate}, ML can alleviate this bottleneck by accelerating numerical optimization while retaining the underlying solving procedure, as will be discussed in Section~\ref{sec:accelerate_solving}. Beyond conventional policy learning, the generative models for decision synthesis discussed in Section~\ref{sec:gen_decision} use ML to compress knowledge from simulation, optimization, and operational experience into expressive conditional distributions over actions, trajectories, or plans, enabling multiple feasible decision alternatives to be generated from current system information. At a higher level, agentic AI can extend conventional ML from producing individual decisions to reasoning about the decision process itself by interpreting engineering objectives, decomposing problems, coordinating computational resources, and revising plans based on physical feedback, which will be discussed in Section~\ref{sec:agentic_ai}. Together, these roles allow Digital Twins to be scalable to increasingly complex decision-making problems.

\subsection{Accelerating Online Numerical Optimization}
\label{sec:accelerate_solving}

Whereas Section~\ref{sec:surrogate} focuses on reducing the cost of individual model evaluations, online decision synthesis can remain computationally expensive because numerical solvers must repeatedly evaluate candidate decisions, propagate trajectories, and enforce constraints until an acceptable solution is obtained. ML can reduce this solver-level burden by exploiting recurring structure across problem instances and shifting part of the required computation offline. Three relevant mechanisms to Digital Twin decision-making are reviewed: parallelizing computations that would otherwise be performed sequentially, learning reusable information that guides the online search, and constructing model representations that are compatible with computationally efficient solvers.

\subsubsection{Parallelizing Online Solution}

A fundamental bottleneck in real-time numerical optimization is the sequential computation required to evaluate candidate decisions and their associated future trajectories. ML, particularly with neural networks, can offer solutions to parallelize the computation in two practical ways. First, many neural network architectures evaluate inputs through a fixed sequence of tensor operations that can be efficiently batched across candidate decisions and executed concurrently on GPUs or other parallel hardware. Once trained, the same model can therefore evaluate many candidate states or trajectories simultaneously with relatively little additional computational overhead. Second, sequential or time-series models can replace computations on trajectory rollout with direct mappings, for example by predicting an entire future trajectory rather than recursively propagating the state one step at a time. 

These advantages can be exploited at several levels. In dynamic programming (Section~\ref{sec:dp}), value propagation across feasible future trajectories can be performed concurrently, while learned approximations can reduce the cost of evaluating large numbers of candidate states~\cite{wang2017pdp}. In MPC and trajectory optimization, learned surrogate models and ensemble recurrent predictors allow multiple candidate state evolutions to be evaluated in batches~\cite{wu2019machine,wu2019real}. For large-scale systems, learned subsystem models can similarly support distributed MPC, allowing local optimization problems to be solved concurrently while coordination is maintained through information exchange among subsystems~\cite{chen2020machine}. Across the prediction horizon, multi-step ahead models such as TiDE, Transformers, or diffusion models directly generate future state sequences, avoiding recursive rollout and enabling horizon-wide objective and constraint evaluations~\cite{das2023long,chen2025real, park2023simultaneous, zhou2024diffusion}. Neural operators can also support long-horizon prediction by reducing error accumulation in one-shot multi-step prediction~\cite{de2025deep}, while $\pi$MPC exploits parallel in-horizon variable splitting to reduce sequential dependence within the MPC solver itself~\cite{wu2026pi}. The practical benefits, however, may depend on the specific resources for parallelization, such as the available hardware for parallel computing, or problem formulations that support distributed optimization. This highlights the need to consider computational hardware, model architecture, and optimization formulation jointly when designing the closed-loop Digital Twin, rather than treating it as a plug-and-play solution for accelerating an otherwise fixed computational pipeline.

\subsubsection{Leveraging Offline Learning for Online Solving}

A complementary strategy is to learn recurring solution structures offline and reuse them during online optimization. Historical optimization results and virtual simulations from the Digital Twin can reveal how favorable initializations, active constraints, value information, or promising regions of the decision space vary with the system state. With proper use of ML, these relationships can be captured so that the computationally intensive exploration of solution structures is shifted offline, while the online solver remains responsible for satisfying the objectives and constraints of the current decision problem.

The most direct use is warm starting and search guidance. NNs can predict initial control trajectories, feasible solutions, active constraint sets, or fixed points, reducing the iterations required by nonlinear programming, active set methods, and other initialization-sensitive solvers~\cite{vaupel2020accelerating,klauvco2019machine,sambharya2024learning}. Generative models can serve a similar auxiliary role by learning problem-conditioned proposal distributions that concentrate sampling-based optimization in regions favored by previously obtained solutions~\cite{power2022variational}. Beyond initialization, learned value functions and heuristics can direct dynamic programming, tree search, and rollout-based planning toward high-value alternatives~\cite{kool2022deep,anthony2017thinking}. Learned value information can also augment MPC by representing consequences beyond its finite prediction horizon while retaining explicit online optimization and constraint handling~\cite{romero2024actor}.

Having offline learning prior to the deployment of Digital Twins can further capture reusable solution structure beyond initialization. For example, in high-dimensional dynamic programming, a DNN can be trained offline to approximate the value function over representative system states. During online decision-making, this learned value approximation can be used to guide the search toward promising actions or to provide a terminal cost for finite-horizon optimization, thereby reducing the amount of online computation while retaining an explicit optimization step. Compared with grid-based state-space discretization, the learned approximation provides a compact global representation that can be evaluated efficiently at runtime ~\cite{fernandez2020solving}. 

The offline pretrained models nevertheless remain dependent on the problem distribution from which they were developed. In practice, the data distribution encountered during closed-loop operation may differ from that represented by offline training data \cite{celestini2024transformer}, and the benefit of the pretrained model may therefore diminish as the operating domain shifts or the physical system evolves, such as poor warm-starts or inaccurate active set predictions. Unlike online adaptation of the system model (Section~\ref{sec:epistemic_uncertainty}) or on-policy fine-tuning of the decision policy (Section~\ref{sec:peft}), adapting such pretrained models can be more challenging because system feedback may not provide direct supervisory information for updating the quantities learned offline, such as the value function. Consequently, additional mechanisms may be required to monitor the validity of the pretrained model during deployment and determine when retraining or online correction is necessary.

\subsubsection{Linearization for Fast Online Optimization}

Even when a learned surrogate is inexpensive to evaluate, embedding it within online optimization can remain costly if the resulting problem is strongly nonlinear or nonconvex. For solving-based decision algorithms, surrogate design should therefore consider not only predictive accuracy but also compatibility with tractable numerical solvers. With a learned surrogate model, there are two major approaches to support convex optimization. The first is via a local approximation of the nonlinear model, and the second is to learn a solver-compatible representation (which is usually linear). 

Local linearization retains the expressive capability of nonlinear learned models while reducing their computational burden within online optimization. This can be achieved in three main ways. First, \emph{instantaneous linearization} approximates the first-order linearization of the nonlinear model around the current operating condition at each control step and uses this local linear model, i.e., through automatic differentiation for NN models, for optimization~\cite{yang2022machine}. Second, \emph{iterative prediction and linearization} first predicts the future trajectory using the nonlinear model, then linearizes the model around the predicted trajectory and solves the optimization problem. The resulting solution is used to update the predicted trajectory, and this process can be repeated to improve the local approximation of the surrogate model~\cite{schwedersky2022nonlinear}. Third, the nonlinear model can be locally approximated at multiple points along the predicted trajectory so that these approximations can be directly used within a real-time optimization solver~\cite{salzmann2023real}. These approaches trade exact nonlinear optimization for locally accurate representations that can be solved more efficiently online.

Alternatively, the model architecture can be designed to produce solver-compatible dynamics directly, enabling formulating the problem with linear constraints and solving via convex optimization. For example, a Transformer-based ARX model learns state-dependent coefficients of a linear time-varying representation that can be supplied directly to the optimizer without explicitly linearizing a nonlinear network~\cite{peng2025transformer}. Another example is to use the Koopman operator (Discussed in Section~\ref{sec:surrogate}) to directly represent a nonlinear system in a high-dimensional linear space. 

The computational benefit of these approaches comes with a tradeoff in model fidelity, as the local or state-dependent linear representations may become inaccurate when the system moves through strongly nonlinear operating regimes, and their effectiveness can depend on sufficiently frequent sensing, synchronization, and model updates. Therefore, whether these methods should be adopted should be evaluated jointly by the reduction in decision latency and the fidelity retained over the operating region relevant to online decision-making.

\subsection{Generative Models for Decision Synthesis}
\label{sec:gen_decision}

Section~\ref{sec:policy_based} introduced policy-based decision-making as a way to replace repeated online optimization with a learned or precomputed decision rule. A conventional learned policy is typically designed to directly produce an action, or a relatively simple distribution over actions, conditioned on the current system information. Generative models extend this idea by learning an expressive conditional distribution over structured decisions such as control sequences, trajectories, or plans,
\begin{equation}
\mathbf{u} \sim p_\theta(\cdot \mid \mathbf{z}_k),
\end{equation}
where $\mathbf{u}$ denotes a decision generated from the current system information $\mathbf{z}_k$, which may be a single action, a sequence of actions over a horizon, or a structured plan.

The distinction between generative policies and conventional policy models is not that conventional policies must be deterministic. Rather, generative models are designed to represent and sample from complex, potentially multi-modal distributions over high-dimensional decision spaces. Instead of compressing decision-making into a single preferred action, they can preserve multiple feasible or near-optimal solution modes and generate several alternatives for the same operating condition. This capability is valuable when distinct actions or trajectories can achieve comparable objectives, or when the preferred solution depends on constraints, objectives, or information that becomes available only at runtime.

\begin{figure}[h!]
    \centering
    \includegraphics[width=0.7\linewidth]{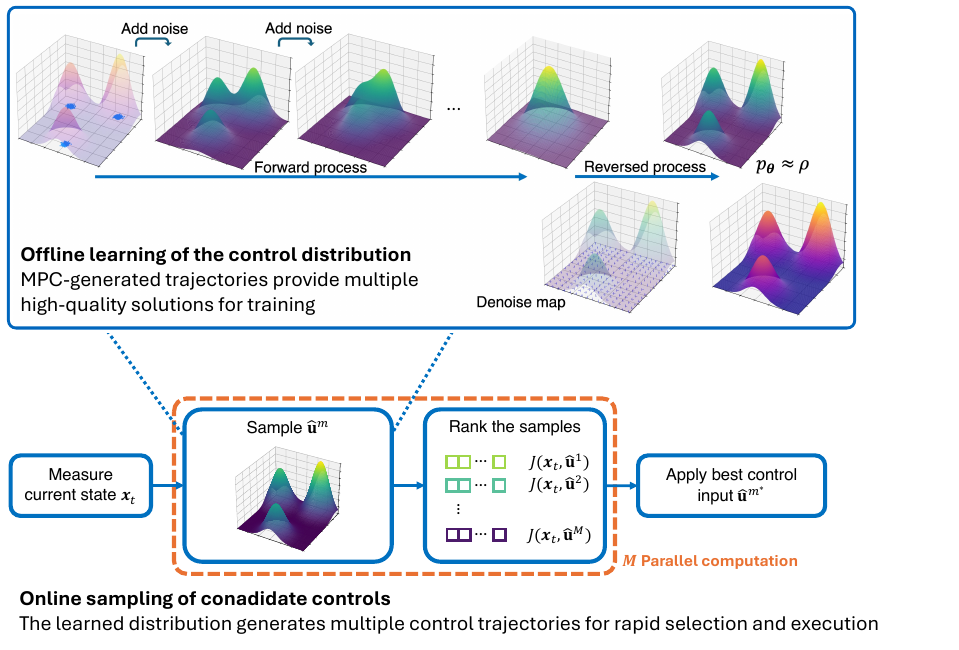}
    \caption{Illustration of generative policy via a diffusion model. (a) Offline training of a generative model using MPC-generated trajectories that capture multiple locally optimal and globally competitive solutions across operating conditions. (b) Online generation of candidate control trajectories from the learned conditional distribution, enabling multiple solution modes to be sampled and evaluated without repeatedly solving the original nonlinear optimization problem. (Courtesy of Tzu-Yuan Huang)}
    \label{fig:placeholder}
\end{figure}

These distinct features of generative models have been leveraged for real-time decision-making, such as applications in control and trajectory planning. For example, motion planning through diffusion models formulates planning as a generative inference problem and produces goal-conditioned trajectories through a reverse denoising process, allowing multiple modes of robot motion to be represented~\cite{carvalho2023motion}. Similarly, diffusion models trained on locally optimized nonlinear MPC trajectories can generate multiple candidate control sequences for the same operating condition, thereby representing distinct locally optimal solutions that would otherwise require repeated numerical searches~\cite{huang2024toward}. GFlowNets provide another distributional approach by learning to sample decision sequences according to their reward rather than concentrating exclusively on a single optimum~\cite{bengio2021flow}. In these settings, it does not merely approximate one optimized policy, but learns the structure of a broader decision space from which alternatives can be generated online.

This flexibility, however, introduces an additional requirement for \emph{physical grounding}. A generated action or trajectory is not inherently feasible simply because it has high probability under the learned distribution. State and input constraints, collision avoidance, conservation laws, and other engineering requirements must therefore be incorporated during generation or checked before execution. Existing approaches introduce such information through guidance~\cite{liu2026guideflow,utkarsh2026physics}, safety filtering~\cite{gadginmath2026provably}, or projection onto feasible regions~\cite{christopher2024constrained}. More recently, SAD-Flower integrates control-theoretic guidance with flow matching to enforce state and action constraints and dynamic consistency at test time without retraining the generative model~\cite{huang2025sad}. For Digital Twins, the virtual model can also provide a verification layer in which generated candidates are simulated and screened according to predicted physical consequences before a decision is transferred to the physical system.

At a higher level, LLMs and vision-language models (VLMs) extend generative decision synthesis from numerical control trajectories to symbolic actions, subgoals, and task plans derived from textual, visual, and multi-modal information~\cite{zhong2025survey,cao2025large}. DriveVLM, for example, generates hierarchical driving plans that are subsequently grounded through conventional perception and trajectory planning~\cite{tian2024drivevlm}, while VLM MPC uses a vision-language model to propose symbolic maneuvers and MPC to convert them into physically feasible control actions~\cite{atsuta2025lvlm}. Similar architectures use VLM-generated subgoals together with task and motion planning for robotic manipulation~\cite{yang2025guiding}. When generative models additionally select tools, coordinate multiple models, maintain memory, and revise multi step plans toward a sustained objective, their role extends beyond decision generation toward the agentic orchestration discussed in Section~\ref{sec:agentic_ai}.

Two limitations should be considered when generative models are used within the decision loop. First, their online adaptation is considerably harder than that of conventional models, since learning a multi-modal distribution online requires more data and computational cost than refitting a single action per condition, and incremental updates risk collapsing the distribution onto a single mode. Second, their inference latency is substantially higher. Diffusion models require multiple denoising passes per decision, and VLMs rely on large autoregressive networks, so both are slower than a single forward pass of a conventional neural network policy. Although diffusion-based policy generation offers a computationally faster replacement for MPC without iterative solving, the forward speed of using conventional neural networks still remains faster~\cite{huang2024toward}. This latency is therefore not negligible and must be accounted for in the time budget of the loop at the system design level, rather than treated as an implementation detail.

\subsection{Agentic AI for Goal-Directed Orchestration}
\label{sec:agentic_ai}

The preceding subsections considered ML as a means of learning decision policies and generating candidate decisions. Agentic AI extends this role from synthesizing individual decisions to orchestrating the broader decision process. Agentic AI systems are defined as \emph{"systems that can pursue complex goals with limited direct supervision"}~\cite{shavit2023practices}. In Digital Twin applications, such systems can interpret high-level engineering objectives, decompose them into operational subgoals, construct and execute multi-step plans, coordinate models, solvers, and knowledge resources, retain relevant context in memory, and revise their plans according to feedback from the physical system~\cite{bandi2025rise,abouali2025agentic,wang2024survey,du2026survey}. ML therefore serves not only as a predictive or decision model, but increasingly as an enabling component for reasoning, planning, and coordination across the Digital Twin decision loop. These capabilities complement the vision of Cognitive Digital Twins introduced in Section~\ref{sec:CDT} by organizing their cognitive and computational functions around sustained, human-defined engineering objectives.

\begin{figure}[h!]
\centering
\includegraphics[width=0.95\linewidth]{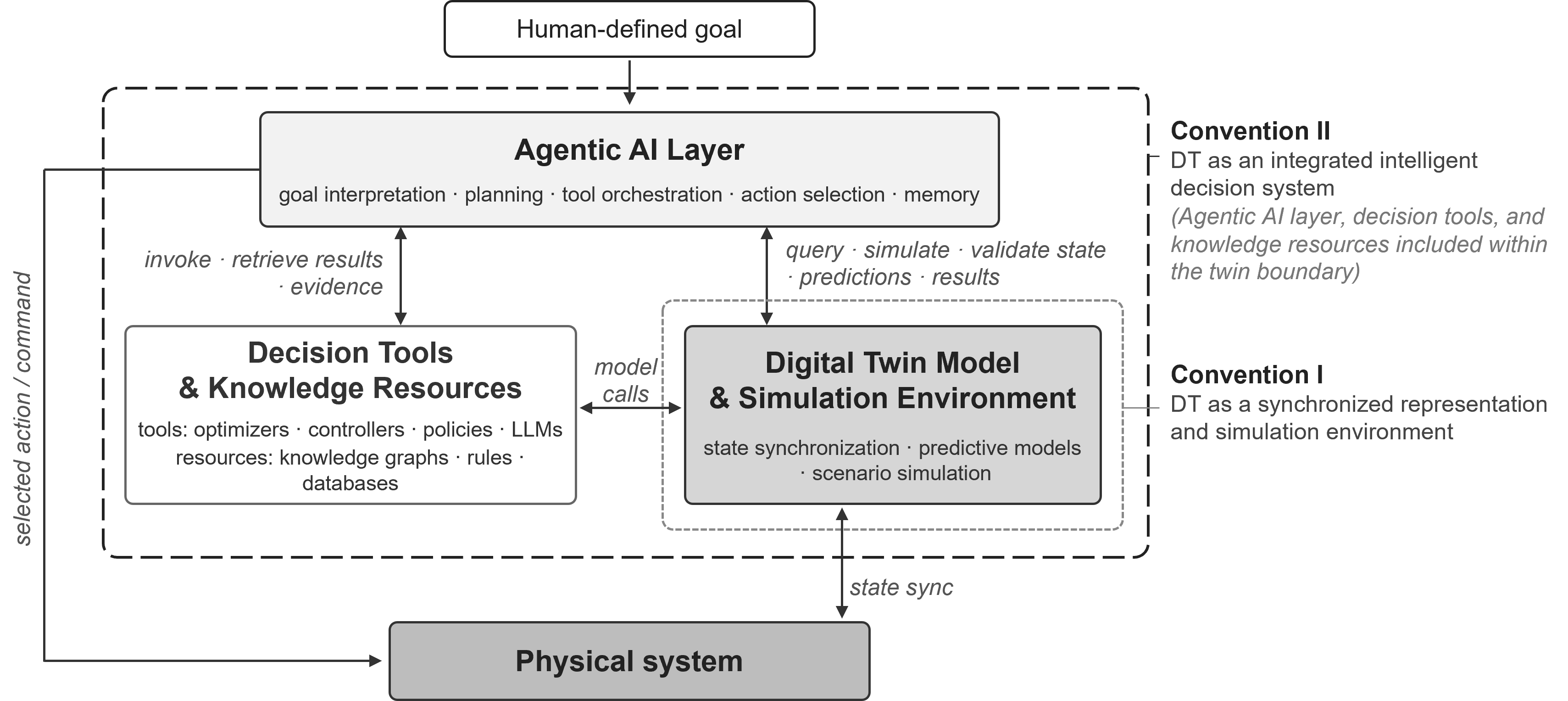}
\caption{Functional architecture for integrating agentic AI into Digital Twin decision-making. Dashed lines indicate alternative conventions for defining the Digital Twin system boundary.}
\label{fig:agentic_ai}
\end{figure}

Two broad conventions for incorporating agentic AI into Digital Twins have emerged in the literature. As illustrated in Figure~\ref{fig:agentic_ai}, the inner dashed boundary represents a model-centered convention, in which the Digital Twin is limited to the synchronized model and simulation environment and is used by a separate agentic layer for reasoning and action selection \cite{chen2025integrating,goyal2025empowering,kalyani2024role}. The outer dashed boundary represents a broader system-centered convention, in which the agentic decision layer and the Digital Twin model are jointly framed as an integrated Digital Twin or agentic AI--DT system \cite{nicoletti2026digital,hasan2026integrating}. Despite this difference in system boundary, both conventions share the same functional loop, in which human-defined goals are translated into plans, decision tools and knowledge resources are coordinated with the Digital Twin model and simulation environment, and selected actions are evaluated through feedback from the physical system.

The agentic process begins by interpreting a human-defined engineering objective in relation to the current Digital Twin state and engineering constraints. The objective is then decomposed into operational sub-goals and ordered tasks, with their dependencies and priorities organized into a multi-step plan. Memory supports this process by retaining the current task state, intermediate computational results, prior actions, and observed outcomes across successive decision steps. This persistent context allows the agent to maintain continuity over extended decision horizons and to adapt subsequent decisions to both the evolving physical state and progress toward the overall engineering objective \cite{wang2024survey}. A technical illustration is provided by an agentic grid-balancing architecture, in which the system-level objective of maintaining supply--demand equilibrium is operationalized through coordinated forecasting, resource-dispatch, charging-scheduling, and constraint-evaluation tasks supported by a shared grid twin \cite{hasan2026integrating}.

Once high-level objectives have been decomposed into operational sub-goals and multi-step plans, their execution depends on coordinating the decision tools, knowledge resources, and Digital Twin models available to the decision process. The agentic layer organizes the information and computational operations required by each task and returns intermediate evidence to the reasoning process before a decision is produced. A representative implementation for bridge crane maintenance distributes this process across a resource agent, a generation agent, multiple local reasoning agents, and a supervisory agent. These components transform sensor sequences and maintenance records into semantic evidence, combine inferences across different data dimensions, revise them using domain knowledge, and ultimately produce fault diagnosis and prognosis decisions~\cite{SUN202483}.

Execution and physical feedback complete the agentic decision loop. Selected actions are transferred to the physical system through existing controllers or operational interfaces, after which the resulting system state is synchronized with the Digital Twin model and simulation environment. The updated Digital Twin state provides new evidence for subsequent reasoning, memory updates, and plan revision. Agentic AI therefore establishes a continuous loop among goal interpretation, planning, computational orchestration, action, and feedback, allowing the decision process itself to evolve as the physical system and engineering objective change.

Overall, ML progressively expands its role in Digital Twin decision synthesis from learning to optimize, to learning to control, and ultimately to learning to orchestrate these capabilities toward higher-level engineering objectives. This progression highlights that the evolution of Digital Twins toward autonomous systems is not driven by surrogate modeling alone, but by the integration of learning and reasoning capabilities with models, decision tools, and domain knowledge. Such integration transforms the Digital Twin from a predictive representation into an intelligent decision system capable of coordinating heterogeneous resources, pursuing complex goals, and responding to emerging engineering needs.

\section{Uncertainty Quantification, Continual Validation and Adaptation for Digital Twins}
\label{sec:uncertainty}

UQ is a crucial component of Digital Twins because inherent variability in physical operation and incomplete knowledge in the virtual twin can make both future system response and state prediction uncertain, leading to suboptimal or unsafe decisions if this uncertainty is not accounted for. Different from existing work that reviews UQ methods for Digital Twins \cite{thelen2023comprehensive, battula2024uncertainty}, this review deliberately selects and structures the related topics in UQ that are most relevant to real-time decision-making, especially when ML models are implemented. The reliability of these components, however, cannot be established only during model development, but requires to be carefully monitored and maintained as the operating conditions, system properties, sensor characteristics, data distributions, decision objectives, control policies, and interactions among coupled Digital Twin components may change after deployment. A model or policy that was previously valid may therefore lose its fidelity and become unsuitable throughout the system life-cycle.

Maintaining the robustness of Digital Twins under uncertainty requires three key elements working together: 1) Uncertainty-aware decision-making, 2) continual monitoring and validation of the active models, and 3) continual learning and adaptation of the Digital Twins. These three elements together pinpoint how uncertainty can be quantified and leveraged in Digital Twins and decision-making. Therefore, this section first examines two operational roles of UQ in Digital Twins for supporting model adaptation and uncertainty-aware decision-making in Section~\ref{sec:uq}. It then reviews continual validation (Section~\ref{sec:continual} and drift detection (Section~\ref{sec:drift_detection}) as mechanisms for identifying when the current Digital Twin may no longer be adequate, followed by continual adaptation strategies ranging from recursive parameter estimation and bias correction to efficient fine-tuning of learned representations in Section~\ref{sec:continual_Adaptation}.
 
\subsection{Uncertainty Quantification for Real-Time Digital Twins}
\label{sec:uq}

The methodological foundations of UQ for ML and Digital Twin modeling have been reviewed extensively~\cite{fridkin2026physics,he2026survey,thelen2023comprehensive}. Existing reviews provide broad treatments of uncertainty characterization, calibration, propagation, validation, and model updating. However, these elements are often discussed as separate methodological components rather than organized around how uncertainty should be acted upon during closed-loop decision-making, and their application with ML models in particular is not widely investigated. In this review, UQ is therefore considered from an operational perspective, encompassing the representation, estimation, propagation, and management of uncertainty associated with the Digital Twin. The objective is not to provide another comprehensive taxonomy of uncertainty sources or UQ algorithms, but to examine how different forms of uncertainty inform different actions. Specifically, we consider how virtual models should be updated as new information becomes available, how predictive uncertainties should be considered in decision-making, and how the validity and computational cost of these uncertainty treatments should be assessed during operation.

A useful categorization of the types of uncertainty is whether uncertainty can, in principle, be reduced by acquiring additional information, which can lead to different purposes of UQ strategies. \emph{Epistemic uncertainty} represents incomplete knowledge/hypothesis about the physical system or its approximation in the Digital Twin and can be reduced when sufficiently informative observations become available. For example, uncertainty on estimated parameters can be reduced once more observation data are available. \emph{Aleatoric uncertainty} instead represents variability that is irreducible when the model is well-described, such as disturbances, environmental variability, manufacturing variation, or measurement noise \cite{arendt2012quantification}. Additional data may improve the characterization of this variability, but do not generally eliminate the underlying randomness. Importantly, this distinction depends on the adopted modeling assumptions rather than on the mathematical quantity through which uncertainty is represented. For example, uncertainty in a material property may be epistemic when the property has a fixed but unknown value, but aleatoric when the same parameter is used to describe physical variation among components or operating conditions.

For real-time decision-making in Digital Twins, the epistemic--aleatoric distinction characterizes the nature of uncertainty and methods to reduce it, but does not by itself prescribe how UQ is integrated with decision-making. From a decision-making perspective, UQ serves two primary operational roles \cite{agrell2023optimal}. The first is to support model adaptation by representing reducible epistemic uncertainty in a form that identifies which components of the Digital Twin remain uncertain and can be updated as new information becomes available. The second role is to support uncertainty-aware decision-making by characterizing and propagating uncertainty in future decision-relevant quantities. Accordingly, Section~\ref{sec:epistemic_uncertainty} examines UQ for model adaptation, with emphasis on how different representations of epistemic uncertainty correspond to different adaptation mechanisms. Section~\ref{sec:aleatoric_uncertainty} examines UQ for uncertainty-aware decision-making, with emphasis on how predictive uncertainty is quantified and propagated to support risk-aware decisions.

\subsubsection{Uncertainty Quantification for Model Adaptation}
\label{sec:epistemic_uncertainty}

For model adaptation, UQ focuses on epistemic uncertainty that reflects incomplete knowledge of the physical system as represented by the Digital Twin and can, in principle, be reduced as additional information becomes available. To enable online model adaptation of the surrogate models, it requires identifying which components of the Digital Twin remain insufficiently known and can be updated during operation. In this review, reducible uncertainty is organized according to the adopted model representation into three categories: model-parameter uncertainty, model-form uncertainty, and approximation uncertainty. Model-parameter uncertainty arises from incomplete knowledge of the parameters governing a physics-based model, whereas model-form uncertainty reflects inadequacies in the assumed model structure or missing physics. For data-driven models, approximation uncertainty refers to uncertainty in the learned approximation caused by limited training information. This usage follows the notion of approximation uncertainty described by Hüllermeier and Waegeman \cite{hullermeier2021aleatoric}, where the learned predictor remains uncertain because finite data provide incomplete information about the underlying function. In a Digital Twin, this can appear as disagreement among plausible predictions of state transitions in rarely encountered regimes, spatial fields at unsensed locations, or degradation trajectories beyond observed time horizons. These three forms of uncertainty naturally lead to different adaptation mechanisms through parameter estimation, model correction, and adaptation of the learned data-driven model, respectively.

This attribution of epistemic uncertainty to model parameters, model form, or the learned approximation then becomes a modeling choice. The same mismatch between predicted and observed behavior may arise from uncertain physical parameters, an inadequate model form, or an insufficiently learned representation. The appropriate representation depends on the available physical knowledge, the identifiability of candidate quantities from online measurements, and which components can feasibly be updated in real time. Accordingly, the following discussion considers model parameter uncertainty, model discrepancy, and approximation uncertainty as three common ways of assigning reducible epistemic uncertainty within an adaptive Digital Twin, rather than as intrinsically distinct physical sources of uncertainty.

\begin{table}[h!]
\centering
\footnotesize
\caption{Comparison of epistemic model uncertainty representations and their corresponding online adaptation approaches.}
\label{tab:model_uncertainty_adaptation}
\renewcommand{\arraystretch}{1.35}
\setlength{\tabcolsep}{5pt}
\begin{tabular}{
>{\raggedright\arraybackslash}p{0.1\textwidth}
>{\raggedright\arraybackslash}p{0.23\textwidth}
>{\raggedright\arraybackslash}p{0.20\textwidth}
>{\raggedright\arraybackslash}p{0.17\textwidth}
>{\raggedright\arraybackslash}p{0.18\textwidth}}
\hline

\textbf{Type of Epistemic Model Uncertainty}
&
\textbf{Representation}
&
\textbf{Corresponding Online Adaptation Methods}
&
\textbf{Strengths}
&
\textbf{Weaknesses}
\\
\hline

\textbf{Model parameter uncertainty}
&
Parametric physics-based or system identification models,
$
\mathbf{x}_{k+1}
=
\hat{f}(\mathbf{x}_k,\mathbf{u}_k;\boldsymbol{\theta})
$
&
Recursive parameter estimation, Bayesian calibration, Kalman filtering, and moving horizon estimation
&
Low-dimensional, computationally efficient, and physically interpretable updates
&
Limited by the predefined model structure and requires sufficiently identifiable parameters
\\[1ex]

\textbf{Model-form uncertainty}
&
Nominal physics-based or hybrid models with an explicit discrepancy,
$\mathbf{x}_{k+1}
=
\hat{f}_m(\mathbf{x}_k,\mathbf{u}_k)
+
\hat{\boldsymbol{\delta}}(\mathbf{x}_k,\mathbf{u}_k)
$
&
Recursive bias or discrepancy correction using Gaussian processes, NNs, RNNs, or other residual models
&
Corrects unmodeled dynamics while retaining the knowledge and stability of the nominal model
&
Requires a sufficiently reliable nominal model, and parameter and discrepancy effects may be difficult to distinguish
\\[1ex]

\textbf{Approximation uncertainty}
&
Data-driven surrogate models,
$
\mathbf{x}_{k+1}
=
\hat{f}_{\boldsymbol{\phi}}
(\mathbf{x}_k,\mathbf{u}_k)
$
&
Model fine-tuning, selective or head fine-tuning, and parameter-efficient adaptation
&
Provides greater plasticity for adapting learned dynamics and internal representations
&
Higher computational and data requirements, with risks of unstable updates and catastrophic forgetting
\\

\hline
\end{tabular}
\end{table}

When a physics-based or system identification model has a trusted functional form but contains uncertain or time-varying parameters, epistemic uncertainty can be assigned directly to those parameters as \emph{model parameter uncertainty}. Consider
\begin{equation}
\label{eq:par_uncertainty}
\mathbf{x}_{k+1}
=
\hat{f}(\mathbf{x}_k,\mathbf{u}_k;\boldsymbol{\theta}_k),
\qquad
\boldsymbol{\theta}_k \sim p(\boldsymbol{\theta}_k),
\end{equation}
where $\boldsymbol{\theta}_k$ may represent physically meaningful quantities that are uncertain or not directly observable during operation, such as material properties, degradation coefficients, heat transfer parameters, moments of inertia, or others. Here, uncertainty in $\boldsymbol{\theta}_k$ is interpreted as incomplete knowledge about quantities governing the current system dynamics rather than intrinsic variability across future realizations. The functional form of $\hat{f}$ is assumed to remain adequate, and new observations are used primarily to reduce uncertainty in $\boldsymbol{\theta}_k$. Because adaptation is restricted to a relatively small set of parameters, this approach can provide computationally efficient and interpretable model updating. Its effectiveness, however, depends on whether the selected parameters are sufficiently identifiable from the available measurements and whether the assumed model structure remains valid. This representation therefore leads naturally to recursive parameter estimation, as reviewed later in Section~\ref{sec:param_estimation}.

When a nominal model is available, but its functional form cannot fully represent the physical system, epistemic uncertainty can instead be associated with \emph{model-form uncertainty}, represented through a discrepancy or bias function,
\begin{equation}
\label{eq:model_form_uq}
\mathbf{x}_{k+1}
=
\hat{f}_m(\mathbf{x}_k,\mathbf{u}_k;\boldsymbol{\theta})
+
\hat{\boldsymbol{\delta}}(\mathbf{x}_k,\mathbf{u}_k),
\end{equation}
where $\hat{f}_m$ denotes the nominal model and $\hat{\boldsymbol{\delta}}$ represents unresolved structural discrepancy. Here, the discrepancy is interpreted as a reducible deficiency in the assumed model rather than irreducible stochastic variability. This formulation is appropriate when parameter calibration alone cannot account for missing physics, neglected interactions, or systematic modeling error. Rather than modifying the nominal model directly, adaptation targets the discrepancy while retaining the existing model as a stable physical backbone. The discrepancy can be represented using Gaussian processes, NNs, or other data-driven models and continually corrected as new observations become available. This approach provides greater flexibility than parameter estimation while preserving prior physical knowledge, although separating parameter error from structural discrepancy may be difficult when both can explain the same observed behavior. These methods are discussed further in Section~\ref{sec:bias_model}. In the Bayesian calibration framework proposed by Kennedy and O'Hagan~\cite{kennedy2001bayesian}, for example, model parameter uncertainty and model-form discrepancy can be estimated jointly through probabilistic representations of the simulator and discrepancy function. However, because the framework relies on repeated Bayesian inference and uncertainty propagation, it has predominantly been used for offline calibration and is considerably more difficult to incorporate into continuous online adaptation of dynamical systems.

A different situation arises when the dynamics are approximated primarily by a parametric data-driven surrogate model, such as an NN. In this case, epistemic uncertainty may be associated with the learned input-output representation itself and arises from finite or insufficiently informative training data, limited coverage of the operating domain, and the inductive assumptions imposed by the selected model architecture,
\begin{equation}
\mathbf{x}_{k+1}
=
\hat{f}_{\boldsymbol{\phi}}(\mathbf{x}_k,\mathbf{u}_k),
\qquad
\boldsymbol{\phi}
\sim
p(\boldsymbol{\phi}\mid\mathcal{D}_{0:k}),
\end{equation}
where $\boldsymbol{\phi}$ denotes the learned model parameters and $\mathcal{D}_{0:k}$ denotes the available training data. The uncertainty represented by $p(\boldsymbol{\phi}\mid\mathcal{D}_{0:k})$ describes incomplete knowledge of the learned mapping and can be propagated to quantify how predictions vary across plausible model realizations. Unlike a parametric physics-based model, however, there may be no small set of physically interpretable quantities that can simply be recalibrated as the operating regime changes. If new observations reveal dynamics that are inadequately represented by the pretrained model, adaptation may therefore require updating selected parameters, internal representations, or a broader portion of the model. Such adaptation provides substantially greater plasticity but generally requires more data and computation and introduces additional concerns such as unstable updates and catastrophic forgetting. This motivates the fine-tuning and parameter-efficient adaptation strategies discussed in Section~\ref{sec:peft}.

These representations are not mutually exclusive and may not be distinguishable from operational data alone. The same prediction discrepancy may be explained by changing $\boldsymbol{\theta}$, updating $\boldsymbol{\delta}$, or modifying $\hat{f}_{\boldsymbol{\phi}}$, while hybrid Digital Twins may contain several such uncertain components simultaneously. Effective UQ for model adaptation therefore requires not only representing reducible uncertainty, but also determining which uncertain components are identifiable and can be reliably updated from available online information. Model parameter estimation, discrepancy correction, and learned-model adaptation consequently provide different trade-offs between physical interpretability, flexibility, data requirements, and computational cost when maintaining an accurate Digital Twin under evolving operating conditions, which will be elaborated in Section~\ref{sec:continual_Adaptation}.

\subsubsection{UQ for Uncertainty-Aware Decision-Making}
\label{sec:aleatoric_uncertainty}

Beyond guiding model adaptation, one of the primary reasons for incorporating UQ into a Digital Twin is to support uncertainty-aware decision-making by characterizing the possible outcomes of candidate decisions and the associated performance and risk. Not only is uncertainty unavoidable due to sources of variability in Digital Twins \cite{karkaria2025optimization}, but it is also consequential for data-driven models, whose predictions depend on the coverage, distribution, and the stochasticity of the training data. Hence, both epistemic and aleatoric uncertainty can contribute to this predictive uncertainty. Epistemic uncertainty reflects incomplete knowledge of the Digital Twin model, while aleatoric uncertainty reflects variability that remains for a specified model realization. Their effects on the decision may therefore differ, but both can alter the predicted objective, constraint satisfaction, and resulting optimal action.

A general risk-aware decision problem can be expressed as
\begin{equation}
\begin{aligned}
\label{eq:robust_decision}
\mathbf{u}_k^{\star}
=
\arg\min_{\mathbf{u}_k\in\mathbb{U}}
\quad & J(\mathbf{u}_k) \\
\mathrm{s.t.}\quad
& g_j(\mathbf{x}_{k+1},\mathbf{u}_k)\leq 0,
\quad
\forall\,\mathbf{x}_{k+1}\in\mathcal{X}_{1-\alpha}(\mathbf{x}_{k},\mathbf{u}_k),
\quad \forall j,
\end{aligned}
\end{equation}
where $\mathcal{X}_{1-\alpha}(\mathbf{x}_k,\mathbf{u}_k)$ denotes a prediction region designed to contain the future response with probability at least $1-\alpha$. This provides the interface between the predictive model and the robust design formulation. Its usefulness depends on whether the resulting distribution or uncertainty bounds can be explicitly considered in the downstream decision.

In the literature, we identify two approaches that have been widely used to leverage epistemic uncertainty for risk-aware decision-making. One approach is to explicitly characterize uncertain quantities and propagate them through the Digital Twin. Consider a predictive model
\begin{equation}
\mathbf{x}_{k+1}
=
\hat{f}(\mathbf{x}_k,\mathbf{u}_k,\boldsymbol{\zeta}_k),
\qquad
\boldsymbol{\zeta}_k
\sim
p(\boldsymbol{\zeta}_k\mid\mathcal{D}_{0:k}),
\end{equation}
where $\boldsymbol{\zeta}_k$ may include uncertain model parameters, initial states, disturbances, boundary conditions, or other uncertain quantities. Through sampling or Bayesian inference, one can propagate the uncertainty into the predictive distribution of $\mathbf{x}_{k+1}$ \cite{kapteyn2021probabilistic}. Note that this formulation can incorporate epistemic and aleatoric uncertainty when both are explicitly represented. The resulting predictive distribution can then be summarized through decision-relevant risk measures and incorporated into the optimization problem. For example, Chaudhuri et al.~\cite{chaudhuri2023predictive} propagated uncertainty in patient-specific tumor-model parameters, which includes epistemic uncertainty stemming from Bayesian calibration, through a Digital Twin built on a probabilistic graph to obtain distributions of future tumor response. An $\alpha$-superquantile was then used to characterize the upper tail of the response distribution and support risk-aware optimization of radiotherapy treatments. The main challenge with this approach is that uncertainty propagation can become computationally demanding. Over a multi-step prediction horizon or long causal chain in the probabilistic graph, the distribution may become increasingly non-Gaussian and require numerical methods such as Markov Chain Monte Carlo (MCMC) to sample from the posterior over uncertainty parameters and propagate through the models, which requires many model evaluations. These costs become consequential when UQ is embedded within real-time decision-making.

\begin{figure}[h!]
\centering
\includegraphics[width=0.8\linewidth]{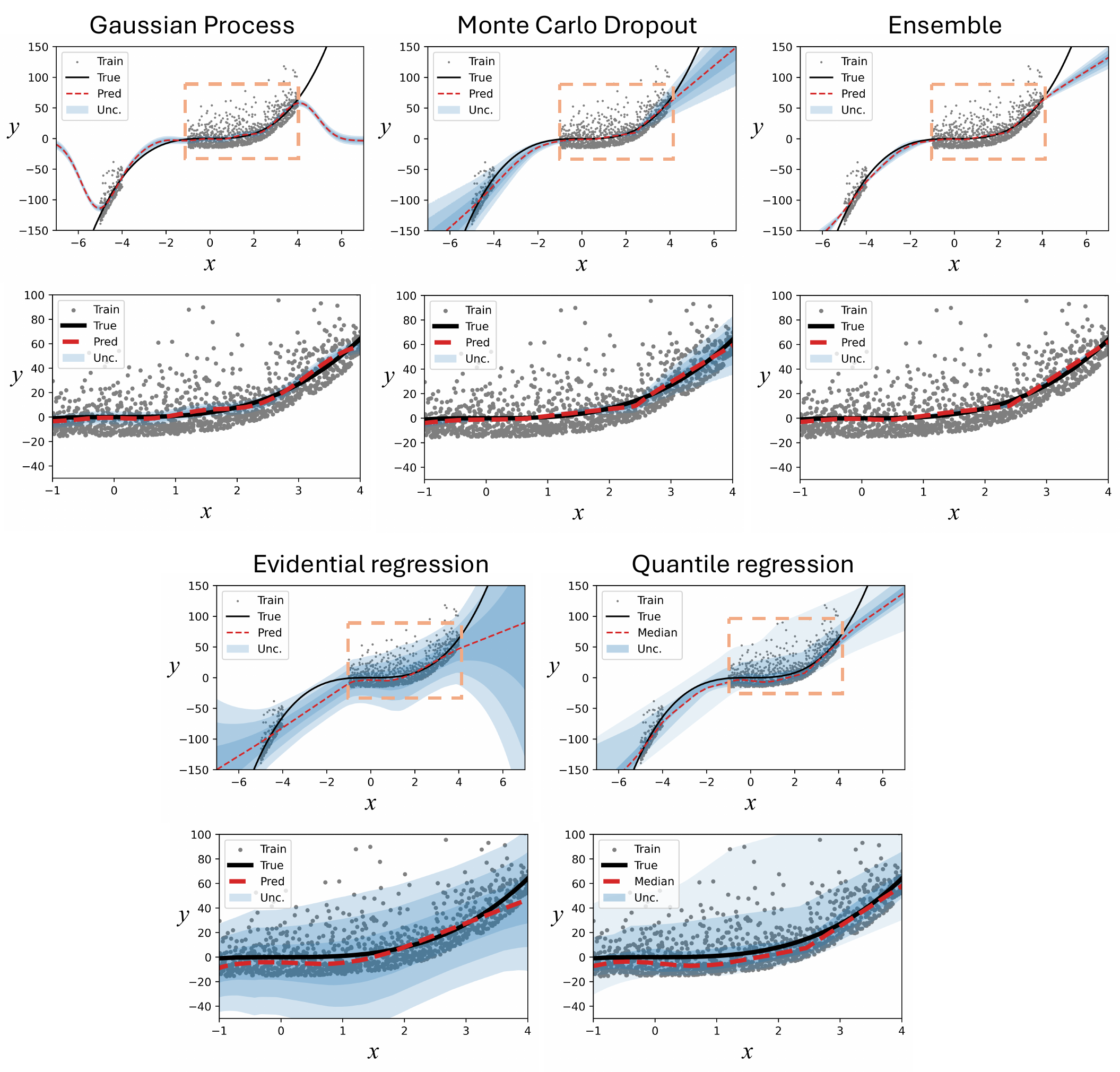}
\caption{Illustrative comparison of uncertainty estimates obtained using Gaussian processes, Monte Carlo dropout, ensembles, evidential regression, and quantile regression for synthetic regression problems with asymmetric Gamma-distributed noise, modified from \cite{chen2025uncertainty}. The purpose of this figure is to demonstrate that different UQ methods with NNs may reflect different types of uncertainty, and therefore, the selection of  UQ methods can be important depending on different applications. Gray points denote training observations, the black curve represents the underlying response, the red dashed curve denotes the predicted mean or median, and the blue regions represent the estimated uncertainty. Gaussian processes, Monte Carlo dropout, and ensembles primarily characterize epistemic uncertainty, with larger uncertainty in regions with limited training data. Their estimated uncertainty also differs substantially in magnitude because it depends on the adopted model and inference assumptions. Evidential regression additionally models aleatoric uncertainty but relies on an assumed Gaussian observation model and therefore produces symmetric intervals. Quantile regression instead estimates conditional response quantiles and captures the asymmetric variability of the response distribution in this example.}
\label{fig:ML_UQ}
\end{figure}

Another approach is to estimate predictive uncertainty directly from the data-driven surrogate model. For NN-based models, Bayesian NNs, deep ensembles, and Monte Carlo (MC) dropout generate multiple plausible model realizations and use the resulting prediction distribution to characterize uncertainty~\cite{gawlikowski2023survey, chen2025uncertainty}. Gaussian processes and evidential regression \cite{amini2020deep} assume the prediction distribution follows a Gaussian distribution and can provide the predictive variance, while the linearized Laplace Approximation (LLA) calculates model uncertainty by treating the network's parameters as a Gaussian distribution centered at the trained weights, using a simplified Hessian matrix to define the width and shape of that uncertainty \cite{eschenhagen2023kronecker}. These methods can therefore inform uncertainty-aware decisions by taking the confidence of the model into account. However, the predictive uncertainty should not automatically be interpreted as the distribution of future physical responses. First, the aforementioned UQ methods mostly quantify the \emph{model} uncertainty, indicating how confident the prediction could be based on the distribution of the training set, but not the \emph{data} uncertainty that describes the possible future state distribution; see Figure~\ref{fig:ML_UQ} for illustrations. Second, the magnitude of the estimated uncertainty depends on the hyperparameters such as model architecture, regularization assumptions, dropout rate, and therefore the distribution of these uncertainty estimates depends on modeling and inference choices, and their intervals do not necessarily provide calibrated coverage (from a frequentist perspective) unless additional calibration is performed. Thus, the resulting intervals primarily describe uncertainty of the learned predictor rather than the stochastic variability of future observations. 

When inherent variability in the physical system materially affects downstream decisions, quantifying aleatoric uncertainty can provide a predictive distribution that reflects the dispersion of possible future responses under the current operating condition. To serve this purpose, quantile regression and conformal prediction provide two approaches for constructing prediction intervals or sets directly from data. Quantile regression estimates selected conditional quantiles without prescribing a parametric observation distribution, and a sufficiently rich set of quantiles can characterize conditional response distributions of arbitrary form~\cite{fan2020deep, chen2025uncertainty}. For example, estimating the $\alpha/2$ and $1-\alpha/2$ quantiles provides the lower and upper bounds of a nominal $(1-\alpha)$ conditional prediction interval. The resulting bounds can capture asymmetric and state-dependent variability. Further, by leveraging the quantiles, the probabilistic constraints in Equation~\ref{eq:robust_decision} can be reformulated as a deterministic constraint and can be solved without using robust design optimization algorithms. Its reliability depends on sufficient training data, with estimation becoming more difficult for extreme quantiles and sparsely sampled operating regimes. Conformal prediction provides another approach for constructing predictive sets by calibrating model residuals or conformity scores against observed data. Under exchangeability, conformal prediction provides finite-sample marginal coverage without requiring a predefined parametric distribution~\cite{angelopoulos2023conformal}. It calibrates a predictive set against the observed prediction error. This makes it useful when the downstream robust decision requires coverage guarantees without an explicit decomposition of uncertainty sources \cite{hu2026conformal}.

These representations of aleatoric uncertainty can be incorporated directly into constrained optimization, such as through the form of RMPC or SMPC as introduced in Section~\ref{sec:rmpc}. For example, in the RMPC framework for additive manufacturing in~\cite{chen2025uncertainty}, a multi-step-ahead surrogate simultaneously predicts lower, median, and upper quantiles of melt pool temperature and depth. Probabilistic constraints on melt pool depth are then reformulated as deterministic constraints on the predicted quantiles. Because the uncertainty bounds are predicted over the complete control horizon, the MPC can account for aleatoric uncertainty learned from data without performing iterative uncertainty propagation during online optimization. 

The approaches above also reveal an important gap in uncertainty-aware decision-making. A physical system generally contains both aleatoric variability and epistemic uncertainty regarding its model representation~\cite{agrell2023optimal}. Yet these uncertainties are often quantified and incorporated into decisions separately. Methods based on ensembles, MC dropout, or Bayesian model inference commonly focus on epistemic uncertainty, while methods such as quantile regression on aleatoric uncertainty. This also raises two main research questions. The first concerns how epistemic and aleatoric uncertainty should be represented and propagated jointly while preserving their different interpretations, as these two types of uncertainty are usually represented with different statistical meanings and treated differently in decision-making. Simply adding independently estimated epistemic and aleatoric uncertainty bounds does not, in general, produce a calibrated predictive distribution. For instance, \cite{fan2020deep} attempts to perform a more conservative uncertainty estimate for robust MPC by adding the learned quantiles with epistemic uncertainty computed from the NN, while it acknowledges that the ratio between the addition of the two still relies on heuristics. The second concerns how a decision-making framework should use both forms of uncertainty simultaneously. A decision may need to remain robust to incomplete model knowledge while also controlling the risk associated with stochastic future outcomes. Developing computationally tractable formulations that preserve this distinction while supporting real-time optimization remains an important direction for uncertainty-aware Digital Twins.

\subsection{Continual Validation}
\label{sec:continual}
 
\begin{figure}[h!]
    \centering
    \includegraphics[width=0.8\linewidth]{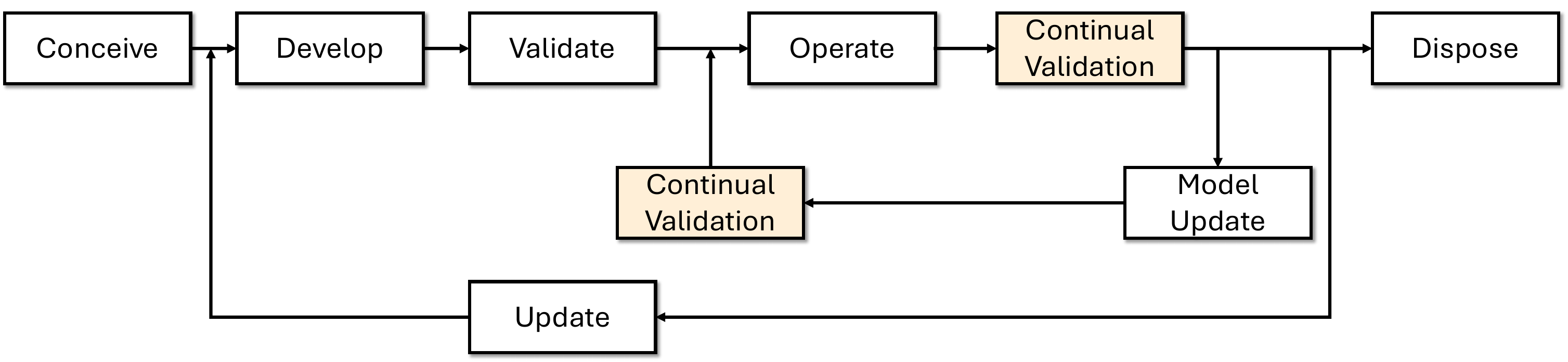}
    \caption{The life cycle of Digital Twin, with the emphasis on continual validation. The flowchart is modified from \cite{ferrari2024digital}.}
    \label{fig:validation}
\end{figure}

Because model accuracy may deteriorate as the physical system evolves, a Digital Twin must continuously assess whether its virtual model remains faithful to the physical system and valid under the current data distribution~\cite{national2023foundational}. Continual validation is therefore essential for determining when a Digital Twin should be updated and for verifying that each update preserves or improves its reliability \cite{ferrari2024digital}. Despite its importance, existing reviews/perspectives provide limited evidence of how continual validation can be incorporated into automated online Digital Twin operation, leaving several fundamental questions unresolved. As illustrated in Figure~\ref{fig:validation}, continual validation occupies two positions in the life cycle. The first validates the model during operation and determines whether a model update should be activated. The second validates the updated model after adaptation has been performed before it is admitted into the operational decision loop.
 
In existing work, continual validation generally follows the same principle as conventional model validation, i.e., comparing model predictions with corresponding ground-truth observations under the same inputs. Existing studies commonly replay the control actions applied to the physical system in the virtual model and compare the resulting simulated and measured responses \cite{lugaresi2023online, mertens2025reusing, hua2022validation}. Validation metrics may include root mean square error (RMSE) and normalized Euclidean distance for regression tasks \cite{mertens2025reusing}, as well as accuracy, precision, and recall for classification tasks \cite{abdoune2025digital}. The main distinction between offline and online validation lies in the available evaluation data. Offline validation can use design of experiments to systematically cover the operating regimes of interest, whereas online validation is restricted to streaming observations generated along the realized operating trajectory and may therefore provide evidence only within a narrow regime. Directly transferring offline validation methods to online operation consequently introduces several limitations that require dedicated methodological development.

First, there is a lack of principled criteria for determining if a model is \emph{good enough} or requires an update. Different from classification tasks, such as defect detection, where fixed or hard thresholds may be easier to interpret and determine, the decisions on validation are more difficult to define for regression models because the relationship between validation metrics, such as the RMSE of a surrogate model used in MPC, and system-level performance, such as reference-tracking error, may be weak. This makes it difficult to justify a single threshold for triggering an update. Even when such thresholds can be established, one threshold may not remain applicable throughout the system life cycle because the scale of the validation metric may change after model updates or across operating conditions. Second, labeled data may not be available during autonomous operation, preventing direct model evaluation. For example, classification metrics can only be computed after the data are correctly labeled, which often requires human intervention \cite{trovato2026continual}. Validation may therefore be delayed until operation is paused. Third, conventional validation metrics such as RMSE can be computed for temporally correlated data, but interpreting their variation and establishing statistically meaningful thresholds often relies on independence or stationarity assumptions that may not hold for streaming data from dynamical systems. Because these observations are also noisy, evaluating such metrics at high frequency may make validation overly sensitive to transient deviations, while batch-based aggregation may conceal early indicators of persistent change. We therefore argue that continual validation should not rely solely on repeated computation of conventional validation metrics. Formal drift-detection methods, introduced in the next section, provide a complementary mechanism for monitoring changes in streaming operation and determining when model validation or adaptation should be initiated. 
 
\begin{figure}
    \centering
    \includegraphics[width=0.9\linewidth]{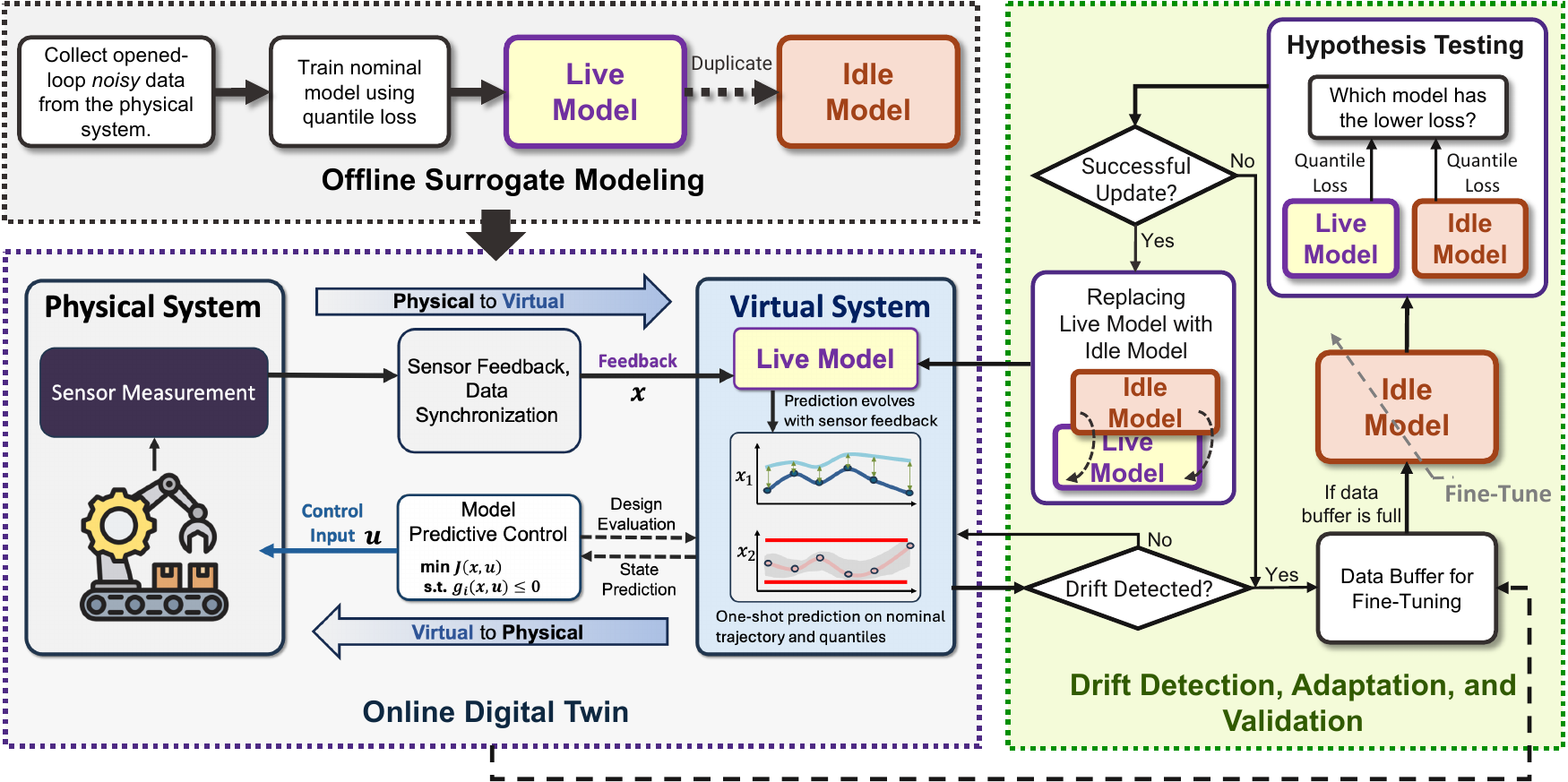}
    \caption{An online framework integrating continual drift detection, adaptation, and validation through an A/B testing-inspired live and idle model structure, modified from \cite{chen2026continual}. The integrated module can be executed in parallel to the online Digital Twin so that the updated model will only be deployed to the decision loop once it is validated. }
    \label{fig:adaptive_DT}
\end{figure}

Two additional challenges remain for continual validation. First, while most of the efforts on model validation focus on monitoring the prediction of the model (i.e., the validation block in Figure~\ref{fig:validation} after Operation), few studies emphasize the validation of the updated model for near-real-time adaptation (i.e., the validation block in Figure~\ref{fig:validation} after Model Update) in a continuous, autonomous setting \cite{ferrari2024digital}. Most existing approaches either require stopping the machine for additional fine-tuning and offline validation \cite{trovato2026continual, gupta2024twin}, or deploy the fine-tuned model without explicit validation under the assumption that its performance has improved \cite{hedjar2013adaptive}. The latter is risky because an inadequate update to a model embedded in the decision loop may further degrade predictions and harm the physical system. A potential solution to this challenge is introduced in \cite{chen2026continual} through an A/B testing-inspired live and idle model architecture, as illustrated in Figure~\ref{fig:adaptive_DT}. The framework maintains the model currently active in the decision loop as the \emph{live} model and creates a duplicate \emph{idle} model that remains outside the operational decision pathway. When adaptation is triggered, only the idle model is updated, while the live model continues to support ongoing operation. The updated idle model is subsequently evaluated using newly available operational data and replaces the live model only after satisfying the prescribed validation criteria. This separation provides a dedicated pathway for post-update validation and prevents partially adapted or insufficiently validated models from being introduced directly into the decision loop. However, fundamentally, even when the surrogate model is validated and accepted after fine-tuning, it does not provide a guarantee that the model will remain accurate in the future, especially when the system is operating under non-stationary conditions. Therefore, it also highlights the need for adaptation and validation to be continuously executed throughout the system life cycle.

The second challenge arises when a Digital Twin comprises multiple coupled ML components, such as data assimilation, surrogate modeling, and decision layers connected within the same closed-loop architecture. Each component supports the overall functionality, while errors may propagate across layers and affect system-level performance. The central challenge is therefore to validate both each constituent component and the Digital Twin as an integrated system. This challenge mirrors the attribution problem described in Section~\ref{sec:uq}, since component-level validation evidence does not compose into a system-level guarantee when the layers are coupled. Addressing these challenges requires continual-validation methods that link model-level evidence to system-level performance and support safe model updates during operation.

\subsection{Drift Detection}
\label{sec:drift_detection}
 
Unlike validation, which evaluates model accuracy against ground truth using predefined metrics, drift detection assesses whether deployment behavior remains consistent with previously validated operation \cite{pan2009survey, redko2020survey}. Compared to online validation, it can be more compatible with Digital Twins with streaming data available, which can be noisy, temporally correlated, trajectory-dependent, and may lack immediate labels. Rather than reacting to individual prediction errors, drift detection accumulates evidence over time to identify persistent changes in inputs, outputs, residuals, or input--output relationships. By operating sequentially and distinguishing transient fluctuations from abrupt or gradual changes, it provides a more suitable mechanism for triggering further validation or model adaptation during online operation.
 
The persistent statistical change identified by these methods is referred to as \emph{drift}. It describes a departure between previously validated and current operation in the distribution of model inputs, outputs, or their underlying relationship \cite{pan2009survey, redko2020survey}. Formally, drift occurs when
\begin{equation}
    \exists\, t_0 < t_1 \ \text{ such that } \ P_{t_0}(x,y) \neq P_{t_1}(x,y).
    \label{eq:concept_drift}
\end{equation}
The change may affect the input distribution \(P_t(x)\), the conditional relationship \(P_t(y|x)\), or both. Of particular importance to Digital Twin adaptation is \emph{concept drift}, where \(P_t(y|x)\) changes and the previously validated model may no longer represent the current system. Because such changes are not always directly observable, practical drift detectors may monitor inputs, outputs, residuals, or model-derived statistics as indicators of an evolving data-generating process. In Digital Twins, concept drift may be induced by changing process settings, environmental disturbances, equipment aging, material variability, sensor degradation, or new operating regimes. Concept drift may also be sudden, gradual, incremental, or recurring, and its temporal pattern influences both detection and adaptation \cite{gama2014survey}.
 
Drift detection methods can be organized according to how monitored information is processed to detect changes. \emph{Error-based} methods track prediction performance and interpret deviations from historical error statistics as evidence of drift. Examples include the Drift Detection Method \cite{gonccalves2014comparative}, exponentially weighted moving average (EWMA) control charts \cite{crowder1992ewma}, and cumulative sum (CUSUM) tests \cite{biau2007quality}. These methods are suitable for sequential monitoring because they operate on low-dimensional statistics and can be updated as observations arrive. For example, \cite{wu2019real} leverages a Lyapunov function to monitor the stability and the predictive error of the model in MPC, and develops an event trigger and an error trigger to initiate model updating. Their main limitation is that prediction error may increase only after the underlying distribution has already changed, resulting in late drift detection, especially under gradual drift or covariate shift \cite{zhang2023concept}. Scalar error statistics also provide limited information about the source of degradation. Nevertheless, implementing error-based approaches is relatively straightforward and computationally cheap for monitoring streaming data.
 
\emph{Distribution-based} methods compare incoming data with a historical reference distribution. Adaptive Windowing \cite{bifet2007learning}, Maximum Mean Discrepancy \cite{borgwardt2006integrating}, and the Kolmogorov--Smirnov test are representative examples. These methods can identify changes before prediction error becomes substantial and have been applied in process-monitoring settings such as semiconductor manufacturing \cite{li2025sudden}. An example can be seen in \cite{karkaria2025ai} on monitoring concept drift in tires' remaining useful life (RUL) prediction. Here, the maximum concept discrepancy (MCD) is implemented to project high-dimensional, temporal prediction and feedback into a lower-dimensional latent space, and leverage the distances between deep temporal embeddings for drift detection. Their cost, however, can increase with data dimensionality because they may require large reference windows, repeated kernel evaluations, or density estimation. Window-based methods may also respond slowly to incremental changes when successive observations remain locally similar \cite{desale2023concept, yang2021lightweight}.
 
\emph{Model-based} methods use information encoded within the predictive model. The score-based framework of Zhang and Apley \cite{zhang2023concept}, for example, monitors the Fisher score vector, defined as the gradient of the log-likelihood with respect to model parameters. The score measures how strongly new observations conflict with the assumptions represented by the current model. Because the monitored score components can be associated with individual parameters or parameter groups, model-based drift detection can provide not only a drift signal but also attribution of where the learned representation has become inconsistent with new observations. This attribution can support diagnosis across multiple machine-learning components of a Digital Twin and can help determine which models or parameter groups require adaptation. This method is demonstrated in a case study in additive manufacturing in \cite{chen2026continual}. The consistency of incoming observations with the current NN-based surrogate, which provides multi-step, multi-level quantile predictions for melt pool temperature and melt pool depth, is monitored through the score vector, followed by model adaptation and continual validation.
 
The deployment of drift detectors in Digital Twins remains challenging because some existing monitoring approaches assume that operation can be paused while anomalies are investigated or monitoring limits are recalibrated. Such assumptions are poorly aligned with Digital Twins that must continue issuing decisions during operation. Some adaptive monitoring schemes, for example, update control limits only after newly collected data have been classified as normal by an operator \cite{zhao2007adaptive}. When the operating regime is continuously evolving, distinguishing normal variation from meaningful drift becomes difficult, while repeatedly resetting detection thresholds may introduce delays or anchor the detector to an outdated reference. Methods that require large data windows or repeated statistical estimation may also conflict with the latency and data constraints of real-time deployment. Drift detection should therefore be treated as a trigger within a broader validation and adaptation loop rather than as a standalone monitoring task. A detected change indicates that the data-generating process may have shifted, but does not by itself establish that the current model is no longer adequate for its decision task. Continual validation remains necessary to assess the operational consequence of the detected change and determine whether adaptation is warranted. 

\subsection{Continual Adaptation}
\label{sec:continual_Adaptation}

Once drift detection identifies a persistent change and continual validation confirms that the affected Digital Twin component no longer provides sufficient predictive fidelity or decision performance, adaptation should be initiated~\cite{ferrari2024digital}. As model parameters, model discrepancy, and learned representations have been introduced in Section~\ref{sec:epistemic_uncertainty} as three common representations for reducible epistemic uncertainty, here we introduce the corresponding adaptation methods according to which component of the Digital Twin must be modified. When the existing model structure remains valid, adaptation can be limited to recursive estimation of a compact set of parameters. When a nominal model remains useful but exhibits systematic discrepancy, a learned bias model can provide corrective dynamics. When the learned representation itself is no longer adequate, adaptation may require fine-tuning of the model parameters or internal representations. The following sections elaborate these three mechanisms, which provide increasing flexibility at the cost of generally higher data and computational requirements.
 
\subsubsection{Recursive Parameter Estimation and Adaptation}
\label{sec:param_estimation}

Parameter estimation has been studied extensively for physics-based and system identification models and remains relevant to Digital Twins because it can support both offline calibration and online model updating~\cite{ouyang2020bayesian,nath2022probabilistic}. It is suitable when the model can be characterized by a manageable number of uncertain or time-varying parameters, in contrast to NNs with a massive amount of tunable parameters. Offline estimation commonly relies on Bayesian inference or maximum likelihood methods, providing either a posterior distribution of the parameters or a point estimate such as the maximum a posteriori estimate~\cite{jalal2021baycann,arendt2012quantification,kennedy2001bayesian,li2016integrating}. For continual adaptation, recursive parameter estimation represents the lowest-cost mechanism because it preserves the existing model structure and updates only a small set of parameters. It is therefore a natural choice when drift can be attributed to parameter changes rather than inadequacy of the model form.

Formally, the uncertain parameters can be updated using Bayesian model calibration \cite{chaudhuri2023predictive, kapteyn2021probabilistic}. Prior distributions are assigned to the uncertain parameters, and each time new measurements arrive, the posterior is obtained through Bayes rule. The strength of this approach is that the posterior is not restricted to any particular shape, where it can be skewed, correlated across parameters, and confined to a physically admissible range, which matters when only a handful of measurements are available and the priors carry much of the information. Propagating this posterior forward yields a full predictive distribution of the quantity of interest, so risk-aware decisions can be made \cite{chaudhuri2023predictive}. The cost is that the posterior has no closed form for a nonlinear model and must be sampled, typically by MCMC, which requires thousands to millions of forward simulations. This is acceptable when calibration is performed once before a decision is made, but prohibitive for Digital Twins whose decision-making and model updates require high frequency. 

To estimate parameters online, a widely adopted approach is to augment the unknown parameters as additional states and estimate them recursively using Kalman filtering (KF). This is the same Bayesian update carried out under linear and Gaussian assumptions, which make the update available in closed form. By treating the uncertain model parameters as the augmented states, KFs can provide efficient, probabilistic estimates of the unknown parameters effectively. For example, augmented-state Extended Kalman Filters (EKFs) have been used to maintain Digital Twin fidelity by continuously updating model parameters from nonintrusive sensor measurements~\cite{baldassarre2024digital} and to update dynamic models of power systems from streaming data~\cite{Fan2013EKFPMU}. These methods are attractive for real-time Digital Twins because of their low per-step computational cost, recursive updates that reuse the previous estimate rather than reprocessing all past data, and an uncertainty estimate carried in the state error covariance. The compromise is that this covariance summarizes a Gaussian approximation of the posterior, so the estimate is most trustworthy near its mean; it is an acceptable trade-off when measurements arrive frequently enough to keep the estimate well informed, and the downstream decision depends on the parameter estimate. When system nonlinearities are pronounced, noise distributions are heavy-tailed, or hard constraints must be imposed, purely recursive filters can become unreliable. Moving horizon estimation (MHE) instead solves a finite window optimization problem at each time step, providing greater flexibility for handling constraints, outliers, and drifting parameters~\cite{Rangegowda2022RNK, Valluru2018RobustJPC}.

\begin{figure}[h!]
    \centering
    \includegraphics[width=0.9\linewidth]{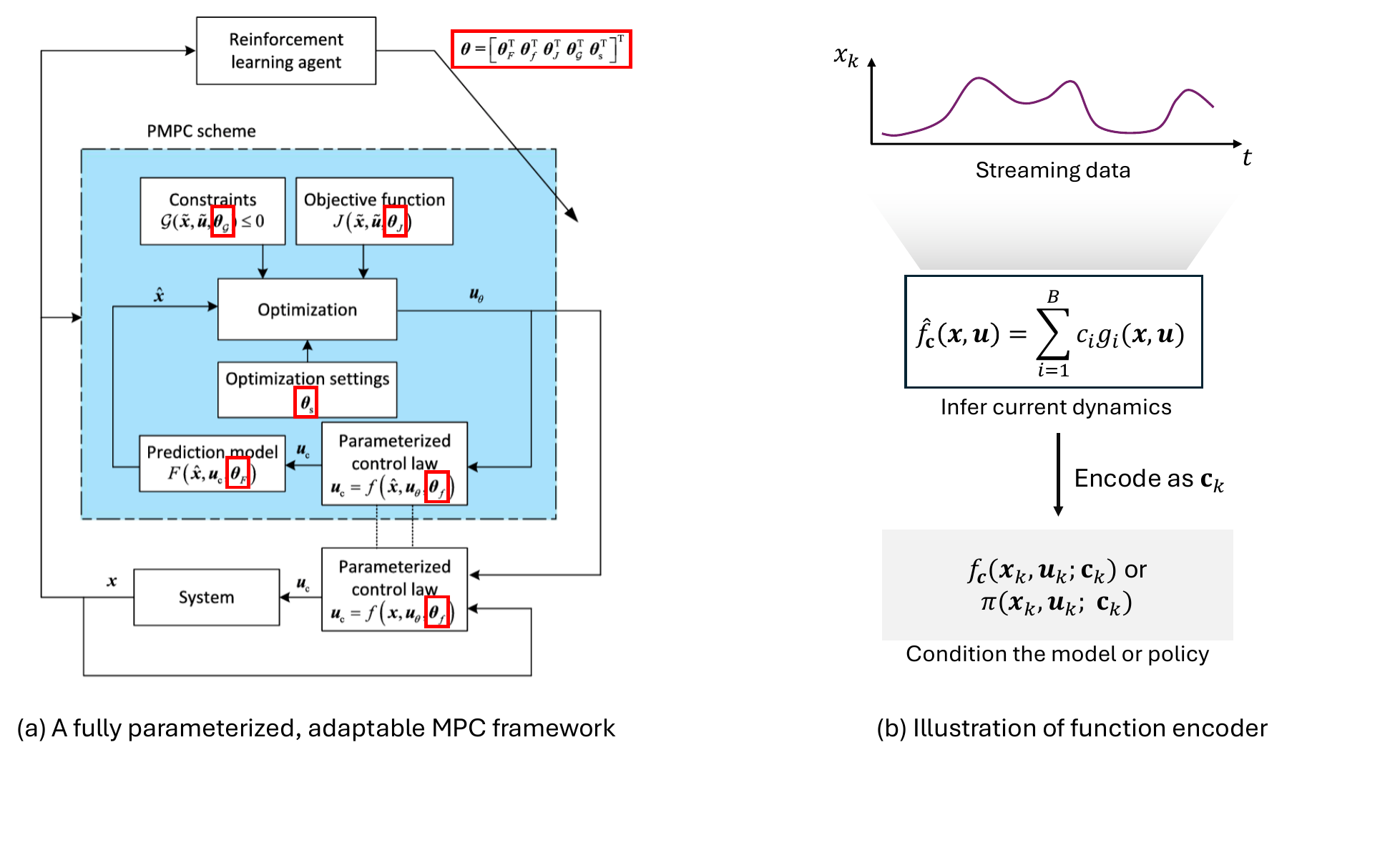}
    \caption{Two examples of leveraging recursive parameter estimation for system adaptation. (a) Adaptive parameterized model predictive control based on reinforcement learning, modified from \cite{sun2024adaptive}. Each module within the MPC, such as the objective function, constraints, model, or control law setup, is parameterized by its own parameter $\theta_i$, where $i$ indicates labels for different modules in the system. The joint parametric space is sampled offline to train an RL agent, which is used to adapt the system by treating the parameter values as the actions.  (b) Illustration of function-encoder-based adaptation, where streaming data are used to identify the current dynamics $\hat{f}_\mathbf{c}$ through coefficients $\mathbf{c}_k$ over fixed learned basis functions $g_i$ (usually by NNs), and the resulting low-dimensional representation conditions the predictive model or policy.}
    \label{fig:parameter_estimation}
\end{figure}

Moreover, online parameter estimation is not restricted to physics-based models but can also be applied to data-driven models with fixed, parameterized representations. The same principle can be extended to learning-based models by treating the adaptive parameters as additional conditioning variables. For example, Equation~\ref{eq:par_uncertainty} can be rewritten as
\begin{equation}
\mathbf{x}_{k+1}=\hat{f}(\mathbf{x}_k,\mathbf{u}_k,\boldsymbol{\theta}_k),
\end{equation}
such that the predicted dynamics depend not only on the state $\mathbf{x}_k$ and decision $\mathbf{u}_k$, but also on a recurrently estimated parameter vector $\boldsymbol{\theta}_k$. By training the model across the joint state, input, and parameter domain, the learned mapping can represent a family of system behaviors and adapt to a new operating condition once the corresponding parameters are estimated, without modifying the model weights. Such zero-shot adaptation has been demonstrated by embedding physics-inspired parameters directly into learned models~\cite{Thosar2025PINNParamMaintenance,jeong2026ppie}. Other architectures expose quantities such as stiffness or defect size as identifiable variables that can be inferred from operational measurements~\cite{Song2025BearingDTparam}. The same idea can also be applied beyond surrogate modeling. For example, multiple elements of an MPC framework, including objectives, constraints, control laws, and optimization settings, can be represented through a unified set of tunable parameters and adjusted online by a reinforcement learning agent~\cite{sun2024adaptive}, as shown in Figure~\ref{fig:parameter_estimation}(a). In this case, recurrent estimation extends from model parameters to system-level quantities that determine how the Digital Twin makes decisions. A special case under this category is to update the learned linear Koopman operator. By assuming the lifting functions remain fixed, the update of the Koopman operator can be recursively estimated using computationally efficient methods through closed-form solutions, such as recursive least squares \cite{chen2025adaptivedigitaltwinsheet} or Bayesian updates that also provide a posterior predictive distribution that captures both epistemic and aleatoric uncertainty \cite{selim2026metakoopman}. 

Following the same principle, identifiable parameters can be generalized to low-dimensional latent representations that do not necessarily have direct physical meaning. Function encoders provide one realization~\cite{ingebrand2025function,hu2025manifold}, representing a family of functions using shared learned basis functions and a compact coefficient vector $\mathbf{c}_k$ that characterizes the current system or task, as illustrated in Figure~\ref{fig:parameter_estimation}. For dynamical systems, these coefficients can be inferred from streaming observations and supplied to the model as
\begin{equation}
\mathbf{x}_{k+1}=\hat{f}(\mathbf{x}_k,\mathbf{u}_k, \mathbf{c}_k),
\end{equation}
allowing adaptation through estimation of $\mathbf{c}_k$ while keeping the learned basis functions fixed. For example, \cite{iqbal2025zero} infer function encoder coefficients from limited state transition data to identify unseen dynamics and enable zero-shot adaptation of both the dynamics model and control policy. Similarly, \cite{li2026zero} represent optimal control policies using shared basis functions and task-specific coefficients, enabling zero-shot adaptation to new objectives from limited trajectory data or problem specifications. Recursive parameter estimation can therefore extend beyond physically interpretable parameters to compact latent coordinates that identify the current dynamics or task within a learned function space, although additional efforts on offline simulation and design of experiments are required for generating training samples. 

A fundamental concern for physically interpretable parameters is \textit{structural identifiability}: the mapping from observations to parameters may not be unique, such that multiple parameter configurations produce equally plausible model outputs. This problem becomes significant in high-dimensional or multiphysics systems, where parameters interact nonlinearly and available measurements provide only partial information about the system state. Even when the parameters are theoretically identifiable, estimations may converge to local solutions or physically implausible values, such as negative stiffness coefficients or heat transfer rates outside admissible ranges, particularly when the estimation problem is poorly constrained or the observed trajectory provides insufficient excitation. Potential remedies include imposing physical admissibility constraints, using sensitivity analysis to identify informative parameters~\cite{chen2021unknown}, and incorporating multiple measured responses to improve identifiability~\cite{arendt2012improving}. Scalability presents an additional challenge as Digital Twins expand to multicomponent, hierarchical, or multiphysics systems, where the number of quantities to be estimated may increase faster than the available measurements. Recursive parameter estimation is therefore most effective when adaptation can be represented through a compact and sufficiently identifiable parameterization. When this assumption no longer holds, more flexible model correction or fine-tuning mechanisms become necessary.

\subsubsection{Continual Model Correction on Model-Form Uncertainty}
\label{sec:bias_model}
 
When estimating uncertain model parameters alone is insufficient to recover the required model fidelity because important physics are missing or inadequately represented, adaptation must account for model-form uncertainty. A common strategy is to decompose the system dynamics into a nominal physics-based model and a learned residual dynamics model through a bias function \cite{ebers2024discrepancy, li2024digital, thelen2023comprehensive}, consistent with Equation~\ref{eq:model_form_uq}:
\begin{equation}
    \mathbf{x}_{k+1}
    =
    \hat{f}_m(\mathbf{x}_k,\mathbf{u}_k;\boldsymbol{\theta}^*)
    +
    \hat{\boldsymbol{\delta}}(\mathbf{x}_k,\mathbf{u}_k;\boldsymbol{\eta})
    +
    \boldsymbol{\epsilon}_k ,
\end{equation}
where $\hat{\boldsymbol{\delta}}$ is a black-box function approximator for the unresolved residual dynamics with parameters $\boldsymbol{\eta}$, and $\boldsymbol{\epsilon}_k$ represents Gaussian noise. The nominal model $\hat{f}_m(\mathbf{x}_k,\mathbf{u}_k;\boldsymbol{\theta}^*)$ is typically constructed before deployment using the best available parameter estimate $\boldsymbol{\theta}^*$ and encodes the known first-principle dynamics. During operation, $\hat{\boldsymbol{\delta}}$ is updated recursively from the residuals between model predictions and observations, allowing the hybrid model to adapt to unmodeled behavior while preserving the nominal model. This correction-based formulation constrains the hypothesis space of the bias model through the physics-based backbone and therefore improves the balance between stability and plasticity during online adaptation. Importantly, the bias term is not merely an additive correction to the final model output. It modifies the state transition itself, so even a small residual error can accumulate and substantially alter the predicted dynamics, especially in nonlinear systems. This highlights the importance of bias modeling when parameter correction alone cannot compensate for incomplete physics.

In real-time Digital Twin applications, Gaussian processes have been widely used to represent $\hat{\boldsymbol{\delta}}(\mathbf{x})$ because they provide uncertainty estimates and support recursive updating \cite{gardner2021learning, brynjarsdottir2014learning}. Their uncertainty estimates can also be propagated into downstream uncertainty-aware decision methods such as stochastic MPC \cite{hewing2019cautious}. Sparse approximations, including inducing point methods, can further reduce the computational cost of GP prediction and updating to support real-time implementation \cite{hewing2019cautious, maupin2020model}. In \cite{hewing2019cautious}, an iteratively updated GP bias model is combined with a simplified bicycle model for race car control, improving MPC performance by correcting model mismatch while maintaining computational feasibility. Neural network (NN)-based bias models have also been explored for more complex residual dynamics \cite{kaheman2019learning}. For example, \cite{ren2022bias} uses an RNN bias model on top of a linear bicycle model and updates it online through stochastic gradient descent to compensate for changing road conditions and improve MPC predictions.

\begin{figure}[h!]
    \centering
    \includegraphics[width=0.6\linewidth]{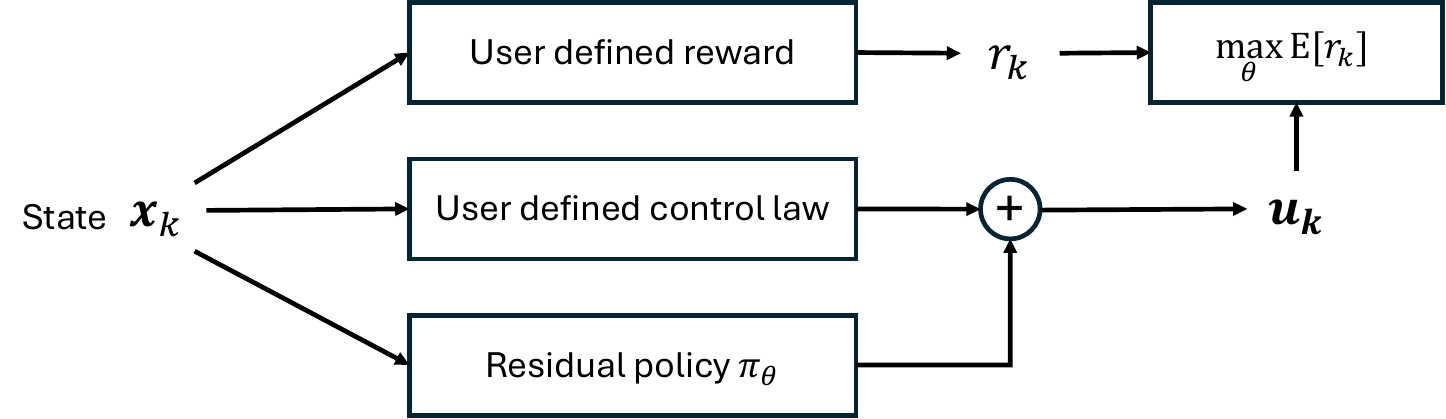}
    \caption{Illustration of residual policy learning, modified from \cite{johannink2019residual}. A learned residual policy provides a corrective action on top of an existing base control law to improve the realized reward.}
    \label{fig:residual_policy}
\end{figure}

The same principle can also be applied to the decision layer. Just as a predictive model can be corrected by combining a nominal dynamics model with a learned residual dynamics model, residual policy learning combines a trusted base policy with a learned corrective policy,
\begin{equation}
    \bf{u}_k
    =
    \pi_{\mathrm{base}}(\mathbf{x}_k)
    +
    \pi_{\mathrm{res}}(\mathbf{x}_k),
\end{equation}
as illustrated in Figure~\ref{fig:residual_policy}. The base policy may be an MPC controller, an RL policy, or a hand-designed feedback law, while the residual policy is learned from operational data to compensate for behavior not captured by the original controller \cite{silver2018residual, johannink2019residual}. This structure preserves the existing decision policy and restricts learning to the corrective component, which can improve data efficiency and reduce the risk associated with learning a controller from scratch. In \cite{silver2018residual}, residual policy learning is applied to robotic manipulation tasks where direct reinforcement learning is data inefficient or infeasible, and the learned residual improves hand-designed controllers and MPC policies under partial observability, sensor noise, model mismatch, and controller miscalibration. Similarly, \cite{rana2023residual} introduce residual skill policies that adapt pretrained skill representations to task variations that cannot be handled by the original skills alone, improving transfer to unseen tasks.

Beyond bias function learning, recent advances in system identification provide an alternative approach for addressing model-form uncertainty and incorrect modeling assumptions without introducing an explicit bias function. One example is the online adaptation of dynamical models through Sparse Identification of Nonlinear Dynamics (SINDy) \cite{brunton2016discovering}. SINDy represents the governing dynamics as a sparse combination of candidate nonlinear basis functions and identifies a small set of active terms and their corresponding coefficients from observed system trajectories. In this way, changes in the active terms or their coefficients can modify the effective model structure rather than correcting the prediction error through an additional residual model. By treating the SINDy coefficients as uncertain or time-varying quantities, the identified model can be updated sequentially as new system measurements become available, enabling online identification of evolving dynamics \cite{rosafalco2025online}. Therefore, by reformulating model identification as a sparse parameter estimation problem, SINDy-based adaptation can leverage the sample and computational efficiency of recursive estimation methods such as EKFs to continuously update both the effective model structure and its coefficients. This provides a different route to model-form adaptation in which the governing model itself is corrected rather than supplemented by a learned bias function.

A key limitation of bias-based adaptation is the difficulty of jointly estimating model parameters and model discrepancy in real time. In many online applications, the nominal parameters $\boldsymbol{\theta}^*$ are treated as fixed, and the entire correction is assigned to the bias model. Bayesian calibration methods such as \cite{kennedy2001bayesian} can estimate uncertain parameters and discrepancy functions jointly, as demonstrated in \cite{zhu2020real,titscher2023bayesian}, but their computational cost still limits real-time deployment. The applicability of the gray box formulation also depends on the availability of a sufficiently reliable nominal model. When such a model cannot be constructed, adaptation can no longer rely on correcting a trusted backbone and instead requires direct updating of a flexible learned model. The same situation arises for other learning-based components of a Digital Twin, including NN-based perception, state estimation, and policy models, which motivates the fine-tuning approaches discussed in the following section.

\subsubsection{Efficient Fine-Tuning for Online Adaptation}
\label{sec:peft}

When changes in system behavior cannot be accommodated through recurrent estimation of a compact parameter set or through an external bias model, adaptation may require modification of the internal representations of a learned model. For DNN-based components of a Digital Twin, full online fine-tuning provides this flexibility but can be computationally expensive and may interfere with previously acquired knowledge. A simple alternative is head fine-tuning, where only the final layers of the network are updated using one or a small batch of observations collected during operation \cite{gasparino2023unmatched,wang2026adajepa}. Although effective under incremental drift, head fine-tuning provides limited adaptability when abrupt changes alter system behavior in ways that cannot be represented by modifying the output mapping alone \cite{chen2026continual}. Because the internal representations and temporal features remain fixed, the required correction may lie outside the feature space learned under the original operating conditions. This motivates efficient adaptation of selected internal components of the network while retaining most of the previously learned representation.

\begin{figure}[h!]
    \centering
    \includegraphics[width=1\linewidth]{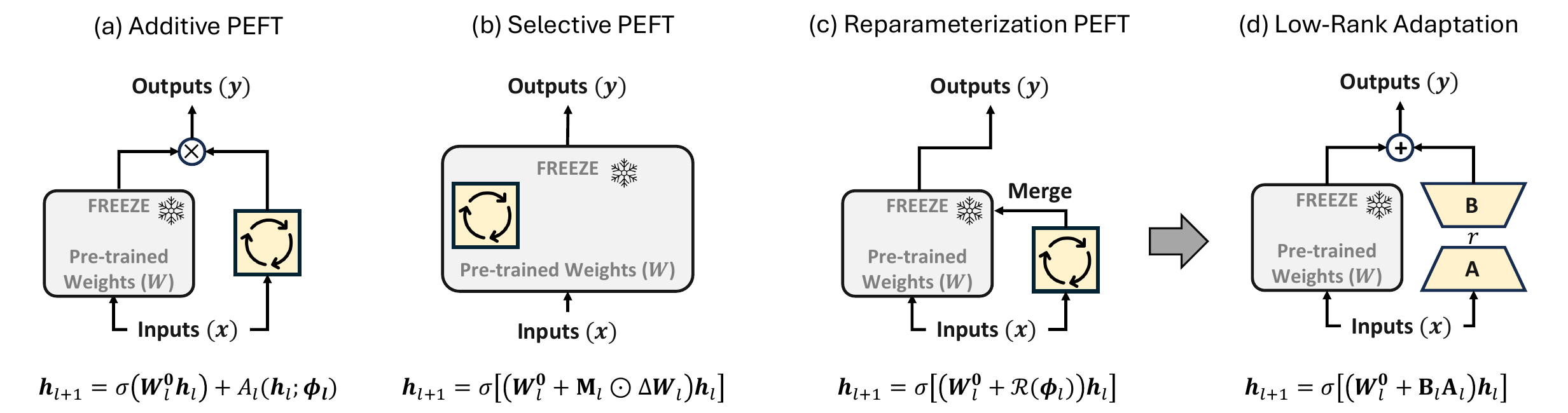}
    \caption{Layer-level illustration of representative parameter-efficient fine-tuning (PEFT) strategies. Here, $\mathbf{h}_l$ and $\mathbf{h}_{l+1}$ denote the input and output of layer $l$, $\mathbf{W}_l^0$ denotes the pretrained weight matrix, and $\sigma(\cdot)$ denotes the nonlinear activation function. The figure illustrates how PEFT modifies the computation of an individual neural network layer rather than the input-output mapping of the complete model. (a) Additive PEFT introduces a trainable module $A_l(\mathbf{h}_l;\boldsymbol{\phi}_l)$ into the layer computation while keeping the pretrained weights fixed. (b) Selective PEFT updates only a selected subset of the pretrained parameters through the mask $\mathbf{M}_l$, while the remaining weights remain frozen. (c) Reparameterization PEFT represents the weight update through a compact trainable mapping $\mathcal{R}(\boldsymbol{\phi}_l)$ that can be merged with the pretrained weights. (d) Low-Rank Adaptation (LoRA) realizes reparameterization by expressing the weight update as $\Delta\mathbf{W}_l=\mathbf{B}_l\mathbf{A}_l$, where the rank $r$ is much smaller than the dimensions of the original weight matrix.}
    \label{fig:PEFT}
\end{figure}

Parameter-efficient fine-tuning (PEFT) provides a family of methods for achieving this objective by freezing most pretrained parameters and adapting only a small number of selected or additional parameters \cite{han2024parameter,ding2023parameter}. Although PEFT has been developed extensively for large pretrained neural networks and has gained prominence through its use in LLMs, its underlying principle is directly applicable to DNN-based Digital Twin components that must adapt continually under limited computational resources and streaming data. Head fine-tuning can be viewed as a simple selective strategy in which trainable parameters are restricted to the final layers. More general PEFT methods extend this principle by introducing compact trainable modules, selecting parameters throughout the network, or constraining parameter updates through compact reparameterizations. As illustrated in Figure~\ref{fig:PEFT}, these methods can be broadly grouped into additive, selective, and reparameterization approaches. Additive methods introduce small trainable modules into the computation of a frozen network, as in adapter layers \cite{houlsby2019parameter}. Selective methods update only a subset of the existing parameters. For example, FishMask identifies parameters with large Fisher information and constructs a mask that determines which parameters are updated for a new task \cite{sung2021training}. Reparameterization methods instead constrain the update of pretrained parameters through a compact parameterization that can often be merged with the original weights for inference. A representative method is Low-Rank Adaptation (LoRA) \cite{hu2022lora}, which represents the update of a weight matrix through two trainable low-rank matrices while keeping the pretrained weights frozen. This substantially reduces the number of trainable parameters and the memory required during adaptation. In \cite{chen2026continual}, LoRA is integrated with model predictive control to update an NN-based surrogate after drift detection, demonstrating superior predictive accuracy under both sudden and incremental shifts compared with head fine-tuning.

The same principle can be extended from predictive models to learned decision components. When a policy, value function, or critic is represented by a DNN, updating the entire network online can be computationally expensive and sample inefficient, whereas PEFT restricts adaptation to a compact set of trainable parameters while preserving most of the existing representation. For example, \cite{zhuang2026low} apply LoRA to critic learning in reinforcement learning by freezing the base weight matrices and optimizing only low-rank updates, demonstrating that parameter-efficient adaptation can also be introduced within decision learning architectures. Direct application of PEFT to continually adapting control policies in Digital Twins remains relatively underexplored, providing an opportunity to extend efficient adaptation beyond predictive models to the decision layer.

Beyond computational efficiency, PEFT can also reduce interference with previously learned representations because most of the pretrained backbone remains unchanged while newly acquired information is absorbed by a compact set of trainable parameters. Maintaining multiple regime-specific adaptation modules further provides a natural connection to continual learning in nonstationary environments \cite{parisi2019continual,cossu2021continual}. When previously encountered operating regimes recur, previously validated modules could be retrieved and reused rather than learned again from newly collected data. 

Realizing this capability in Digital Twins, however, requires reliable mechanisms for identifying the active operating regime, selecting or configuring the appropriate adaptation module, and validating the adapted model before it is deployed in the decision loop. More broadly, the integration of PEFT with UQ, drift detection, and continual validation remains largely unexplored. Important open questions include how to determine which layers or parameters should be adapted, how much adaptation is sufficient under limited streaming data, how predictive uncertainty changes after an update, and how predictive and decision models can be adapted without compromising latency, stability, generalization, and safety.

Across these adaptation mechanisms, the appropriate update should depend on the source and structure of the detected mismatch rather than computational efficiency alone. Recursive parameter estimation is preferable when the required adaptation can be represented through a compact and identifiable parameterization, bias correction provides additional flexibility when a trusted nominal model remains available, and fine-tuning becomes necessary when the learned representation itself must change. Regardless of the mechanism used, an adapted component should not be returned directly to operation without reassessing its validity and its effect on the coupled Digital Twin. Continual adaptation therefore closes the operational assurance loop introduced in this section, linking UQ, drift detection, model updating, and post-update validation into a continuous process for maintaining reliable decision-making as the physical system evolves.

\section{Perspective and Future Research Opportunities}
\label{sec:future}

Although individual Digital Twin components have been widely investigated, a system-centric perspective on their design and evaluation is lacking in most of the literature. The future of Digital Twins therefore lies in moving from isolated algorithmic advances toward integrated, adaptive, and trustworthy systems for real-time decision-making. Building on the challenges identified throughout this review, we highlight four interconnected directions: surrogate modeling for decision support, continual adaptation under evolving conditions, scalable and intelligent decision-making, and system-level design and trustworthiness. Across these directions, we emphasize the need for methods that co-develop models, data acquisition, decisions, and system architectures so that Digital Twins remain reliable, adaptive, and operational throughout their lifecycle. Each open challenge is summarized as a research question (RQ).

\subsection{Surrogate Modeling}
\label{sec:future_surrogate}

Future surrogate models for Digital Twins should be designed around their role in the complete decision loop rather than their standalone predictive accuracy.

\subsubsection{From Prediction-Focused to System-Focused}

\begin{rqbox}
\textbf{RQ:} How can data generation, surrogate training, and evaluation be co-designed to balance predictive fidelity with decision performance, uncertainty calibration, computational efficiency, and continual adaptability within an operational Digital Twin?
\end{rqbox}

Although substantial research on Digital Twins has focused on improving the accuracy of surrogate models, predictive accuracy alone does not ensure that a surrogate is suitable for decision-making in Digital Twins. From a \emph{system-centric} perspective, the requirements of a surrogate are defined by its role, interfaces, and influence on the broader Digital Twin, such as quantifying uncertainty in a form useful for downstream decisions, remaining amenable to continual validation and adaptation, and satisfy real-time computational constraints.

This perspective is closely related to \emph{decision-focused learning} (DFL)~\cite{mandi2024decision}. In conventional prediction-focused, or two-stage, learning, the model $\hat{f}_{\boldsymbol{\theta}}$ is trained independently of the downstream optimization problem by minimizing a prediction loss such as the mean squared error. Minimizing prediction error, however, does not necessarily minimize the degradation of the resulting decisions. DFL instead trains the model on the regret induced by its predictions,
\begin{equation}
    \mathcal{L}_{\mathrm{DFL}}
    =
    \frac{1}{N}\sum_{i=1}^{N}
    \left[
    J\!\left(
    \mathbf{u}^{\star}\!\left(\hat{f}_{\boldsymbol{\theta}}(\mathbf{x}_i)\right),
    \mathbf{y}_i
    \right)
    -
    J\!\left(
    \mathbf{u}^{\star}(\mathbf{y}_i),
    \mathbf{y}_i
    \right)
    \right],
\end{equation}
where $J$ is the decision objective and $\mathbf{u}^{\star}(\cdot)$ denotes the solution of the downstream optimization problem. Prediction errors with little influence on the decision are thus tolerated, whereas errors that substantially alter decision quality are penalized.

We suggest extending this principle toward \emph{system-centric learning}, in which the development of the surrogate model is posed as a multi-objective or constrained learning problem that balances predictive fidelity against decision performance, UQ, continual validation, and adaptability, with computational cost entering as an additional term or as a constraint set by the real-time requirement. The appropriate terms and weights depend on the role of the surrogate within the Digital Twin. This view further implies that data generation, training, and evaluation should be co-designed according to the operational requirements of the Digital Twin rather than developed as independent activities, a principle extended to the full physical--digital architecture in Section~\ref{sec:co_design}.

\subsubsection{From Interpretability to Mechanistic and Causal Surrogate Modeling}

\begin{rqbox}
\textbf{RQ:} How can mechanistic and causal structure be integrated with data-driven surrogate models so that interpretability directly supports Digital Twin decision-making, validation, diagnosis, and adaptation?
\end{rqbox}

Interpretability should serve an operational role within a Digital Twin rather than only provide post hoc explanations of model predictions. Engineers need to know which variables drive predictions and decisions, whether these relationships are physically plausible, and how they change as the system evolves. Hybrid models that embed known physical structure, conservation laws, or causal dependencies into the learned representation can ground predictions in physically meaningful quantities, while data-driven components capture unresolved behavior. Such models support validation, fault diagnosis, drift attribution, and the selection of appropriate adaptation mechanisms.

Structured probabilistic models already represent dependencies among physical states, observations, model states, actions, and outcomes~\cite{kapteyn2021probabilistic, chaudhuri2023predictive}. Extending them toward causal models would enable interventional and counterfactual reasoning about why an outcome occurs and how a proposed action would change it. However, physical and causal explanations currently remain largely analytical outputs that are weakly connected to the real-time decision loop, and methods to translate them quantitatively into decisions, validation, and adaptation are still lacking.

\subsubsection{Data Acquisition and Reuse}

\begin{rqbox}
\textbf{RQ:} How can simulation, physical experiments, and knowledge accumulated across assets and tasks be jointly exploited to acquire and transfer decision-relevant information under differences in cost, fidelity, and observability, while preserving data ownership and real-time deployability?
\end{rqbox}

Acquiring sufficiently informative data is often the most expensive part of developing ML-enabled Digital Twins, while data collection from physical experiments are often restricted by sparsity, data quality, and is often restricted by its reachability. The central challenge is therefore not to collect more data, but to acquire the \emph{right} data efficiently, and to organize or accumulate them into reusable knowledge. Simulation provides scalable coverage of the operating and design space, while physical experiments ground the Digital Twin in actual system behavior and reduce the sim-to-real discrepancy arising from simplified physics, uncertain parameters, and idealized boundary conditions. Multi-fidelity learning, active learning, and optimal experimental design offer complementary mechanisms for deciding which information to generate computationally and which experiments justify their cost. Integrating these sources nevertheless remains difficult when they differ in modality, fidelity, observability, and uncertainty, often under highly biased data distributions.

We suggest that knowledge accumulated across operating regimes and tasks should be reused rather than rebuilt from limited data for individual models. For instance, shared representations through joint embeddings \cite{assran2025v}, multi-task learning \cite{DEramo2020SharingKI},  and foundation models \cite{bommasani2021opportunities} can be potential solutions to learn and reuse features from heterogeneous data sources, and federated learning \cite{zhuang2023foundation} can aggregate information across assets or organizations without centralizing data. Along this direction, the remaining open challenges include learning unified representations/feature extraction from heterogeneous engineering data with the least information lost, accommodating continual change across distributed systems, and data provenance. Moreover, model distillation, parameter-efficient adaptation, and few-shot learning may further be needed to deploy compact, task-specific models under real-time constraints.

\subsection{Continual Adaptation}
\label{sec:challenge_in_continual_adaptation}

Future research on continual adaptation should move beyond isolated model updating toward maintaining reliability as a continuous, system-level process within the Digital Twin decision loop.

\subsubsection{Online Adaptation and Validation with Limited Streaming Data}

\begin{rqbox}
\textbf{RQ:} How can online adaptation and validation be conducted to enable rapid, data-efficient model updates while maintaining stability and sufficient confidence under limited and temporally correlated streaming data?
\end{rqbox}

As new observations arrive, a Digital Twin must determine whether adaptation is needed, update the model efficiently, and validate the update before it affects downstream prediction or decision-making. Depending on model structure, parameter identifiability, and feedback quality, adaptation can require substantial data, and the resulting trade-off among adaptation speed, data requirements, and model stability is still commonly managed heuristically. Validation is further limited because the updated model must often be evaluated on the same short, temporally correlated data stream, without the independent holdout data available offline.

We suggest that adaptation and validation should therefore be constructed for data-limited operation. Physics-based knowledge can constrain the admissible model changes, reformulating broad uncertainty in a learned representation as lower-dimensional parameter or model-discrepancy uncertainty, so that costly fine-tuning can be replaced by parameter estimation or discrepancy correction when appropriate. Validation metrics should account for temporal dependence and limited coverage rather than directly reusing the metrics developed for offline model validations. Live--idle or champion--challenger architectures \cite{chen2026continual} can further safeguard deployment by evaluating an updated model alongside the deployed one before admitting it into the decision loop. A key open problem is establishing statistically grounded criteria for when sufficient evidence has accumulated to accept an update.

\subsubsection{Continual Adaptation under Evolving Observation Spaces}

\begin{rqbox}
\textbf{RQ:} How can Digital Twin models continually adapt to evolving observation spaces while preserving learned knowledge and maintaining observability, uncertainty characterization, validation, and decision performance?
\end{rqbox}

Continual adaptation commonly assumes that the model input structure remains fixed while the data distribution changes. In long-term operation, however, the observation space itself may evolve as sensors are added or removed, new process variables become measurable, sensor malfunctions occur, or new operating modes expose previously unavailable information. Data-driven surrogates are difficult to adapt to such changes because their input structure is fixed at training time, and changes in the observation space can also alter the observability and identifiability of states and parameters, weakening previously established uncertainty estimates and validation evidence.

To address this gap, we suggest that the input for the surrogate model needs to be defined over a variable set of inputs rather than a fixed vector, or to perform multi-view learning \cite{yu2025review}. Architectures that treat each measurement as an element of a set, trained with masking techniques \cite{assran2025v}, with an embedding describing its physical meaning, location, or units, can accommodate a changing number of inputs without altering the model structure, and allow a new sensor to be interpreted from its description rather than learned from scratch.

\subsubsection{Drift and Operational-Domain Monitoring under Partial Observability}
\label{sec:domain_monitoring}

\begin{rqbox}
\textbf{RQ:} How can drift and operational-domain monitoring accumulate evidence from limited, partially observable, and decision-influenced feedback to identify consequential changes and determine whether conservative operation, information acquisition, validation, or component-level adaptation should be triggered?
\end{rqbox}

Concept drift monitoring and OOD detection (e.g., covariate drift monitoring) often assume that sufficient feedback is available to judge whether the current model remains valid. In practice, many states and quantities of interest are only partially observable, so that the ground-truth labels may be delayed or unavailable. Thus, drift in the physical system may go undetected, and adapting directly from incomplete feedback can introduce further error. Moreover, even seemingly minor distributional shifts that move operational data outside the training distribution can substantially degrade the performance of data-driven models used across multiple Digital Twin components. Because such conditions cannot be fully anticipated or validated before deployment, monitoring is needed that targets the operational domain of each component and of the integrated system.

We suggest that the monitoring mechanisms should be included while designing the continual validation process that accumulates evidence from multiple indirect sources, such as state prediction and decision outcomes, while accounting for their respective uncertainty and observability. Information acquisition and redundant sensor placement can be integrated so that sensing and operating actions also improve the evidence available for drift diagnosis. The objective is not only to detect that the system has drifted, but to determine whether it compromises the validated operating domain, which model component can be reliably adapted, and whether robust decision-making, additional data collection, validation, or adaptation should be triggered.

\subsection{Challenges in Real-Time Decision-Making}
\label{sec:future_decision}

As engineering systems grow more complex, real-time decision-making faces challenges in computational scalability, in keeping learned policies coupled to the evolving virtual twin, and in orchestrating agentic and human decision authority.

\subsubsection{Scalable Online Optimization}

\begin{rqbox}
\textbf{RQ:} How can optimization be scaled for high-dimensional, nonlinear problems alongside evolving Digital Twin models to maintain solution quality and convergence under real-time computational requirements?
\end{rqbox}

Optimization-based decision-making can become computationally prohibitive as Digital Twins grow in dimensionality, coupling, and decision frequency. We suggest that scalable optimization can be improved in two directions. The first is to accelerate online solving through learning-based initialization paired with the surrogate model for optimization, as reviewed in Section~\ref{sec:accelerate_solving}: learned models provide high-quality initial solutions that reduce solution time, while solving using surrogates retains the constraint handling and convergence properties of the underlying optimizer. However, warm-start models are typically trained offline and may lose effectiveness as the physical system, surrogate model, or feasible region evolves, so the predictive model, optimization problem, and warm-start model must be adapted together without excessive online computation. The second direction is to exploit the subsystem or hierarchical structure of complex systems. Decomposition and multidisciplinary design optimization (MDO) divide a coupled system-level problem into smaller subproblems while coordinating shared variables, but how to partition and coordinate a problem to balance solution quality, convergence, communication, and real-time requirements remains open.

\subsubsection{Integrating Policy-Based Decision-Making with Adaptive Digital Twins}

\begin{rqbox}
\textbf{RQ:} How can the adaptation of surrogate models be efficiently integrated with a policy so that decision behavior remains consistent with the updated Digital Twin without requiring risky or data-intensive online policy learning?
\end{rqbox}

Policy-based methods, such as model-free reinforcement learning, generative policies, and approximate MPC, reduce online computational cost by mapping states or observations directly to actions. Many such approaches, however, use the physical or simulated model only during offline training, so the virtual representation is completely removed from the online decision loop. This creates a fundamental tension with the definition of a Digital Twin as a continuously synchronized coupling between virtual and physical systems, and it becomes critical when the Digital Twin must adapt during operation. While the surrogate can be updated directly from sensor feedback, as reviewed in Section~\ref{sec:continual_Adaptation}, updating a policy may require new interactions, reward observations, or on-policy exploration that is data-intensive and potentially unsafe on the physical system.

A key direction is to leverage model adaptation into the policy without extensive online policy learning. Two mechanisms reviewed in this work are promising. In model-based RL, the adapted virtual model serves as the \emph{gym} in which the policy is updated before redeployment, moving policy adaptation from the physical to the virtual space (Section~\ref{sec:RL}). Alternatively, a policy trained offline across a family of system models can be conditioned on estimated physical parameters or operating conditions, reducing policy adaptation to parameter estimation and enabling zero- or few-shot adaptation~\cite{iqbal2025zero}. When these strategies provide sufficiently rapid and reliable adaptation remains to be established.

\subsubsection{Adaptive Goal-to-Action Decision-Making with Agentic AI}

\begin{rqbox}
\textbf{RQ:} How can agentic AI in Digital Twins revise goals and plans as system states and constraints evolve, and translate them into computational workflows that can be validated and repaired as the availability and validity of models and tools change?
\end{rqbox}

Agentic systems are given greater freedom to interpret goals, organize computation, and determine actions within a Digital Twin, which leaves open gaps in coordination, validation, and safe execution. Current agentic Digital Twins coordinate local agents, forecasting and optimization modules, and central decision agents, but largely rely on predefined workflows and provide limited evidence of autonomous goal revision or long-horizon replanning. Two capabilities are therefore needed. First, engineering goals should be grounded in measurable Digital Twin states, so that changes in states, constraints, or priorities can trigger plan revision while preserving system-level feasibility across agents. Second, revised plans must be compiled into computational workflows over tools that differ in input requirements, operating domains, and reliability. Machine-readable descriptions of tool capabilities and validity conditions, together with validation of intermediate results, would allow an agent to repair a workflow by requesting additional data, substituting a model, or returning to the planning layer when the original plan is no longer executable. Whether and when an agent should act on the physical system is discussed jointly with human oversight in Section~\ref{sec:decision_authority}.

\subsubsection{Decision Authority and Human-in-the-Loop Operation}
\label{sec:decision_authority}

\begin{rqbox}
\textbf{RQ:} How can decision authority be adaptively allocated among human operators, agentic components, and automated execution according to model validity, uncertainty, safety constraints, and the consequences of proposed actions?
\end{rqbox}

Human expertise remains essential for decisions involving incomplete information, competing objectives, safety concerns, and contextual knowledge that is difficult to encode mathematically. This requires the translation from human knowledge and preferences to design objectives, constraints, or candidate actions, while predictions, uncertainty, and recommendations must be communicated back so that operators can evaluate and revise them. LLMs provide a promising interface by connecting natural-language instructions, maintenance records, failure modes, and operating procedures with Digital Twin models and decision tools, provided their outputs remain grounded in verified models, optimization algorithms, and uncertainty estimates rather than serving as an independent source of physical decisions.

A central challenge is how decision authority should be distributed among human operators, agentic components, and automated execution. Fully manual approval limits the speed and autonomy of real-time operation, whereas unrestricted execution is inappropriate when model validity is uncertain. Current systems can screen candidate actions in the Digital Twin before execution, but the conditions governing whether to act are generally predefined. Adaptive architectures are needed in which the choice to execute or escalate to a human is determined adaptively.

\subsection{Challenges in System Design and Trustworthiness}
\label{sec:future_system}

Future research should treat the Digital Twin as an integrated engineering system rather than a collection of independently developed components, preserving system-level traceability and rigor as its components become increasingly adaptive, and as AI becomes a component of the system. Operational-domain monitoring, a key element of trustworthiness, is discussed in Section~\ref{sec:domain_monitoring}. This subsection focuses on architecture, emergent behavior, and lifecycle assurance.

\subsubsection{System Design and Integration through Co-Design}
\label{sec:co_design}

\begin{rqbox}
\textbf{RQ:} How can the physical and digital components of a Digital Twin be co-designed, both before deployment and during operation, to balance system performance, sensing, model fidelity, computation, resilience, and decision performance throughout the lifecycle?
\end{rqbox}

As Digital Twins connect physical assets, sensing, computation, models, communication, and decision-making, their architectures should be co-designed rather than developed sequentially, component by component. Decisions on physical configuration, sensor placement, model fidelity, computational resources, information flow, redundancy, and control architecture are strongly coupled, so improving one component in isolation may yield limited system-level benefit or introduce additional computational cost, communication requirements, or unprecedented failure modes.

We suggest that co-design should therefore extend beyond established couplings, such as plant--control design, toward the broader physical--digital architecture. Systems engineering provides a structured foundation for requirement definition and interface specification, while multi-objective optimization supports trade-space exploration across subsystem performance, resource allocation, and lifecycle constraints. A further direction is to extend co-design into online operation, dynamically allocating sensing and computational resources across subsystems and determining which components remain fixed, inactive, or reconfigurable, so that computational budget, information flow, and the trade-offs between model fidelity and decision-making, and even the objective functions/constraints, can adapt as conditions and objectives evolve.

\subsubsection{Emergent Behavior Handling}

\begin{rqbox}
\textbf{RQ:} How can Digital Twins detect and attribute emergent system-level behavior arising from interactions among individually validated components, and determine appropriate mitigation or reconfiguration actions during operation?
\end{rqbox}

As Digital Twins become more tightly coupled and autonomous, unexpected behavior may emerge from interactions among components, even when each component operates within its individually validated domain. Such \emph{emergent behavior} differs from OOD events because it may not originate from a novel input or an identifiable component failure, but from feedback across the integrated system. The challenge intensifies when multiple adaptive components evolve simultaneously, since local updates can change system-level behavior in ways not observed during component-level validation. As emphasized by the National Academies, the safety of an ML-enabled system cannot be inferred from the performance of its individual ML components alone, and requires application-centric evaluation of the integrated system across its lifecycle~\cite{nationalacademies2025machine}.

We suggest that more methods should be developed that focus on detecting, attributing, and mitigating emergent behavior at the system level. By monitoring interactions among components and subsystems, a Digital Twin may identify unexpected behavior, diagnose which components or interactions are responsible, and select an appropriate response. LLM-enabled reasoning may support this process by combining system context, operational history, and domain knowledge to generate mitigation strategies for scenarios not anticipated during development, extending adaptation beyond low-level model or control updates toward coordinated diagnosis, adaptation, and reconfiguration.

\subsubsection{A Systems Engineering-Oriented Approach toward Lifecycle Assurance and Certification}

\begin{rqbox}
\textbf{RQ:} How can systems engineering be extended to AI-enabled Digital Twins, and supported by MBSE and AI-assisted design automation, to maintain traceability among requirements, architecture, component performance, and lifecycle assurance when component behavior and failure modes are data-dependent?
\end{rqbox}

The development of a trustworthy Digital Twin should be treated as a systems engineering problem rather than the ad hoc integration of sensing, models, computation, and decision-making modules. System-level requirements must be allocated across components, and interfaces and dependencies among physical and digital elements must be verified, validated, and, where required, certified. Systems engineering thus provides a foundation for defining the Digital Twin architecture and establishing the verification, validation, test, and evaluation plan needed for system-level assurance.

Conventional systems engineering, however, becomes more difficult to apply when ML models are system components. Unlike mechanical components specified through explicit functional requirements, ML models are shaped by training data, statistical assumptions, and deployment distributions, so their behavior and failure modes are difficult to specify completely through evaluation protocols. Systems engineering for AI (SE4AI)~\cite{mcdermott2020ai4se} is therefore needed to establish systematic verification, validation, testing, and evaluation of AI-enabled Digital Twins.

At the same time, AI offers an opportunity to make Digital Twin development itself more automated and iterative. It is observed that integrating Model-Based Systems Engineering (MBSE) with Digital Twin development allows requirements, documentation, computational tools, and verification artifacts to be represented as machine-readable system models and ontologies rather than disconnected documents. Combined with the cross-domain reasoning of LLMs, the orchestration capabilities of agentic AI, and design automation, these tools could support an end-to-end pipeline spanning requirement extraction, architecture generation, interface checking, model selection, test generation, and traceability of design changes. The central challenge is to accelerate development in this way without weakening traceability, verification, or engineering accountability.

\section{Closure}
\label{sec:closure}

This review set out to understand how Digital Twins can support reliable online operation and real-time decision-making. Following the National Academies' definition, a Digital Twin is dynamically updated by its physical counterpart, possesses predictive power, and drives decisions that create real-world value through two-way interaction. We therefore treated decision-making as the defining function of a Digital Twin rather than an application built on top of it. From this perspective, we categorized the major decision-making approaches that close the physical-to-virtual-to-physical loop and identified the requirements each imposes on sensing, state estimation, predictive modeling, and computation. We then showed that ML enables this loop well beyond surrogate modeling, from transforming physical feedback into decision-relevant information to accelerating or directly generating decisions. Finally, we examined how UQ, continual validation, drift detection, and model adaptation jointly determine whether a Digital Twin remains valid and when it must be updated as the physical system evolves. Across these findings, a common conclusion emerges: a Digital Twin is not a model but a system, whose value arises from how its components work together in the closed decision loop rather than from any component alone. Therefore, each component should be designed and evaluated by its contribution to the decisions of the integrated system, rather than by its standalone performance. We also acknowledge that this review focuses on the capabilities and their integration required for reliable closed-loop decision-making, rather than comprehensively addressing the broader systems engineering challenges of Digital Twin implementation, interoperability, cybersecurity, and lifecycle management.

Looking ahead, the progress of Digital Twins will depend on shared foundations as much as on new methods. Because a Digital Twin spans a wide range of disciplines, a common testbed, whether virtual or low-cost hardware, that is simple enough to be widely accessible yet rich enough to reflect real engineering systems would allow new capabilities to be developed and benchmarked at the system level. Such foundations would, in turn, enable convincing demonstrations that Digital Twins can be designed and maintained systematically throughout their lifecycle. Further ahead, as LLMs grow in reasoning and learning capability, Digital Twins may interact not only with surrogate models but with an ontology of the system itself, gaining the flexibility to reformulate models, reason about decisions, and propose new hypotheses. In this regard, the future Digital Twins would then evolve from autonomous systems that execute decisions into true intelligent systems.

\section*{Acknowledgment}

We are grateful for the grant support from the National Science Foundation's Engineering Research Center for Hybrid Autonomous Manufacturing: Moving from Evolution to Revolution (ERC-HAMMER), under Award Number EEC-2133630, the Manufacturing ADvanced Electronics through Printing Using Biobased and Locally Identifiable Compounds (MADE PUBLIC) Future Manufacturing Research Grant Program under Award Number CMMI-2037026, the support from NASA’s Minority University Research and Education Project Institutional Research Opportunity (MIRO) through the Center for In-Space Manufacturing (CISM-R2): Recycling and Regolith Processing (Award 80NSSC24M0176), and the support from the NSF Boosting Research Ideas for Transformative and Equitable Advances in Engineering (BRITE) Fellow Program (CMMI-2227641). Yi-Ping Chen acknowledges the Taiwan-Northwestern Doctoral Scholarship, and Seul Lee acknowledges the Samsung Electronics Doctoral Scholarship.

\bibliography{reference}
\bibliographystyle{elsarticle-num-names}

\end{document}